\documentclass{paper}
\usepackage{url}
\usepackage{tikz}
\usepackage{multirow}
\usepackage{pgf}
\usepackage{pgfplots}
\usepackage{colortbl}
\usepackage{tabularx} 
\usepackage{caption}
\usepackage{subfig} 
\usepackage{mathtools}
\usepackage{appendix}
\usepackage{pdflscape}
\usepackage{xtab}
\usepackage[named]{algo}
 \usepackage{amsmath}
 \PassOptionsToPackage{pdftex}{graphicx}
\usepackage{graphicx}

\begin{document}
\title{Dependencies: Formalising Semantic Catenae for Information Retrieval}
\subtitle{Disputats for the Degree of Doctor Scientiarum (dr.scient.)}
\author{Christina Lioma \\
Department of Computer Science,\\ University of Copenhagen, Denmark\\ March 2017}
\maketitle

\begin{abstract}
Building machines that can understand text like humans is an AI-complete problem. A great deal of
research has already gone into this, with astounding results, allowing everyday people to \textit{discuss} with their
telephones, or have their reading materials analysed and classified
by computers. A
prerequisite for processing text semantics, common to the above examples, is having some computational
representation of text as an abstract object. Operations on this representation practically correspond to
making semantic inferences, and by extension simulating \textit{understanding text}. The complexity and
granularity of semantic processing that can be realised is constrained by the mathematical and computational
robustness, expressiveness, and rigour of the tools used. 

This dissertation contributes a series of such tools, diverse in their mathematical formulation, but common in their application to model semantic inferences when machines process text. These tools are principally expressed in nine distinct models that capture aspects of semantic dependence in highly interpretable and non-complex ways. This dissertation further reflects on present and future problems with the current research paradigm in this area, and makes recommendations on how to overcome them. 

The amalgamation of the body of work presented in this dissertation advances the complexity and
granularity of semantic inferences that can be made automatically by machines.

\end{abstract}

\section*{Foreword}
This document is a \textit{doktordisputats} - a dissertation within the Danish academic system required to obtain the degree of \textit{Doctor
Scientiarum}, in form and function equivalent to the French and German
Habilitation and the Higher Doctorate of the Commonwealth.

The dissertation contains my work in the field of semantic dependence for information retrieval,
realised in the period January 2009 - April 2017. 

The first chapter of this dissertation consists of an executive summary that introduces the general area and gives a compact overview of the specific contributions of this work. This chapter is aimed at readers with a broad background in computer science or related disciplines.
The remainder of the dissertation consists of nine technical chapters that are slightly reformatted versions of previously published papers.

\section*{Publications} 

The following published papers have been included in the text of
this dissertation. The papers are listed in the order they appear in the dissertation:

\begin{enumerate}
\item Roi Blanco and Christina Lioma. \textbf{Graph-based Term Weighting for Information Retrieval}. In: \textit{Information Retrieval}. Springer, 2012. Vol. 15, no. 1, pp. 54--92. issn: 1386-4564. \\doi: 10.1007/s10791-011-9172-x. \\url: \url{http://dx.doi.org/10.1007/s10791-011-9172-x}.

\item Christina Lioma, Jakob Grue Simonsen, Birger Larsen, and Niels Dalum Hansen. \textbf{Non-Compositional Term Dependence for Information Retrieval}. In: \textit{Proceedings of the 38th International ACM SIGIR Conference on Research and Development in Information Retrieval, Santiago, Chile, August 9-13, 2015}. Ed. by Ricardo A. Baeza-Yates, Mounia Lalmas, Alistair Moffat, and Berthier A. Ribeiro-Neto. ACM, 2015, pp. 595--604. \\isbn: 978-1- 4503-3621-5. \\doi: 10.1145/2766462.2767717. \\url: \url{http://doi.acm.org/10.1145/2766462.2767717}

\item Christina Lioma and Niels Dalum Hansen. \textbf{A Study of Metrics of Distance and Correlation Between Ranked Lists for Compositionality Detection}. In: \textit{Journal of Cognitive Systems Research}. Elsevier, 2017, in press. 
\\doi: 10.1016/j.cogsys.2017.03.001. 

\item Christina Lioma, Birger Larsen, and Wei Lu. \textbf{Rhetorical Relations for Information Retrieval}. In: \textit{The 35th International ACM SIGIR conference on research and development in Information Retrieval, SIGIR'12, Portland, OR, USA, August 12--16, 2012}. Ed. by William R. Hersh, Jamie Callan, Yoelle Maarek, and Mark Sanderson. ACM, 2012, pp. 931--940. \\isbn: 978-1-4503-1472-5. \\doi: 10.1145/2348283.2348407. \\url: \url{http://doi.acm.org/10.1145/2348283.2348407}.

\item Casper Petersen, Christina Lioma, Jakob Grue Simonsen, and Birger Larsen. \textbf{Entropy and Graph Based Modelling of Document Coherence using Discourse Entities: An Application to IR}. In: \textit{Proceedings of the 2015 International Conference on The Theory of Information Retrieval, ICTIR 2015, Northampton, Massachusetts, USA, September 27-30, 2015}. Ed. by James Allan, W. Bruce Croft, Arjen P. de Vries, and Chengxiang Zhai. ACM, 2015, pp. 191--200. \\isbn: 978-1-4503-3833-2. \\doi: 10.1145/2808194.2809458. \\url: \url{http://doi.acm.org/10.1145/2808194.2809458}.

\item Christina Lioma, Fabien Tarissan, Jakob Grue Simonsen, Casper Petersen, and Birger Larsen. \textbf{Exploiting the Bipartite Structure of Entity Grids for Document Coherence and Retrieval}. In: \textit{Proceedings of the 2016 ACM on International Conference on the Theory of Information Retrieval, ICTIR 2016, Newark, DE, USA, September 12--6, 2016}. Ed. by Ben Carterette, Hui Fang, Mounia Lalmas, and Jian-Yun Nie. ACM, 2016, pp. 11--20. \\isbn: 978-1-4503-4497-5. \\doi: 10.1145/2970398.2970413. \\url: \url{http://doi.acm.org/10.1145/2970398.2970413}.

\item Christina Lioma, Roi Blanco, Raquel Mochales Palau, and Marie-Francine Moens. \textbf{A Belief Model of Query Difficulty That Uses Subjective Logic}. In: \textit{Advances in Information Retrieval Theory, Second International Conference on the Theory of Information Retrieval, ICTIR 2009, Cambridge, UK, September 10-12, 2009, Proceedings}. Ed. by Leif Azzopardi, Gabriella Kazai, Stephen E. Robertson, Stefan M. R{\"u}ger, Milad Shokouhi, Dawei Song, and Emine Yilmaz. Vol. 5766. Lecture Notes in Computer Science. Springer, 2009, pp. 92--103. 
\\isbn: 978-3-642-04416-8. 
\\doi: 10.1007/978-3-642-04417-5\_9. 
\\url: \url{http://dx.doi.org/10.1007/978-3-642-04417-5_9}.

\item Christina Lioma, Birger Larsen, Hinrich Sch{\"u}tze, and Peter Ingwersen. \textbf{A Subjective Logic Formalisation of the Principle of Polyrepresentation for Information Needs}. In: \textit{Information Interaction in Context Symposium, IIiX 2010, New Brunswick, NJ, USA, August 18--21, 2010}. Ed. by Nicholas J. Belkin and Diane Kelly. ACM, 2010, pp. 125--134. \\isbn: 978-1-4503-0247-0. \\doi: 10.1145/1840784.1840804. \\url: \url{http://doi.acm.org/10.1145/1840784.1840804}.

\item Christina Lioma, Birger Larsen, and Peter Ingwersen. \textbf{Preliminary Experiments using Subjective Logic for the Polyrepresentation of Information Needs}. In: \textit{Information Interaction in Context: 2012, IIix'12, Nijmegen, The Netherlands, August 21--24, 2012}. Ed. by Jaap Kamps, Wessel Kraaij, and Norbert Fuhr. ACM, 2012, pp. 174--183. \\isbn: 978-1-4503-1282-0. \\doi: 10.1145/2362724. 2362755. \\url: \url{http://doi.acm.org/10.1145/2362724.2362755}.

\end{enumerate}

The papers presented below are published in the period January 2009 -- April 2017 and are unrelated or marginally related to my research on semantic dependence for information retrieval. For this reason, they are \textit{not} included in the dissertation. The list is shown here to outline the context of the research presented in this dissertation and to indicate
the broadness of my research. The papers are presented in increasing chronological order:

\begin{enumerate}
\item Roi Blanco and
               Christina Lioma. \textbf{Mixed Monolingual Homepage Finding in 34 Languages: the Role of Language
               Script and Search Domain}. In: \textit{Journal of Information Retrieval, Special Issue on non-English Web Retrieval}. Springer, 2009. vol. 12, no. 3, pp. 324--351. Full reference: \cite{DBLP:journals/ir/BlancoL09}.

\item Christina Lioma and
               Roi Blanco. \textbf{Part of Speech Based Term Weighting for Information Retrieval}. In: \textit{Advances in Information Retrieval, 31th European Conference on {IR}
               Research, {ECIR} 2009, Toulouse, France, April 6-9, 2009. Proceedings}. 
               Springer, 2009, pp. 412--423. Full reference: \cite{LiomaB:2009ecir}.

\item Christina Lioma,
               Roi Blanco, and
               Marie{-}Francine Moens. \textbf{A Logical Inference Approach to Query Expansion with Social Tags}. In: \textit{Advances in Information Retrieval Theory, Second International Conference
               on the Theory of Information Retrieval, {ICTIR} 2009, Cambridge, UK,
               September 10-12, 2009, Proceedings}. 
Lecture Notes in Computer Science. Springer, 2009, pp. 358--361. Full reference: \cite{DBLP:conf/ictir/LiomaBM09}.

\item Charles Jochim,
               Christina Lioma, 
               Hinrich Sch{\"{u}}tze, Steffen Koch, and Thomas Ertl. \textbf{Preliminary Study into Query Translation for Patent Retrieval}. In: \textit{Proceedings of the 3rd International Workshop on Patent Information Retrieval {(PaIR '10)}}. ACM, 2010. pp. 57--66. Full reference: \cite{Jochim:2010:PSQ:1871888.1871899}.

\item Lukas Michelbacher,
               Alok Kothari,
               Martin Forst,
               Christina Lioma, and
               Hinrich Sch{\"{u}}tze. \textbf{A Cascaded Classification Approach to Semantic Head Recognition}. In: \textit{Proceedings of the 2011 Conference on Empirical Methods in Natural
               Language Processing, {EMNLP} 2011}, 
               pp. 793--803. Full reference: \cite{MichelbacherKFLS11}.

\item Charles Jochim,
               Christina Lioma, and
               Hinrich Sch{\"{u}}tze. \textbf{Expanding Queries with Term and Phrase Translations in Patent Retrieval}. In: \textit{Multidisciplinary Information Retrieval - Second Information Retrieval
               Facility Conference, {IRFC} 2011}, 
               pp. 16--29. Full reference: \cite{DBLP:conf/irfc/JochimLS11}.

\item Radu Dragusin,
               Paula Petcu,
               Christina Lioma,
               Birger Larsen,
               Henrik J{\o}rgensen and
               Ole Winther. \textbf{Rare Disease Diagnosis as an Information Retrieval Task}. In: \textit{Advances in Information Retrieval Theory - Third International Conference,
               {ICTIR} 2011}, 
               pp. 356--359. Full reference: \cite{DBLP:conf/ictir/DragusinPLLJW11}.

\item Christina Lioma,
               Alok Kothari, and
               Hinrich Sch{\"{u}}tze. \textbf{Sense Discrimination for Physics Retrieval}. In: \textit{Proceeding of the 34th International {ACM} {SIGIR} Conference on Research
               and Development in Information Retrieval, {SIGIR} 2011}, 
               pp. 1101--1102. Full reference: \cite{LiomaKS11}.

%

\item Christina Lioma,
               Birger Larsen, and
               Hinrich Sch{\"{u}}tze. \textbf{User Perspectives on Query Difficulty}. In: \textit{Advances in Information Retrieval Theory - Third International Conference,
               {ICTIR} 2011}, 
               pp. 3--14. Full reference: \cite{DBLP:conf/ictir/LiomaLS11}.

\item Wei Lu,
               Qikai Cheng, and
               Christina Lioma. \textbf{Fixed Versus Dynamic Co-Occurrence Windows in TextRank Term Weights for Information Retrieval}. In: \textit{The 35th International {ACM} {SIGIR} conference on research and development
               in Information Retrieval, {SIGIR} '12}, 
               pp. 1079--1080. Full reference: \cite{DBLP:conf/sigir/LuCL12}.
               
\item Birger Larsen,
               Christina Lioma,
               Ingo Frommholz, and
               Hinrich Sch{\"{u}}tze. \textbf{Preliminary Study of Technical Terminology for the Retrieval of Scientific Metadata Book Records}. In: \textit{The 35th International {ACM} {SIGIR} conference on research and development
               in Information Retrieval, {SIGIR} '12}, 
               pp. 1131--1132. Full reference: \cite{LarsenL12}.

\item Peter Ingwersen,
               Christina Lioma,
               Birger Larsen, and
               Peiling Wang. \textbf{An Exploratory Study into Perceived Task Complexity, Topic Specificity and Usefulness for Integrated Search}. In: \textit{Information Interaction in Context: 2012, IIix'12}, 
               pp. 302--305. Full reference: \cite{DBLP:conf/iiix/IngwersenLLW12}.

\item Raf Guns, Christina Christina Lioma, and Birger Larsen. \textbf{The Tipping Point: F-score as a Function of the Number of Retrieved Items}. In: \textit{Journal of Information Processing \& Management}. Elsevier, 2012. Vol. 48, no. 6, pp. 1171--1180. Full reference: \cite{DBLP:journals/ipm/GunsLL12}.

\item Radu Dragusin, Paula Petcu, Christina Lioma, Birger Larsen, Henrik J{\o}rgensen, Ingemar J. Cox, Lars K. Hansen, Peter Ingwersen, and Ole Winther. \textbf{Specialised Tools are Needed when Searching the Web for Rare Disease Diagnoses}. In: \textit{Rare Diseases}. 2013. Vol. 1, no. 2, pp: e25001-1--e25001-4. Full reference: \cite{DragusinPLLJCHIW13}.

\item Radu Dragusin, Paula Petcu, Christina Lioma, Birger Larsen, Henrik J{\o}rgensen, Ingemar J. Cox, Lars K. Hansen, Peter Ingwersen, and Ole Winther. \textbf{FindZebra: A Search Engine for Rare Diseases}. In: \textit{International Journal of Medical Informatics}. Elsevier, 2013. Vol. 82, no. 6, pp. 528--538. Full reference: \cite{DBLP:journals/ijmi/DragusinPLLJCHIW13}.

\item Casper Petersen, Christina Lioma, and Jakob Grue Simonsen. \textbf{Comparative Study of Search Engine Result Visualisation: Ranked Lists Versus Graphs}. In: \textit{Proceedings of the 3rd European Workshop on Human-Computer Interaction
               and Information Retrieval, Dublin, Ireland, August 1, 2013}, 
               pp. 27--30. Full reference: \cite{DBLP:conf/eurohcir/PetersenLS13}.

\item Roi Blanco, Manuel Eduardo Ares Brea, and Christina Lioma. \textbf{User Generated Content Search}. In: \textit{Mining of User Generated Content and its Applications}, 2013,
pp.167--187. Full reference: \cite{DBLP:books/crc/p/BlancoBL14}.

\item Niels Dalum Hansen,
               Christina Lioma,
               Birger Larsen, and
               Stephen Alstrup. \textbf{Temporal Context for Authorship Attribution - {A} Study of Danish
               Secondary Schools}. In: \textit{Multidisciplinary Information Retrieval - 7th Information Retrieval
               Facility Conference, {IRFC} 2014}, 
               pp. 22--40. Full reference: \cite{DBLP:conf/irfc/HansenLLA14}.

\item Casper Petersen, Jakob Grue Simonsen, and Christina Lioma. \textbf{The Impact of Using Combinatorial Optimisation for Static Caching
               of Posting Lists}. In: \textit{Information Retrieval Technology - 11th Asia Information Retrieval
               Societies Conference, {AIRS} 2015}, 
               pp. 420--425. Full reference: \cite{DBLP:conf/airs/PetersenSL15}.

\item Alessandro Sordoni, Yoshua Bengio, Hossein Vahabi, Christina Lioma, Jakob Grue Simonsen, and Jian-Yun Nie. \textbf{A Hierarchical Encoder-Decoder for Generative Context-Aware Query Suggestion}. In: \textit{Proceedings of the 24th {ACM} International Conference on Information
               and Knowledge Management, {CIKM} 2015}, 
               pp. 553--562. Full reference: \cite{DBLP:conf/cikm/SordoniBVLSN15}.

\item Casper Petersen, Jakob Grue Simonsen, and Christina Lioma. \textbf{Power Law Distributions in Information Retrieval}. In: \textit{Transactions on Information Systems (TOIS)}. ACM, 2016, Vol. 34, no. 2, pp. 1--37. Full reference: \cite{DBLP:journals/tois/PetersenSL16}.

\item Christina Lioma,
               Birger Larsen,
               Wei Lu, and
               Yong Huang. \textbf{A Study of Factuality, Objectivity and Relevance: Three Desiderata in Large-Scale Information Retrieval?}. In: \textit{Proceedings of the 3rd {IEEE/ACM} International Conference on Big
               Data Computing, Applications and Technologies, {BDCAT} 2016}, 
               pp. 107--117. Full reference: \cite{LiomaLLH16}.

\item Brian Brost,
               Ingemar J. Cox,
               Yevgeny Seldin, and
               Christina Lioma. \textbf{An Improved Multileaving Algorithm for Online Ranker Evaluation}. In: \textit{Proceedings of the 39th International {ACM} {SIGIR} conference on
               Research and Development in Information Retrieval, {SIGIR} 2016}, 
               pp. 745--748. Full reference: \cite{DBLP:conf/sigir/BrostCSL16}.

\item Christina Lioma, Birger Larsen, Casper Petersen, and Jakob Grue Simonsen. \textbf{Deep Learning Relevance: Creating Relevant Information (As Opposed to Retrieving It)}. In: \textit{{Neu-IR'16 SIGIR Workshop on Neural Information Retrieval (Neu-IR'16)}}. {CoRR}, 6 pages. Full reference: \cite{DBLP:journals/corr/LiomaLPS16}. 

\item Birger Larsen and Christina Lioma. \textbf{On the Need for and Provision for an 'IDEAL' Scholarly Information
               Retrieval Test Collection}. In: \textit{Proceedings of the Third Workshop on Bibliometric-enhanced Information
               Retrieval co-located with the 38th European Conference on Information
               Retrieval {(ECIR} 2016)}, 
               pp. 73--81. Full reference: \cite{DBLP:conf/ecir/LarsenL16}.

\item Niels Dalum Hansen, Christina Lioma, and K{\aa}re M{\o}lbak. \textbf{Ensemble Learned Vaccination Uptake Prediction using Web Search Queries}. In: \textit{Proceedings of the 25th {ACM} International on Conference on Information
               and Knowledge Management, {CIKM} 2016}, 
               pp. 1953--1956. Full reference: \cite{DBLP:conf/cikm/HansenLM16}.

\item Casper Petersen, Jakob Grue Simonsen, Kalervo J{\"{a}}rvelin, and Christina Lioma. \textbf{Adaptive Distributional Extensions to DFR Ranking}. In: \textit{Proceedings of the 25th {ACM} International on Conference on Information
               and Knowledge Management, {CIKM} 2016}, 
               pp. 2005--2008. Full reference: \cite{DBLP:conf/cikm/PetersenSJL16}.

\item Brian Brost, Yevgeny Seldin, Ingemar J. Cox, and Christina Lioma. \textbf{Multi-Dueling Bandits and their Application to Online Ranker Evaluation}. In: \textit{Proceedings of the 25th {ACM} International on Conference on Information
               and Knowledge Management, {CIKM} 2016}, 
               pp. 2161--2166. Full reference: \cite{DBLP:conf/cikm/BrostSCL16}.

\item Niels Dalum Hansen,
               K{\aa}re M{\o}lbak,
               Ingemar J. Cox, and
               Christina Lioma. \textbf{Time-Series Adaptive Estimation of Vaccination Uptake Using Web Search Queries}. In: \textit{Proceedings of the 26th International Conference on World Wide Web,
               {WWW} 2017}. In press. Full reference: \cite{hansen2017time}. 

\end{enumerate}




\section{Compositionality and Dependence}


It is generally acknowledged that dismantling something is easier that putting it back together. 
From puzzles to LEGO structures, examples abound where it takes less time and mental effort to reduce some structure to its parts, than to combine parts to form a structure. One reason is that, when a structure is dismantled, it is no longer always obvious how its parts fit together. Or, to put it differently, dismantling causes \textit{information loss}.


It follows that, to make a structure, one needs not only its individual components, but also knowledge about how to combine them. This little bit of common sense lies at the core of Frege's \textit{Principle of Compositionality} \cite{frege}\footnote{Friedrich Ludwig Gottlob Frege is widely credited for the first modern formulation of the Principle of Compositionality in his Foundations of Arithmetic (1884), even though he never explicitly stated the principle. Precursors of this idea appear in the work of Plato and Boole, among others.}.  Formulated more than 100 years ago, this principle posits, roughly, that:
\newline

\begin{quote}
The meaning of an expression is a function of the meanings of its constituent expressions and of the ways these constituents are combined \cite{frege}. 
\end{quote}
\begin{flushright}
\noindent \textbf{\textit{Principle of Compositionality}}
\end{flushright}

Initially formulated in the context of mathematics, this principle has been since applied to several other domains, from linguistics (by Montague \cite{frege}) to programming languages \cite{JonesRHHM99,Roscoe:1997:TPC:550448,Reppy93}, software engineering \cite{DaviesGMW12}, geology \cite{dell2013cognition}, biology \cite{SegataB08}, and -- indeed -- LEGO.

The widespread application of the principle of compositionality across different disciplines is partially due to its appeal to our human inclination to dismantle in order to understand. 
Toddlers spend hours disassembling and reassembling structures; zoologists dissect animals; linguists tokenise language; physicists isolate atoms; geologists separate rock minerals. Common throughout these activities is the wish to learn, i.e. to determine the nature of some phenomenon by investigating its composing elements and their relationship to one another. 

The role of this \textit{decomposing} process to scientific analysis has been acknowledged at least since Aristoteles' time, who referred to it as the \textit{dissecting} method of achieving a solution (\textit{--lysis} [$\lambda\acute{\upsilon}\sigma\iota s$]) through several layers (\textit{--ana} [$\alpha\nu\alpha$]) of processing (\textit{analysis} [$\alpha\nu\acute{\alpha}\lambda\upsilon\sigma\iota s$]).



There are at least two interesting implications to this decomposing paradigm of scientific analysis:
\newline
\begin{itemize}
\item[\textbf{I.}] The constituent parts of a structure that is being decomposed do not necessarily have the same properties as that structure. 
\end{itemize}

\begin{itemize}
\item[\textbf{II.}] Each constituent part of a structure can be itself a separate structure that is, in its turn, decomposable into its own constituent parts.  
\end{itemize}


%
%

The above two implications practically mean that the task of modelling the properties of a structure by decomposing it into its constituent parts increases in complexity, the more one decomposes. One of the most illustrative examples of this increase in complexity is found in physics, and concerns the behaviour of atoms: the movement of a physical object is decided by its mass and the forces acting on it (understanding this requires basic physics); however, on a microscopic level, the movements of the atoms composing that object can only be given probabilistic estimates of behaving in a particular way, for at the heart of their behaviour is randomness. Understanding this requires quantum physics.

Generally, complex structures tend to be studied in science by restricting their representation to factors that can be reasonably computed, 
hence reducing part of the complexity. One of the most common analytical tools deployed for this is the so-called \textbf{\textit{Assumption of Independence}}, which posits that:   
\newline

\begin{quote}
The constituent parts composing a structure can be assumed to occur independently of one another \cite{Cox:2006}.
\end{quote}
\begin{flushright}
\noindent \textbf{\textit{Assumption of Independence}}
\end{flushright}

Making the assumption of independence bypasses the problem of accounting for the mode and degree of dependence of the constituent parts of a structure. This allows to build considerably less complex and more scalable models of the structure studied, compared to when modelling dependence. Often, to compensate for the information loss incurred by the assumption of independence, heuristics are introduced. However, in practice, theoretical explanations of these heuristics are seldom given or generalisable. 

This dissertation presents a series of studies that replace the assumption of independence (and the heuristics that follow from its adoption), with principled methods for capturing \textit{semantic} dependence in the context of \textit{information retrieval}. We explain this next.

\section{Semantic Dependence in Information Retrieval}

\noindent\textbf{DEPENDENCE} (\textit{noun}):  the quality or state of being influenced or determined by or subject to another.

\noindent\textbf{CATENA} (\textit{noun, plural} \textbf{catenae}):  a connected series of related things.
$\;$\\
$\;$\\
Information retrieval is the scientific discipline studying how machines can \textit{infer} the semantics of some information object, like a document\footnote{In information retrieval terminology, \textit{document} refers to any information object, e.g. text of any type or format (pdf, html, tweet, book, product review, etc.).}, so that they can approximate its relevance to some human or automatic request. The best known application of information retrieval is search engines.

Intuitively, one may expect that automatically inferring text semantics would require complex linguistic representations, expressive enough to capture as many facets of meaning in language as possible. Few such approaches have been presented. However, their high complexity is a serious challenge to both computational efficiency and human interpretation of the underlying processes. 
Consequently, the majority of information retrieval approaches today adopt some form of the assumption of independence when processing text. 
This allows to easily represent text as the multiset (or \textit{bag}) of its words, disregarding grammar, word order, and in general any relation binding words together, but keeping multiplicity. These methods are collectively known as \textit{bag of words} approaches\footnote{The divide between linguistic rigour and computational performance was famously captured by Frederick Jelinek at IBM in the 1980s, who anecdotally said: ``every time we fire a linguist, system performance goes up'' [paraphrased].}. 

As old as bag of words approaches may be (they can be traced at least as far back as Zellig S. Harris' 1954 article on Distributional Structure \cite{harris:1981}), equally old is the criticism against them. 
Empirically, it is easy to see that important semantic distinctions, for instance the difference in meaning between the proposition \texttt{John loves Mary} and the proposition \texttt{Mary loves John}, cannot be captured by bag of words approaches. Nevertheless, despite the long and well argued criticism against these approaches, they continue to be the main paradigm in text processing because the benefits they yield (ease of computation, scalability, interpretability of results, robust processing) outrank their disadvantages (semantic inaccuracy, restricted analytical scope). 

\begin{quote}
This dissertation presents:
\begin{enumerate}
\item[(i)] a series of studies showing that several benefits of bag of words approaches can be preserved when representing aspects of semantic dependence in principled, non-complex ways; 
\item[(ii)] reflections on why the current \textit{modus vivendi} of semantic processing in information retrieval is inadequate and should be revised;
\item[(iii)] recommendations on designing a new \textit{modus operandi} that bridges the two extremes of shallow versus deep semantic processing, leading to more accurate and more expressive information retrieval inferences.
\end{enumerate}
\end{quote}

%


The common underlying objective in the body of work included in this dissertation is to model the dependence of textual constituents on three different levels of semantic analysis: \textit{lexical, discourse,} and \textit{cognitive}.
\newline

\noindent \textbf{Lexical level.} The units of analysis in lexical semantics are words. 

\noindent \textbf{Discourse level.} The units of analysis in discourse semantics are sentences.

\noindent \textbf{Cognitive level.} The units of analysis in cognitive semantics are concepts. 
$\;$\\
$\;$\\

For each of the above three levels, the type and strength of the dependence conjoining the constituent parts of text is examined in a principled manner, and different models are developed for processing the corresponding \textit{semantic catenae} (of words, sentences, or concepts) in the context of information retrieval. 
These models capture part of the information loss that takes place when reasoning about text computationally by dismantling it into individual words, sentences, or concept representations. Collectively, these models replace the assumption of independence with non-complex representations of lexical, discourse, and cognitive dependence, using principles from graph theory, probability theory, logic and statistics. 

A total of nine models of semantic dependence for information retrieval are presented in this dissertation (outlined in Table \ref{tab:overview}). The results contributed and conclusions drawn by each model are discussed next, separately for lexical, discourse and cognitive semantics.


\begin{table}
\centering
\caption{The nine models of semantic dependence presented in this dissertation. The numbers inside square brackets point to the bibliographic references of the articles where each model was published.} 
\label{tab:overview}
\resizebox{120mm}{!}{
\begin{tabular}{| l | c | c | c |}
METHOD				&\multicolumn{3}{c|}{SEMANTIC LEVEL}\\
				&\textit{lexical}			&\textit{discourse}						&\textit{cognitive}\\
				\hline
\textit{graph theory}		&Model I \cite{lioma}		&Models VI-VII \cite{LiomaTSPL16,petersen}	&\\
\textit{probabilistic}	&Model II \cite{LiomaSLH15}	& Models IV-V \cite{Lioma12,petersen}			&\\
\textit{statistics}			&Model III \cite{LiomaH17}	&&\\
\multirow{2}{*}{\textit{logic}}				&&&Models VIII-IX \\
&&&\cite{LiomaB:2009,Lioma:2010,LiomaLI12}\\
\end{tabular}
}
\end{table}

\section{Contributions to Lexical Semantics}

Three models of lexical dependence (referred to as Model I, II, and III) are presented as Chapters 2, 3, and 4 of this dissertation. All three models are unsupervised (they incur no added computational cost for training on pre-annotated data). Each model uses principles of different formalisms: graph theory, probability, and statistics.

\subsection{Model I: Unsupervised graph theoretic lexical dependence \cite{lioma}.}
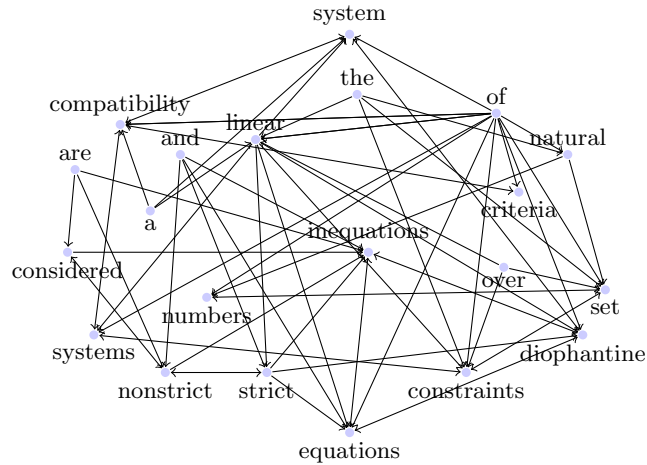
\begin{figure}[]
\beginpgfgraphicnamed{graph-models}
\begin{tikzpicture}
\draw (+2.50,+6.50) node[draw=black,text width=11.5cm,rounded corners,text justified] {\small Compatibility of systems of linear constraints over the set of natural numbers. Criteria of compatibility of a system of linear Diophantine equations, strict inequations, and nonstrict inequations are considered.};
\path 
[inner sep=0.05mm, place/.style={circle,fill=blue!20,thick,inner sep=0pt,minimum size=1.2mm}]
	(-0.30,+3.60) node [place] (compatibility) {}
	(+4.70,+3.75) node [place] (of) {}
	(-0.65,+0.80) node [place] (systems) {}
	(+1.50,+3.40) node [place] (linear) {}
	(+4.30,+0.30) node [place] (constraints) {}
	(+2.85,+4.00) node [place] (the) {}
	(+4.80,+1.70) node [place] (over) {}
	(+0.85,+1.30) node [place] (numbers) {}
	(+6.15,+1.40) node [place] (set) {}
	(+5.00,+2.70) node [place] (criteria) {}
	(+5.65,+3.20) node [place] (natural) {}
	(+0.50,+3.20) node [place] (and) {}
	(+2.75,+4.80) node [place] (system) {}
	(+0.30,+0.30) node [place] (nonstrict) {}
	(+2.75,-0.50) node [place] (equations) {}
	(+1.65,+0.30) node [place] (strict) {}
	(+3.00,+1.90) node [place] (inequations) {}
	(+0.10,+2.45) node [place] (a) {}
	(+5.85,+0.80) node [place] (diophantine) {}
	(-1.00,+1.90) node [place] (considered) {}
	(-0.90,+3.00) node [place] (are) {};
	\draw [->] (of) -- (compatibility);
	\draw [<->] (compatibility) -- (systems);
	\draw [->] (of) -- (systems);
	\draw [->] (of) -- (linear);
	\draw [->] (linear) -- (systems);
	\draw [->] (of) -- (constraints);
	\draw [<->] (constraints) -- (systems);
	\draw [->] (linear) -- (constraints);
	\draw [->] (over) -- (linear);
	\draw [->] (over) -- (constraints);		
	\draw [->] (the) -- (linear);
	\draw [->] (the) -- (constraints);
	\draw [->] (the) -- (set);
	\draw [<->] (constraints) -- (set);
	\draw [->] (over) -- (set);
	\draw [->] (of) -- (set);
	\draw [->] (the) -- (natural);
	\draw [->] (of) -- (natural);
	\draw [->] (natural) -- (set);
	\draw [->] (of) -- (numbers);
	\draw [<->] (set) -- (numbers);
	\draw [->] (natural) -- (numbers);
	\draw [->] (of) -- (compatibility);
	\draw [->] (of) -- (criteria);
	\draw [<->] (criteria) -- (compatibility);
	\draw [->] (a) -- (compatibility);
	\draw [<->] (compatibility) -- (system);
	\draw [->] (of) -- (system);
	\draw [->] (a) -- (system);
	\draw [->] (a) -- (linear);
	\draw [->] (of) -- (linear);
	\draw [->] (linear) -- (system);
	\draw [<->] (system) -- (diophantine);
	\draw [->] (linear) -- (diophantine);
	\draw [->] (of) -- (diophantine);
	\draw [->] (of) -- (equations);
	\draw [->] (linear) -- (equations);
	\draw [<->] (diophantine) -- (equations);
	\draw [->] (linear) -- (strict);
	\draw [->] (strict) -- (diophantine);
	\draw [->] (strict) -- (equations);
	\draw [<->] (diophantine) -- (inequations);
	\draw [<->] (equations) -- (inequations);
	\draw [->] (strict) -- (inequations);
	\draw [->] (and) -- (equations);
	\draw [->] (and) -- (strict);
	\draw [->] (and) -- (inequations);
	\draw [<->] (strict) -- (nonstrict);
	\draw [->] (and) -- (nonstrict);
	\draw [->] (nonstrict) -- (inequations);
	\draw [->] (are) -- (nonstrict);
	\draw [->] (are) -- (inequations);
	\draw [->] (are) -- (considered);
	\draw [<->] (nonstrict) -- (considered);
	\draw [->] (considered) -- (inequations);
	\node [above] at (compatibility) {\small compatibility};
	\node [above] at (of) {\small of};
	\node [below] at (systems) {\small  systems};
	\node [above] at (linear) {\small linear};
	\node [below] at (constraints) {\small constraints};
	\node [above] at (the) {\small the};
	\node [below] at (over) {\small over};
	\node [below] at (set) {\small set};
	\node [below] at (numbers) {\small numbers};
	\node [below] at (considered) {\small  considered};
	\node [above] at (are) {\small  are};
	\node [below] at (diophantine) {\small diophantine};
	\node [above] at (and) {\small  and};
	\node [below] at (a) {\small  a};
	\node [above] at (system) {\small  system};
	\node [below] at (equations) {\small  equations};
	\node [below] at (strict) {\small  strict};
	\node [below] at (nonstrict) {\small  nonstrict};
	\node [above] at (inequations) {\small inequations};
	\node [below] at (criteria) {\small  criteria};
	\node [above] at (natural) {\small  natural};
\end{tikzpicture}
\endpgfgraphicnamed
\caption{[Model I]. Graph representation of the above sample text: vertices denote words, edges denote co-occurrence within a fixed context window, and directionality denotes grammatical modification \cite{lioma}. 
}
\label{fig:00-directed-graph}
\end{figure}

The essential idea of lexical semantics is that the meaning of a word correlates with the semantic entailments associated to this word. 
Following this, 
instead of assuming that words occur independently in text, Model I represents word dependencies as graph edges that connect vertices denoting unique words (see Figure \ref{fig:00-directed-graph} for an illustration). More simply, instead of representing text as a bag of words, text is represented as a graph of interconnected words. Analogies are then made between different aspects of word dependence and aspects of the graph's topology (such as clustering or average path length).

Based on this graph representation of text, Model I contributes a novel term\footnote{\textit{Term} and \textit{word} are used interchangeably in information retrieval and also in this dissertation.} weighting approach for information retrieval. In addition to its theoretical novelty, Model I also makes the following two advances:
\begin{description}
\item [On an algorithmic level] it is not subject to document length bias (because it replaces the currently dominating frequentist practice of word counts by vertex connectivity). Practically this makes the need for additional document length normalisation obsolete (unlike \textit{all} other term weighting approaches).
\item [On a theoretical level] it allows aspects of lexical dependence, such as grammatical type, modification, or non-adjacent transition, to be \textit{seamlessly} incorporated into term weighting (because they are formulated as weights, labels or direction in the graph edges). Practically, this modularity allows generating instances of term weighting methods that capture different aspects of lexical dependence in the approximation of a term's semantic salience. This has not been possible in \textit{any} other term weighting approach. 
\end{description} 

%
%
%

The reliability and validity of Model I is supported by thorough experimental evaluation in two information retrieval tasks (\textit{ad hoc} search and web blog search), using large-scale state of the art benchmark datasets (28.9GB in total), and measuring effectiveness (precision, binary preference), efficiency (overhead in milliseconds, running time of iterations), and parameter sensitivity against competitive state of the art baselines\footnote{Throughout this chapter, \textit{state of the art} refers to the year the corresponding article was first published.}. 


\subsection{Models II \& III: Unsupervised probabilistic \& statistical lexical dependence \cite{LiomaH17, LiomaSLH15}}

The most common approach for approximating the semantic dependence between words in information retrieval is through lexical frequency statistics of their co-occurrence\footnote{The \textbf{Distributional Hypothesis} posits that words used or occurring in the same contexts tend to purport similar meanings. John Rupert Firth famously captured this as: ``you shall know a word by the company it keeps'' \cite{Firth:1968}.}. The underlying rationale is that:
\begin{itemize}
\item If words co-occur often enough in some corpus, they are semantically dependent; 
\item The more often words co-occur, the more semantically dependent they are.
\end{itemize}

Model II reveals (for the first time in information retrieval) that this rationale is not always correct: the frequency of word co-occurrence can be separate from the strength of semantic dependence. Even though the former can be indicative to some extent of the latter, their relation is not symmetric. 
This miscalculation of semantic dependence in words is corrected by two models (one probabilistic, one statistical), which distinguish between frequency of co-occurrence and strength of semantic dependence as follows. 

\begin{figure}
\centering
\resizebox{120mm}{!}{
\includegraphics[]{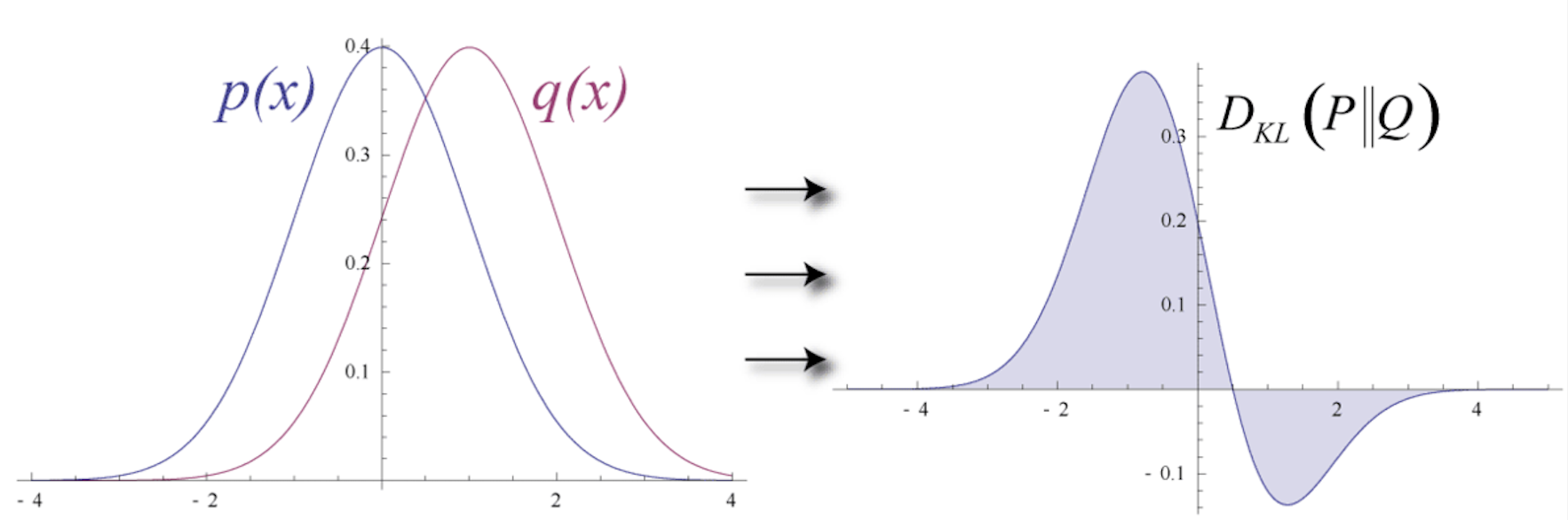}
}
\caption{\label{fig:kl}[Model II]. Illustration of Kullback-Leibler divergence ($D_{KL}$).}
\end{figure}

Model II represents word pairs as probability distributions of their distributional semantics (co-occurring words within fixed context windows extracted from corpora).
 These probability distributions are generated for word pairs and for perturbations of these pairs where one word at a time is replaced by its synonym. This is a modern implementation of Leibniz' \textit{Principle of Intersubstitutivity (salva veritate)} to detect irregular composition of meaning, which posits that words which can be substituted for one another without altering the truth of any statement are the same \textit{(eadem)} or coincident \textit{(coincidentia)}. The divergence between the distribution of the original word pair and the distribution of its perturbation (illustrated as the Kullback-Leibler Divergence of $p(x)$ and $q(x)$ in Figure \ref{fig:kl}\footnote{Image by Nathan Mundhenk. Source: \url{https://upload.wikimedia.org/wikipedia/en/a/a8/KL-Gauss-Example.png}. Licensed under Creative Commons: CC BY-SA 3.0.}) is found by Model II to be an accurate approximation of the semantic strength of the words of the original word pair. 

In addition to the theoretical novelty of Model II, the significance of its findings to information retrieval can be summarised as follows:

\begin{description}
\item [On a conceptual level] it points out a significant error in the estimation of semantic strength between words that has gone so far undetected in this very well studied area of information retrieval (the relevant literature is outlined in Section \ref{00-rw}). 
\item [On an algorithmic level] it proposes a non-complex solution to this error that yields considerable gains in retrieval accuracy and that has increased interpretability\footnote{\textbf{Interpretability} refers to how easy it is for humans to understand a process and its resulting outcome. The more abstract the features used, and the higher the dimension of spaces represented, the harder it is for humans to readily comprehend a process and explain its output.}.
\end{description} 

Model III follows the same rationale as Model II, but uses a different representation. Word pairs are represented not as probability distributions, but as lists of the term weights of their distributional semantics. These lists are ranked by term weight, and the distance or correlation between these lists is found to be an accurate approximation of the semantic strength between the words. The novelty of this approach is that, instead of considering all terms forming the distributional semantics of the input phrase, it ranks these terms by their contribution to the text semantics (approximated by their term weight) and considers \textit{only} the most discriminative terms. This allows for more dense and discriminative representations.  

\begin{figure}
\centering
\resizebox{120mm}{!}{
\includegraphics[]{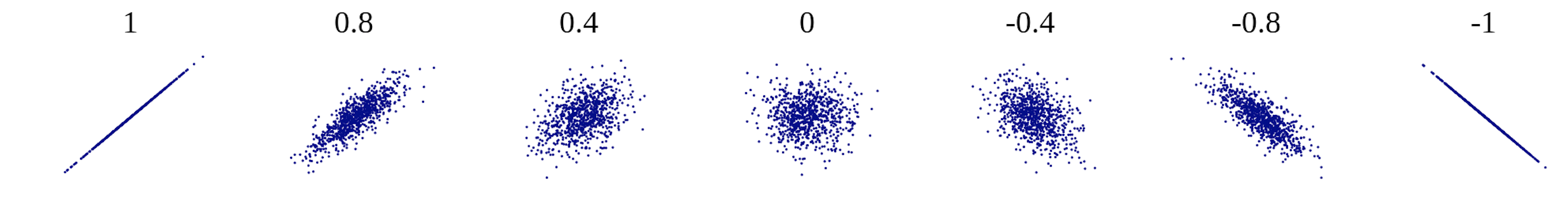}
}
\caption{\label{fig:pearson}[Model III]. Illustration of Pearson correlation coefficient. The correlation reflects the noisiness and direction of a linear relationship between two sets of points. 
}
\end{figure}

In addition to the theoretical novelty of Model III, the significance of its findings can be summarised as follows:
\begin{description}
\item [On a theoretical level] it proposes a simple, unsupervised solution to a problem that has been primarily addressed in increasingly complex ways (more recently deep learning). Any appropriate term weighting method and correlation or distance metric can be plugged into Model III, allowing for different aspects of term salience (and its ranking) to be considered in the computation of lexical dependence.
\item [On an empirical level] it yields the highest performance reported up to the date of publication for this task and benchmarks, outperforming the previous highly tuned supervised deep learning state of the art, while also having high interpretability (elements of which are exemplified in Figure \ref{fig:pearson}\footnote{Image by Denis Boigelot. Source: \url{https://upload.wikimedia.org/wikipedia/commons/thumb/d/d4/Correlation_examples2.svg/400px-Correlation_examples2.svg.png}. Licensed under Creative Commons: CC0.} for the Pearson correlation coefficient).
\end{description} 

Model II is a novel contribution to information retrieval. Model III is a novel contribution to natural language processing.

%
%

%
%
%
The reliability and validity of Models II \& III is supported by thorough experimental evaluation. Model II is evaluated in two information retrieval tasks (\textit{ad hoc} search and web search), using large-scale state of the art benchmark datasets (502.1GB in total), and measuring effectiveness (precision, normalised discounted cumulative gain) both on the whole dataset and also separately per query length. 
Model III is evaluated on the largest manually annotated compositionality dataset publicly available in 2016 (1048 phrases) \cite{farahmand-smith-nivre:2015:MWE} against competitive state of the art baselines.


%
%
%


\section{Contributions to Discourse Semantics}
Four models of discourse dependence (referred to as Models IV, V, VI, and VII) are presented in Chapters 5, 6,  and 7 of this dissertation. All four models are unsupervised (hence they incur no added computational cost for tuning). Models IV \& V are probabilistic, while Models VI \& VII use principles from graph theory.

\subsection{Models IV \& V: Unsupervised probabilistic discourse dependence \cite{Lioma12, petersen}}
The essential idea of discourse semantics is that the meaning of a sentence is bounded by anaphoric elements in that sentence pointing to preceding or succeeding sentences (or elements thereof) that are collectively needed to interpret the discourse context. 
Following this, instead of assuming that text is an unordered set of sentences, Models IV \& V represent the relations and transition between sentences in text when processing text semantics. Two different probabilistic models of discourse dependence are presented. 
\begin{table}
\centering
\caption{\label{tab:rhet}Examples of the rhetorical relations (in bold italics) inferred as part of Model IV \cite{Lioma12}}
\resizebox{120mm}{!}{
\begin{tabular}{l|l} 
attribution&... the islands now known as the Gilbert Islands were settled \\ &\textbf{\textit{by Austronesian-speaking people}} ... \\
&\\
background&... many whites had left the country \textbf{\textit{when Kenyatta divided}} \\ &\textbf{\textit{their land among blacks}} ... \\
&\\
cause-result&... \textbf{\textit{I plugged ``wives'' into the search box and came up with}} \\ &\textbf{\textit{ the following results}} ... \\
&\\
comparison&... so for humans, \textbf{\textit{it is stronger than coloured}} to frustrate\\ & these unexpected numbers ... \\
&\\
condition&... Conditional money \textbf{\textit{based upon care for the pet}} ... \\
&\\
consequence&... voltage drop with the cruise control switch \textbf{\textit{could cause}} \\ &\textbf{\textit{ erratic cruise control operation}} ... \\
&\\
contrast&... \textbf{\textit{Although it started out as a research project}}, the ARPANET\\ & quickly developed into ...\\
&\\
elaboration&... order accutane \textbf{\textit{no prescription required}} ... \\
&\\
enablement&... The project will also \textbf{\textit{offer exercise programs and make eye}} \\ &\textbf{\textit{ care services accessible}} ... \\
&\\
evaluation&... such advances will be reflected in an ever-\textbf{\textit{greater proportion}} \\ &\textbf{\textit{ of grade A recommendations}} ... \\
&\\
explanation&... the concept \textbf{\textit{called as ``evolutionary developmental biology''}} \\ &\textbf{\textit{ or shortly ``evo-devo''}} ... \\
&\\
manner-means&... Fill current path \textbf{\textit{using even-odd rule, then paint the path}} ... \\
&\\
summary&... Safety Last, Girl Shy, Hot Water, The Kid Brother, Speedy\\ & \textbf{\textit{(all with lively orchestral scores)}} ... \\
&\\
temporal&... Take time out \textbf{\textit{before you start writing}} ...\\
&\\
topic-comment&... \textbf{\textit{Director Mark Smith expressed support for greyhound adoption}} ... \\
\end{tabular}
}
\end{table}

Model IV \cite{Lioma12} approximates the retrieval probability of different rhetorical relations (defined by Rhetorical Structure Theory \cite{MannT88} and exemplified in Table \ref{tab:rhet}) between sentences in text, and creates a novel ranking model that includes this probability in its computation. This retrieval probability of different rhetorical relations is found to be highly discriminative of topical relevance. 

In addition to the theoretical novelty of Model IV, the significance of its findings can be summarised as follows:
\begin{description}
\item [On a theoretical level] it proposes a novel ranking model that allows making inferences about \textit{which part of} text is topically relevant to a query \textit{and why}, as opposed to assuming that all parts in text are equally topically relevant to some query. 
\item [On an empirical level] it yields considerable gains in retrieval performance, while also having high interpretability.
\end{description}

\begin{figure}
\centering
\resizebox{60mm}{!}{
\includegraphics[]{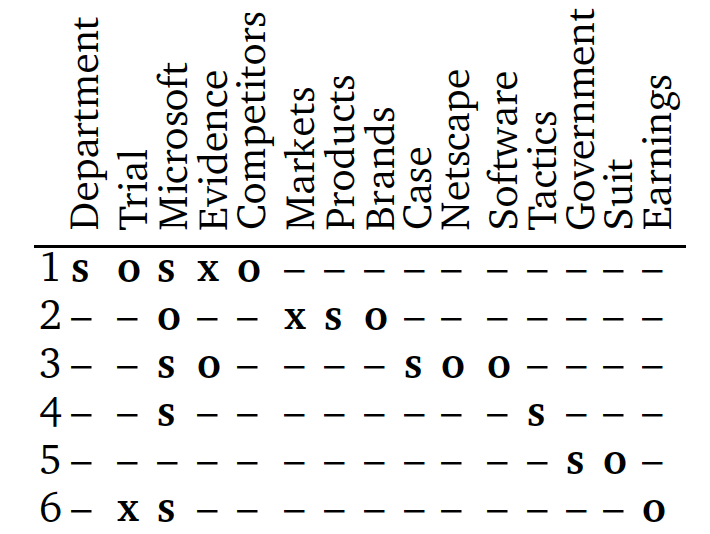}
}
%
%
\resizebox{120mm}{!}{
\includegraphics[]{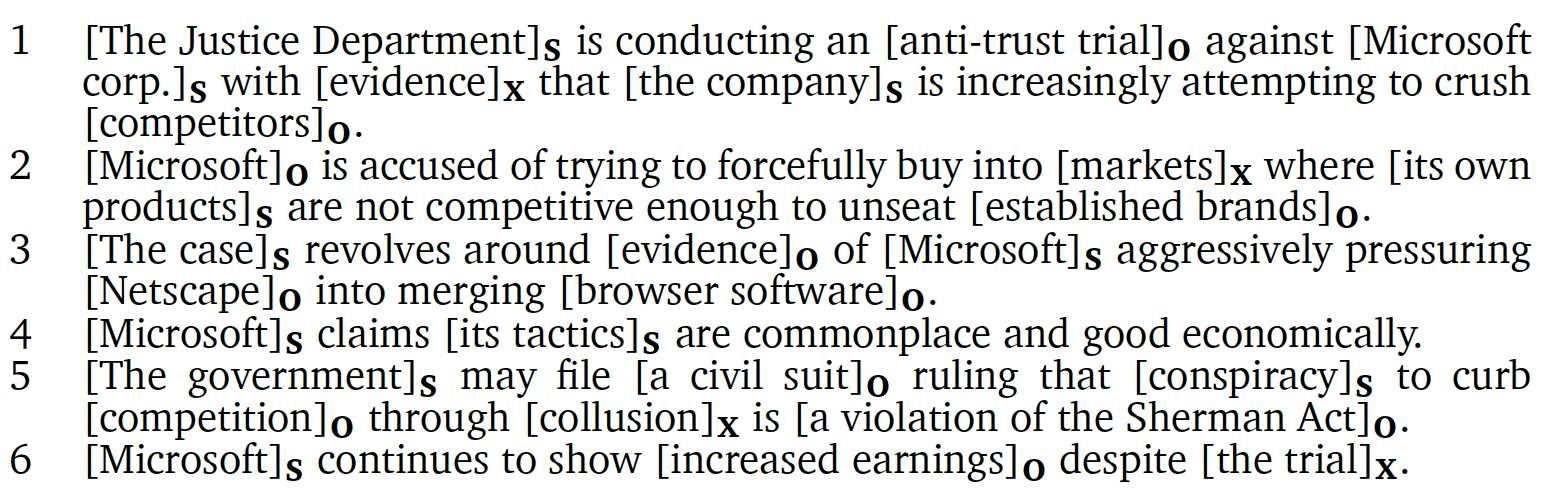}
}
\caption{\label{fig:entitygrid}[Model V]. Catenae of discourse entities (matrix rows) of the sample text shown above (example reproduced from \cite{Barzilay:2008}). \textbf{s,o,x} denote the syntactic role of subject, object, or other.}
\end{figure}

Model V \cite{petersen} represents the relations between core syntactic entities of different sentences in text as catenae of salient discourse entities (see Figure \ref{fig:entitygrid} for an example). Analogies are then made between the entropy of these catenae and the amount of semantic disorder  in text. Entropy is found to be an accurate approximation of text coherence, and further, text coherence is found to be highly discriminative of topical relevance. 

In addition to the theoretical novelty of this Model V, the significance of its findings can be summarised as follows:
\begin{description}
\item [On a theoretical level] it proposes the first ever coherence ranking model for information retrieval, allowing to make inferences about \textit{which part of} text is topically relevant to a query as a function of \textit{both} its topical relevance \textit{and} its coherence with respect to the rest of the text.  
\end{description}

The reliability and validity of Models IV \& V is supported by thorough experimental evaluation. Model IV is evaluated on \textit{ad hoc} information retrieval, using large-scale state of the art benchmark datasets (500GB in total), and measuring effectiveness (precision, binary preference, normalised discounted cumulative gain) both on the whole dataset and also separately per query length, against competitive state of the art baselines. Model V is evaluated both on text reordering (standard task in coherence modelling) and also on \textit{ad hoc} information retrieval. State of the art benchmark datasets are used for both text reordering and retrieval (500GB in total), and effectiveness (accuracy, mean reciprocal rank, expected reciprocal rank) is measured against state of the art baselines.

Models IV \& V contribute novel, non-complex, unsupervised methods of processing two different aspects of discourse semantics in information retrieval: rhetorical relations between sentences, and entity transition across sentences for text coherence. Prior to this work, none of these two aspects had been integrated into information retrieval in non \textit{ad hoc} ways. Model IV is a novel contribution to information retrieval. Model V is a novel contribution to both natural language processing and information retrieval.
 
 

 \subsection{Models VI \& VII: Unsupervised graph theoretic discourse\\ dependence \cite{LiomaTSPL16, petersen}}


Similarly to Models IV \& V, Models VI \& VII represent the relations and transition between sentences in text when processing text semantics. However, unlike Models IV \& V (that are probabilistic), Models VI \& VII use graph theory.

Two different graph theoretic models of sentence dependence are presented. 
\begin{figure}
\centering
\resizebox{120mm}{!}{
\includegraphics[]{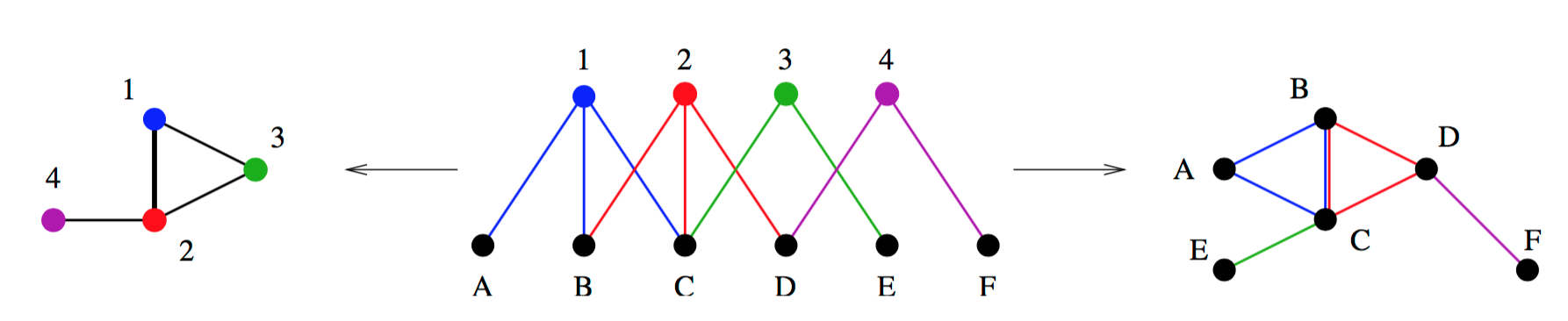}
}
\caption{\label{fig:00-projection}[Model VI]. A bipartite graph (middle) and its one-mode projections (left, right) \cite{lind:2005}.}
\end{figure}

Model VI \cite{petersen} represents the relations between different discourse entities across sentences in text as bipartite graphs whose vertex sets represent sentences and entities respectively. Analogies are then made between different aspects of sentence dependence and aspects of the  topology of this bipartite graph when projected onto a one-mode graph\footnote{\textbf{One-mode projection} is a graph representation of the relation structure among only one of the two set of vertices of a bipartite graph.} (see Figure \ref{fig:00-projection} for an illustration). 

Model VII \cite{LiomaTSPL16} uses the same representation of entities and sentences as a bipartite graph, but makes inferences about discourse semantics directly on the bipartite graph instead of its one-mode projections (this is significantly less trivial).

Both Models VI \& VII are shown to be successful when ranking documents in information retrieval with respect to \textit{both} their topical relevance \textit{and} their coherence.

In addition to the theoretical novelty of these models, the significance of their findings can be summarised as follows:
\begin{description}
\item [On a theoretical level] they propose novel unsupervised modelling of \textit{both global and local} text coherence\footnote{\textbf{Local} versus \textbf{global} coherence refers to the well connectedness of adjacent versus remote text spans.} (very few exist). 
\item [On a theoretical level] Model VII proposes novel graph metrics on the bipartite graph \textit{without one-mode projection} (very few exist in general, and none exist for coherence modelling). 
\item [On an empirical level] both models yield considerable gains in both coherence (Model VII in particular yields the highest accuracy reported at the time of publication, outperforming the previous highly tuned state of the art) and retrieval performance, while also having high interpretability.
\end{description}

The reliability and validity of Models VI \& VII is supported by thorough experimental evaluation both in text reordering (standard task in coherence modeling) and also in \textit{ad hoc} information retrieval. State of the art benchmark datasets are used for both text reordering and retrieval (500GB in total), and effectiveness (accuracy, mean reciprocal rank, expected reciprocal rank) is measured against state of the art baselines.
Models VI \& VII contribute non-complex, unsupervised ways of processing discourse flow in text, which are novel to both natural language processing and information retrieval.

\begin{figure}
\centering
	\begin{tikzpicture}
		\node at (-7,4.6) [black,thick] (state1) {\scriptsize \bf cognitive overlap};
		\node at (-10.0,1.0) [] (A) {\scriptsize \bf concept A};
		\node at (-4.0,1.0) [] (B) {\scriptsize \bf concept B};		
		\draw[-> ,dashed,black,thick] (A) -- (state1);
		\draw[-> ,dashed,black,thick] (B) -- (state1);
		\node at (-7.0,1.0) [blue] (C) {\scriptsize \bf A,B};		
		\node at (-7.0,0.6) [black,thick] () {\bf $\omega^A \oplus \omega^B$};
		\draw[-> ,dashed,blue,thick] (A) -- (C);
		\draw[-> ,dashed,blue,thick] (B) -- (C);			
		\draw[-> ,blue,thick] (C) -- (state1);
		\draw (-4,4.0) node[text width=3.3cm,fill=blue!10,rounded corners,text justified] (arrow-begin) {\scriptsize \bf combined representation};
		\node at (-7.1,4.0) [] (arrow-end) {};
		\draw[- ,black,thick] (arrow-begin) -- (arrow-end);
		\end{tikzpicture}
		\caption{\label{fig:sl1}[Models VIII-IX]. Combining the representations of two independent concepts using consensus.}
\end{figure}
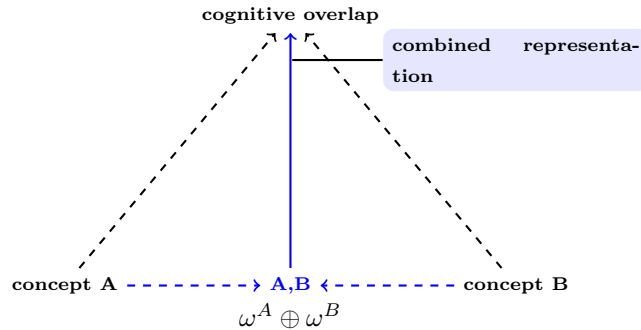
\begin{figure}
\centering
	\begin{tikzpicture}
		\node at (-7,4.6) [black,thick] (state1) {\scriptsize \bf cognitive overlap};
		\node at (-10.0,1.0) [] (A) {\scriptsize \bf concept A};
		\node at (-4.0,1.0) [] (B) {\scriptsize \bf concept B};		
		\draw[-> ,dashed,black,thick] (A) -- (B);
		\draw[-> ,dashed,black,thick] (B) -- (state1);
	
		\draw[-> ,blue,thick] (A) -- (state1);
		\node at (-10.0,3.0) [black,thick] () {\bf $\omega^A \otimes \omega^B$};
		\draw (-4.0,4.0) node[text width=3.3cm,fill=blue!10,rounded corners,text justified] (arrow-begin) {\scriptsize \bf combined representation};
		\node at (-7.6,4.0) [] (arrow-end) {};
		\draw[- ,black,thick] (arrow-begin) -- (arrow-end);
		\end{tikzpicture}
		\caption{\label{fig:sl2}[Models VIII-IX]. Combining the representations of two dependent concepts using recommendation.}
\end{figure}
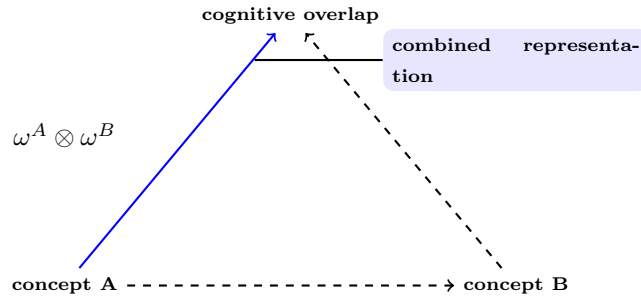
\section{Contributions to Cognitive Semantics}
Two models of cognitive dependence (referred to as Models VIII and IX) are presented as Chapters 8, 9, and 10 of this dissertation. Both models use Subjective Logic\footnote{\textbf{Subjective logic} is a type of probabilistic logic that explicitly takes uncertainty and source trust into account \cite{Josang:2001}.} \cite{Josang:2001} to reason separately about representations of information needs (Model VIII) and polyrepresentation\footnote{The \textbf{Principle of Polyrepresentation} \cite{Ingwersen:1996,IngwersenJ:2005} posits that information retrieval effectiveness may improve through the consideration of multiple and diverse representations of information objects or processes.} (Model IX).
\subsection{Models VIII \& IX: Subjective logic models of cognitive dependence for query difficulty \cite{LiomaB:2009} and polyrepresentation \cite{LiomaLI12, Lioma:2010}}

The essential idea of cognitive semantics is that language is part of a more general human cognitive ability, which describes the world as people conceive it. Different people may have different representations of the semantics of the same object. Modelling these representations is central to information retrieval, where a major challenge is to devise expressive methods for mapping written representations of meaning to their best fitting semantic concepts. 



Models VIII \& IX formally represent different concept representations of query and document semantics, and combine these representations in highly tractable and expressive ways that account for the different types and degrees of dependence between representations (graphically illustrated in Figures \ref{fig:sl1} \& \ref{fig:sl2}).
Based on this, Model VIII contributes a novel query difficulty estimation approach, and Model IX contributes a novel polyrepresentation approach for information retrieval. Both models are novel to information retrieval.

In addition to the theoretical novelty of these models, the significance of their findings can be summarised as follows:
\begin{description}
\item [On a theoretical level] they present a rich calculus for expressing impact, bias, and directionality between cognitive representations. This is the first\footnote{together with the Quantum Model, presented in Section \ref{00-rw}.} mathematical expression of the \textit{Principle of Polyrepresentation}. 
\item [On a theoretical level] they present mathematical means for quantifying the degree of uncertainty in probability estimates of concept representations that are used in information retrieval. Practically, this allows semantic inferences to be made even when the input arguments may be fraught with uncertainty.
\end{description}

The reliability and validity of Models VIII \& IX is supported by thorough experimental evaluation in \textit{ad hoc} information retrieval using rich human-assessed benchmark data and measuring different aspects of effectiveness (precision, normalised discounted cumulative gain, binary preference, mean reciprocal rank) on standard settings.
\linebreak

Collectively Models I - IX show that, in the analytical spectrum of deep (complex) versus shallow (naive) processing, there is a lot to be gained \textit{both} empirically \textit{and} theoretically, from a middle ground of principled, yet non-complex formalisations of text semantics. 

\section{Research Landscape of Semantic Dependence in Information Retrieval}
\label{00-rw}


This section contextualises the findings presented above by contributing a broad, comprehensive and up to date overview of the state of the art and major trends of semantic dependence in information retrieval. This overview includes the literature covered in Chapters 2 - 10, and further extends it with more recent advances. Details on the precise comparison and evaluation of the results presented in this dissertation to prior work are provided in the individual articles.

\subsection{Lexical Dependence}
Broadly speaking, efforts to model dependence on the level of lexical semantics (\textit{term dependence}), also known as \textit{term co-occurrence, adjacency} and \textit{lexical affinities}\footnote{\textit{Dependence, co-occurrence, adjacency} and \textit{lexical affinities} are not synonyms, but in information retrieval they are used interchangeably.} in information retrieval, typically model phrases found in queries and/or documents, motivated by the intuition to consider as \emph{more} relevant those documents in which terms appear in the same order and patterns as they appear in the query, and as \emph{less} relevant those documents in which terms are separated \cite{SmithD:1985}. Lexical dependence is approximated using either statistical or linguistic information. 

Research in this area began with the early work on statistical term associations \cite{Doyle1962, Giuliano1963, Lesk1969, Stiles1961} and syntax-based approaches \cite{Baxendale1958, Earl1972, Salton:1966}, continuing with work on probabilistic term dependence models \cite{HarperK1978, SaltonBY82, CroftT:1991, K1977, Yu1983}, syntactic methods \cite{DillonG:1983, MetzlerN:1984, SmeatonK:1988, Smeaton:1986} and statistical approaches \cite{Fagan1989, Lewis1992, LewisC:1990}. From the mid-1990s onwards, research focused on hybrid methods combining syntactic and statistical approaches of phrase processing \cite{EvansZ:1996}, phrase-based enhancement of the indexed term representations \cite{Zhai1997}, and phrase-based term weighting \cite{NaritaO2000,PedersonS:1997,Str1997}. This was succeeded by a focus on statistical methods, primarily using language modelling \cite{BenderskyMC10, MetzlerC:2005, MishneR2005, NallapatiA:2002, SongC1999, SrikanthS:2003} but not exclusively \cite{Losee:1994}, while attention has also been given to heuristics \cite{TaoZ07} and formalisations of term position \cite{LvZ09}. 

A temporal listing of the major contributions in lexical dependence for information retrieval can be seen in Table \ref{tab:rw-CH00}. Efforts in this area intensified in the late 1990s, and peaked again relatively recently (see Figure \ref{fig:trends}), motivated, among others, by the often stated need to refine the state of semantic processing in information retrieval. However, despite this long and rich literature, no prior\footnote{Prior to the first publication of the articles included in the dissertation.} work on lexical dependence in information retrieval has contributed solutions to lexical dependence that account for semantic aspects such as modification and transitivity (contributed by Model I), or semantic non-compositionality (thus correcting the error of equating frequency to semantic strength) (contributed by Models II \& III).

\begin{figure}
\centering
\resizebox{100mm}{!}{
\pgfplotsset{every axis legend/.append style={
			at={(0.4,0.8)},
			anchor=south}
			}
\begin{tikzpicture}[trim axis left]
			\begin{axis}[
			x tick label style={/pgf/number format/.cd,%
          		set thousands separator={},
          		fixed},
			width = \linewidth,
			title={},
			xlabel=year of publication,
			xtick={1970,1990,2010},
			scaled x ticks = false,
			scale only axis,
			ylabel=number of publications,
			smooth
			]
			\pgfplotstableread{lex.txt}\table 
			\addplot[mark=none, very thick] table[x index=1,y index=2] from \table;
			\addlegendentry{lexical}
			\pgfplotstableread{dis.txt}\table 
			\addplot[mark=+, very thick, blue] table[x index=1,y index=2] from \table;
			\addlegendentry{discourse}
			\pgfplotstableread{cog.txt}\table 
			\addplot[mark=o, very thick, red] table[x index=1,y index=2] from \table;
			\addlegendentry{cognitive}

\end{axis}
\end{tikzpicture}
}
\caption{\label{fig:trends}Number of publications on lexical, discourse, and cognitive dependence in information retrieval (y axis), versus publication year (x axis). The numbers of publications are smoothed by moving 3-year averages.}
\centering
\resizebox{110mm}{!}{
\pgfplotsset{every axis legend/.append style={
			at={(0.4,0.5)},
			anchor=south}
			}
\begin{tikzpicture}[]
			\begin{axis}[
			x tick label style={/pgf/number format/.cd,%
          		set thousands separator={},
          		fixed},
			width = \linewidth,
			title={},
			xlabel=year of publication,
			xtick={1970,1990,2010},
			scaled x ticks = false,
			ylabel= number of publications,
			smooth
			]
			\pgfplotstableread{gra.txt}\table 
			\addplot[mark=none, very thick] table[x index=1,y index=2] from \table;
			\addlegendentry{graphs/networks}
			\pgfplotstableread{heu.txt}\table 
			\addplot[mark=+, very thick, blue] table[x index=1,y index=2] from \table;
			\addlegendentry{heuristics}
			\pgfplotstableread{log.txt}\table 
			\addplot[mark=o, very thick, red] table[x index=1,y index=2] from \table;
			\addlegendentry{logic}
			\pgfplotstableread{pro.txt}\table 
			\addplot[mark=none, very thick, dotted, black] table[x index=1,y index=2] from \table;
			\addlegendentry{probabilistic}
			\pgfplotstableread{vec.txt}\table 
			\addplot[mark=+, very thick, dotted, blue] table[x index=1,y index=2] from \table;
			\addlegendentry{vector space}
			\pgfplotstableread{deep.txt}\table 
			\addplot[mark=o, very thick, dotted, red] table[x index=1,y index=2] from \table;
			\addlegendentry{deep learning}

\end{axis}
\end{tikzpicture}
}
\caption{\label{fig:trends2}Number of publications on the major paradigms used to model semantic dependence in information retrieval (y axis), versus publication year (x axis). The numbers of publications are smoothed by moving 3-year averages.}
\end{figure}

\subsection{Discourse Dependence}
Unlike the long-spanning and rich literature of lexical dependence in information retrieval, discourse dependence has been largely ignored in information retrieval prior to the date of publication of the articles in this dissertation. There was no prior work on automatically inferred discourse transitions for information retrieval, neither for rhetorical relations (contributed by Model IV), nor for coherence (contributed by Models V \& VI). In that sense, Models IV -- VI contribute not only improved semantic processing, but also completely new ways of thinking about text semantics in information retrieval.

Major prior work on discourse semantics \textit{outside} information retrieval is displayed in Table \ref{tab:rw-CH00}. As Figure \ref{fig:trends} graphically shows, advances in this area started emerging in the 1990s, peaked around 2010, and further again more recently. Most of the very recent work uses deep learning (corresponding to its emergence in recent years in Figure \ref{fig:trends2}).

These trends support the reasoning put forward in this dissertation that discourse semantics should not be ignored. Efforts to integrate discourse semantics to information retrieval can benefit from the significant progress made in the field of natural language processing. This line of research has the potential to raise the bar of semantic inferences in search engines, that is required to push information retrieval further beyond searching, in the direction of simulating human intelligence (artificial intelligence - AI).

\subsection{Cognitive Dependence}
 A common starting point when reasoning about cognitive dependence in information retrieval is the \textit{Principle of Polyrepresentation} \cite{Ingwersen:1996}, which posits that the combination of various different cognitive representations of documents and information needs is likely to reveal cognitive overlaps, the semantics of which are discriminative indicators of topical relevance. 
Following the formulation of this principle in the mid-1990s, research in this area has been gaining traction (see Table \ref{tab:rw-CH00}), with a recent peak around 2010 (see Figure \ref{fig:trends}). This peak corresponds to a considerable advance in the field: the \textit{mathematical} formulation of the (up until then) solely \textit{conceptual} formalisation of the Principle of Polyrepresentation. Specifically, two very different mathematical formulations of the principle were presented in the same year and in the proceedings of the same conference: the Quantum model of Frommholz et al. \cite{Frommholz:2010}, and Model IX \cite{Lioma:2010} \cite{LiomaLI12} of this dissertation.

In the Quantum model, cognitive representations are modelled in Hilbert spaces\footnote{\textbf{Hilbert spaces} are a generalisation of Euclidean space to spaces with any finite or infinite number of dimensions.}, and combined by means of their tensor products. This differs from Model IX, where cognitive representations are modelled as subjective beliefs and combined using logical operators of Subjective Logic. Model IX is significantly more efficient and less complex than the Quantum model. 

Overall, the landscape of research in cognitive dependence for information retrieval is different to that of lexical dependence and discourse dependence, in the sense that it is more uniform, characterised mainly by fewer, smaller peaks and relatively steady research interest throughout the years (see Figure \ref{fig:trends}). Table \ref{tab:rw-CH00} highlights the major research in cognitive dependence for information retrieval.
\clearpage

\begin{scriptsize}
\topcaption{Classification of related work on (a) lexical, (b) discourse, and (c) cognitive dependence in information retrieval (second column), according to the approach (third column) used to model dependence. The rows are sorted increasingly by year of publication.} 
\label{tab:rw-CH00}
\tablefirsthead{
\hline 
ARTICLE	&TYPE OF SEMANTIC DEPENDENCE	&TYPE OF APPROACH\\ 
\hline 
}
\tablehead{\multicolumn{3}{c}{
{\captionsize\bfseries \tablename\ \thetable{} -- \textit{continued from previous page}}} \\
\hline  
ARTICLE	&TYPE OF SEMANTIC DEPENDENCE		&TYPE OF APPROACH\\  
\hline 
}
\tablelasthead{
\multicolumn{3}{c}{
{\captionsize\bfseries \tablename\ \thetable{} -- \textit{concluded from previous page}}} \\
\hline   
ARTICLE	&TYPE OF SEMANTIC DEPENDENCE		&TYPE OF APPROACH\\ 
\hline }
 \tabletail{
\hline 
\multicolumn{3}{|c|}{{\textit{Continued on next page}}} \\ 
\hline
}
 \tablelasttail{\hline \hline}
\begin{center}
\begin{xtabular}{lll}
\multirow{2}{*}{Baxendale 1958 \cite{Baxendale1958} }&\multirow{2}{*}{lexical: term associations} &linguistic analysis\\ 
 &&heuristics\\ 
 \arrayrulecolor{gray}\hline
Stiles 1961\cite{Stiles1961} &lexical: term associations&heuristics\\
 \arrayrulecolor{gray}\hline
Doyle 1962 \cite{Doyle1962} &lexical: term associations & graphs/networks\\
 \arrayrulecolor{gray}\hline
\multirow{2}{*}{Guiliano \& Jones 1963 \cite{Giuliano1963}} &\multirow{2}{*}{lexical: term associations} & graphs/networks\\
&&heuristics\\ 
 \arrayrulecolor{gray}\hline
\multirow{2}{*}{Salton 1966 \cite{Salton:1966} }&\multirow{2}{*}{lexical: term associations} &linguistic analysis\\
&&heuristics\\ 
 \arrayrulecolor{gray}\hline
Lesk 1969 \cite{Lesk1969}&lexical: term associations&vector space\\
 \arrayrulecolor{gray}\hline
\multirow{2}{*}{Earl 1972 \cite{Earl1972} }&\multirow{2}{*}{lexical: term associations }&linguistic analysis\\
&&heuristics\\ 
\arrayrulecolor{gray}\hline
Halliday \& Hasan 1976 \cite{HallidayH76}&lexical: term associations&graphs/networks\\
 \arrayrulecolor{gray}\hline
van Rijsbergen 1977 \cite{K1977}&lexical: term associations&heuristics\\
 \arrayrulecolor{gray}\hline
Harper \& van Rijsbergen 1978 \cite{HarperK1978}&lexical: term associations&heuristics\\
 \arrayrulecolor{gray}\hline
\multirow{2}{*}{Hopfield 1982 \cite{Hopfield:1982} }&cognitive: document \&  &\multirow{2}{*}{graphs/networks}\\
& query associations &\\
 \arrayrulecolor{gray}\hline
Salton et al. 1982 \cite{SaltonBY82}&lexical: term dependence&probabilistic\\
 \arrayrulecolor{gray}\hline
Dillon \& Gray 1983 \cite{DillonG:1983} &lexical: term associations &heuristics\\
 \arrayrulecolor{gray}\hline
Yu et al. 1983 \cite{Yu1983}&lexical: term associations&probabilistic\\
 \arrayrulecolor{gray}\hline
Metzler et al. 1984 \cite{MetzlerN:1984} &lexical: term associations &heuristics\\
 \arrayrulecolor{gray}\hline
Sowa 1984 \cite{Sowa:1984} &cognitive: concept associations &graphs/networks\\
 \arrayrulecolor{gray}\hline
Smith \& Devine 1985 \cite{SmithD:1985} &lexical: term associations &heuristics\\
 \arrayrulecolor{gray}\hline
\multirow{2}{*}{Hopfield 1986 \cite{HopfieldT:1986} }&cognitive: document \&  &\multirow{2}{*}{graphs/networks}\\
 &query associations &\\
  \arrayrulecolor{gray}\hline
\multirow{2}{*}{van Rijsbergen 1986 \cite{keith:1986} }& cognitive: document \&  & \multirow{2}{*}{logic}\\
&query associations&\\
 \arrayrulecolor{gray}\hline
Smeaton 1986 \cite{Smeaton:1986} &lexical: term associations &heuristics\\
 \arrayrulecolor{gray}\hline
Huang \& Lippmann 1987 \cite{HuangL:1987} &cognitive: document relations &graphs/networks\\
 \arrayrulecolor{gray}\hline
\multirow{2}{*}{Mann \& Thompson 1988 \cite{MannT88}} &discourse: rhetorical &\multirow{2}{*}{heuristics}\\
&relations&\\
 \arrayrulecolor{gray}\hline
\multirow{2}{*}{Pearl 1988 \cite{Pearl:1988}} &\multirow{2}{*}{cognitive: document \& } &graphs/networks \\
 &query associations&probabilistic\\
 \arrayrulecolor{gray}\hline
\multirow{2}{*}{Saracevic \& Kantor 1988 \cite{ASI:ASI4}}&cognitive: document \&  &\multirow{2}{*}{heuristics}\\
 &query associations&\\
 \arrayrulecolor{gray}\hline
Smeaton \& van Rijsbergen 1988 \cite{SmeatonK:1988} &lexical: term associations &heuristics\\
 \arrayrulecolor{gray}\hline
\multirow{2}{*}{Belew 1989 \cite{Belew:1989} }&cognitive: document \&  &\multirow{2}{*}{graphs/networks}\\
&query associations&\\
 \arrayrulecolor{gray}\hline
Fagan 1989 \cite{Fagan1989} &lexical: term associations &heuristics\\
 \arrayrulecolor{gray}\hline
\multirow{2}{*}{Kwok 1989 \cite{Kwok:1989} }&cognitive: document \&  &\multirow{2}{*}{graphs/networks}\\
&query associations&\\
 \arrayrulecolor{gray}\hline
\multirow{2}{*}{Doszkocs et al. 1990 \cite{DoszkocsR:1990} }&cognitive: document \&  &\multirow{2}{*}{graphs/networks}\\
&query associations&\\
 \arrayrulecolor{gray}\hline
Lewis \& Croft 1990 \cite{LewisC:1990} &lexical: term associations &heuristics\\
 \arrayrulecolor{gray}\hline
Veronis \& Ide 1990 \cite{VeronisI:1990}&lexical: term associations&graphs/networks\\
 \arrayrulecolor{gray}\hline
\multirow{2}{*}{Devlin 1991 \cite{Devlin:1991} }& cognitive: document \&  & \multirow{2}{*}{logic}\\
&query associations&\\
 \arrayrulecolor{gray}\hline
\multirow{2}{*}{Hoey 1991 \cite{Hoey:1991}}	&lexical \& discourse:   &\multirow{2}{*}{graphs/networks }\\
&document cohesion&\\
 \arrayrulecolor{gray}\hline
Lin et al. 1991 \cite{LinS:1991} &cognitive: document relations &graphs/networks\\
 \arrayrulecolor{gray}\hline
Macleod \& Robertson 1991 \cite{MacLeodR:1991} &cognitive: document associations &graphs/networks\\
 \arrayrulecolor{gray}\hline
Sinclair 1991 \cite{Sinclair:1991}&lexical: term associations&graphs/networks\\
 \arrayrulecolor{gray}\hline
Turtle \& Croft 1991 \cite{TurtleC91}&cognitive: document relevance&graphs/networks\\
 \arrayrulecolor{gray}\hline
Wilkinson \& Hingston 1991 \cite{WilkinsonH:1991} &lexical: related terms &graphs/networks\\
 \arrayrulecolor{gray}\hline
\multirow{2}{*}{Bruza \& van der Weide 1992 \cite{BruzaW:1992} }& cognitive: document \& & \multirow{2}{*}{logic}\\
&query associations&\\
 \arrayrulecolor{gray}\hline
\multirow{2}{*}{Chevallet 1992 \cite{Chevallet:1992}} & cognitive: document \& & logic\\
 &query associations&graphs/networks\\
 \arrayrulecolor{gray}\hline
\multirow{2}{*}{Chiaramella \& Chevallet 1992 \cite{ChiaramellaC:1992} }& cognitive: document \&  &\multirow{2}{*}{ logic}\\
 &query associations&\\
 \arrayrulecolor{gray}\hline
  \multirow{2}{*}{Ingwersen 1992 \cite{DBLP:books/tg/Ingwersen92}}&cognitive: document \&  &\multirow{2}{*}{heuristics}\\
&query associations&\\
\arrayrulecolor{gray}\hline
Lewis 1992 \cite{Lewis1992}&lexical: term associations&heuristics\\
 \arrayrulecolor{gray}\hline
\multirow{2}{*}{Nie 1992 \cite{Nie:1992} }& cognitive: document \&  & logic\\
 &query associations &probabilistic\\
 \arrayrulecolor{gray}\hline
\multirow{2}{*}{Belkin et al. 1993 \cite{BelkinCCC93}}&cognitive: document \&  &\multirow{2}{*}{heuristics}\\
&query associations&\\
 \arrayrulecolor{gray}\hline
Kozima 1993 \cite{Kozima:1993}&lexical: term associations&graphs/networks\\
 \arrayrulecolor{gray}\hline
\multirow{2}{*}{Meghini et al. 1993 \cite{MeghiniS:1993} }& cognitive: document \&  & \multirow{2}{*}{logic}\\
&query associations&\\
 \arrayrulecolor{gray}\hline
 \multirow{2}{*}{Ingwersen 1994 \cite{DBLP:conf/sigir/Ingwersen94}}&cognitive: document \&  &\multirow{2}{*}{heuristics}\\
&query associations&\\
\arrayrulecolor{gray}\hline
\multirow{2}{*}{Logan et al. 1994 \cite{LoganR:1994}} & cognitive: document \& & \multirow{2}{*}{logic}\\
&query associations&\\
 \arrayrulecolor{gray}\hline
Losee 1994 \cite{Losee:1994}&lexical: term associations&probabilistic\\
 \arrayrulecolor{gray}\hline
\multirow{2}{*}{Crestani \& van Rijsbergen 1995 \cite{CrestaniK:1995} }& cognitive: document \&  & \multirow{2}{*}{logic}\\
&query associations&\\
 \arrayrulecolor{gray}\hline
Grosz et al. 1995 \cite{grosz:1995} &discourse: document coherence &heuristics\\
 \arrayrulecolor{gray}\hline
\multirow{2}{*}{Muller \& Kutschekmanesch 1995 \cite{MullerK:1995} }& cognitive: document \&  & \multirow{2}{*}{logic}\\
&query associations&\\
 \arrayrulecolor{gray}\hline
Evans \& Zhai 1996 \cite{EvansZ:1996} &lexical: term associations &heuristics\\
 \arrayrulecolor{gray}\hline
\multirow{2}{*}{Huibers et al. 1996 \cite{HuibersL:1996} }& cognitive: document \& & \multirow{2}{*}{logic}\\
&query associations&\\
 \arrayrulecolor{gray}\hline
\multirow{2}{*}{Ingwersen 1996 \cite{Ingwersen:1996}}&cognitive: document \&  &\multirow{2}{*}{heuristics}\\
&query associations&\\
 \arrayrulecolor{gray}\hline
\multirow{2}{*}{Nie et al. 1996 \cite{NieL:1996} }& cognitive: document \&  &\multirow{2}{*}{ logic}\\
&query associations&\\
 \arrayrulecolor{gray}\hline
Pederson et al. 1997 \cite{PedersonS:1997} &lexical: term associations &heuristics\\
 \arrayrulecolor{gray}\hline
Strzalkowski \& Lin 1997 \cite{Str1997} &lexical: term associations &heuristics\\
 \arrayrulecolor{gray}\hline
Tong et al. 1997 \cite{Zhai1997} &lexical: term associations &heuristics\\
 \arrayrulecolor{gray}\hline
\multirow{2}{*}{Crestani \& van Rijsbergen 1998 \cite{CrestaniR:1998} }&cognitive: document \&&graphs/networks\\
&query associations&logic\\
 \arrayrulecolor{gray}\hline
\multirow{2}{*}{Foltz et al. 1998 \cite{FoltzKL98}} &\multirow{2}{*}{discourse: document coherence }&heuristics\\
&&probabilistic\\
 \arrayrulecolor{gray}\hline
\multirow{2}{*}{Lalmas 1998 \cite{Lalmas:1998} }& cognitive: document \& &\multirow{2}{*}{ logic}\\
&query associations&\\
 \arrayrulecolor{gray}\hline
\multirow{2}{*}{van Rijsbergen et al. 1998 \cite{keithC:1998} }& cognitive: document \& &\multirow{2}{*}{ logic}\\
&query associations&\\
 \arrayrulecolor{gray}\hline
\multirow{2}{*}{Ingwersen 1999 \cite{ingwersen99}}&cognitive: document \&  &\multirow{2}{*}{heuristics}\\
&query associations&\\
\arrayrulecolor{gray}\hline
\multirow{2}{*}{Lin 1999 \cite{SongC1999}}&\multirow{2}{*}{lexical: term dependence}&heuristics\\
&&probabilistic\\
 \arrayrulecolor{gray}\hline
Song \& Croft 1999 \cite{SongC1999}&lexical: term dependence&probabilistic\\
 \arrayrulecolor{gray}\hline
Narita \& Ogawa 2000 \cite{NaritaO2000} &lexical: term associations &heuristics\\
 \arrayrulecolor{gray}\hline
Shin \& Stach 2000 \cite{shin:2000} &discourse: document coherence &heuristics\\
 \arrayrulecolor{gray}\hline
Zhu \& Gauch 2000 \cite{ZhuG00} &discourse: document quality &heuristics\\
 \arrayrulecolor{gray}\hline
Dorogovtsev \& Mendes 2001 \cite{DorogovtsevM:2001}&lexical: term associations&graphs/networks\\
 \arrayrulecolor{gray}\hline
Ferrer i Cancho \& Sole 2001 \cite{FerrerS:2001}&lexical: term associations&graphs/networks\\
 \arrayrulecolor{gray}\hline
\multirow{2}{*}{Fujita 2001 \cite{Fujita:2001} }&\multirow{2}{*}{lexical: term associations }& linguistic analysis\\
 &&heuristics\\
 \arrayrulecolor{gray}\hline
Kibble 2001 \cite{Kibble01} &discourse: document coherence &heuristics\\
 \arrayrulecolor{gray}\hline
Lin 2001 \cite{Lin2001} &lexical: term associations &heuristics\\
 \arrayrulecolor{gray}\hline
\multirow{2}{*}{Losada \& Barreiro 2001 \cite{LosadaB:2001} }& cognitive: document \&& \multirow{2}{*}{logic}\\
&query associations&\\
 \arrayrulecolor{gray}\hline
Mikk 2001 \cite{Mikk:2001} &discourse: document quality &heuristics\\
 \arrayrulecolor{gray}\hline
Barzilay et al. 2002 \cite{BarzilayEM02}	&discourse: document coherence &graphs/networks \\
 \arrayrulecolor{gray}\hline
Kehler 2002 \cite{Kehler02} &discourse: document heuristics &heuristics\\
 \arrayrulecolor{gray}\hline
\multirow{2}{*}{Motter et al. 2002 \cite{MotterM:2002}	}&discourse, cognitive: term \& &\multirow{2}{*}{graphs/networks} \\
&concept associations&\\
 \arrayrulecolor{gray}\hline
\multirow{2}{*}{Nallapati \& Allan 2002 \cite{NallapatiA:2002}}&\multirow{2}{*}{lexical: term associations}&graphs/networks\\
&&probabilistic\\
 \arrayrulecolor{gray}\hline
Sigman \& Cecchi 2002 \cite{SigmanC:2002}&lexical: term associations&graphs/networks\\
 \arrayrulecolor{gray}\hline
Teufel \& Moens 2002 \cite{TeufelM02} &discourse: rhetorical relations &heuristics\\
 \arrayrulecolor{gray}\hline
Widdows \& Dorrow 2002 \cite{WiddowsD:2002}&lexical: term associations&graphs/networks\\
 \arrayrulecolor{gray}\hline
Baldwin et al. 2003 \cite{Baldwin:2003} &lexical: term compositionality &vector space\\
 \arrayrulecolor{gray}\hline
Bordag et al. 2003 \cite{BordagH:2003}&lexical: term associations&graphs/networks\\ 
\arrayrulecolor{gray}\hline
\multirow{2}{*}{Lapata 2003 \cite{Lapata03}}&\multirow{2}{*}{discourse: document coherence }&probabilistic\\
&&heuristics\\
 \arrayrulecolor{gray}\hline
McCarthy et al. 2003 \cite{McCarthy:2003} &lexical: phrase compositionality &heuristics\\
 \arrayrulecolor{gray}\hline
Morato et al. 2003 \cite{MoratoL:2003} &discourse: rhetorical relations &heuristics\\
 \arrayrulecolor{gray}\hline
Reddy et al. 2003 \cite{McCarthy:2011}&lexical: term compositionality&vector space\\
\arrayrulecolor{gray}\hline
Srikanth \& Srihari 2003 \cite{SrikanthS:2003}&lexical: term dependence&probabilistic\\
 \arrayrulecolor{gray}\hline
Barzilay \& Lee 2004 \cite{BarzilayL04}&discourse: topic order &probabilistic \\
 \arrayrulecolor{gray}\hline
Blondel et al. 2004 \cite{BlondelG:2004}&lexical: term associations&graphs/networks\\
 \arrayrulecolor{gray}\hline
Erkan \& Radev 2004 \cite{ErkanR:2004}&lexical: term associations&graphs/networks\\
 \arrayrulecolor{gray}\hline
Ferrer i Cancho et al. 2004 \cite{FerrerS:2004}&lexical: term associations&graphs/networks\\
 \arrayrulecolor{gray}\hline
He \& Ounis 2004 \cite{Heo:2004}&lexical: query difficulty &heuristics\\
 \arrayrulecolor{gray}\hline
Ho \& Fairon 2004 \cite{HoC:2004}&lexical: term associations&graphs/networks\\
 \arrayrulecolor{gray}\hline
\multirow{2}{*}{Kibble \& Power 2004 \cite{KibbleP04}}	&\multirow{2}{*}{discourse: document coherence} &linguistic trees \\
&&heuristics\\
 \arrayrulecolor{gray}\hline
Mihalcea \& Tarau 2004 \cite{MihalceaT:2004}&lexical: term associations&graphs/networks\\
 \arrayrulecolor{gray}\hline
Milo et al. 2004 \cite{MiloI:2004}&lexical: term associations&graphs/networks\\
 \arrayrulecolor{gray}\hline
Miltsakaki \& Kukich 2004 \cite{MiltsakakiK04}&discourse: document coherence&heuristics\\
 \arrayrulecolor{gray}\hline
Pang \& Lee 2004 \cite{Pang04} &discourse: &heuristics\\  \arrayrulecolor{gray}\hline
Pedersen et al. 2004 \cite{PedersenP:2004}&lexical: term associations&graphs/networks\\
 \arrayrulecolor{gray}\hline
Poesio et al. 2004 \cite{PoesioSEH04} &discourse: document coherence &heuristics\\
 \arrayrulecolor{gray}\hline
\multirow{2}{*}{van Rijsbergen 2004 \cite{keith:2004} }& cognitive: document \& & \multirow{2}{*}{quantum}\\
&query associations&\\
 \arrayrulecolor{gray}\hline
Tomlinson 2004 \cite{Tomlinson:2004}&lexical: query difficulty &heuristics\\
 \arrayrulecolor{gray}\hline
\multirow{2}{*}{Tsikrika \& Lalmas 2004 \cite{TsikrikaL:2004} }&\multirow{2}{*}{discourse: document relevance }&graphs/networks\\
&&heuristics\\
 \arrayrulecolor{gray}\hline
Caldeira et al. 2005 \cite{CaldeiraP:2006}&lexical: term associations&graphs/networks\\
 \arrayrulecolor{gray}\hline
Ferrer i Cancho 2005 \cite{Ferrer:2005} &discourse: syntactic associations &graphs/networks\\
 \arrayrulecolor{gray}\hline
\multirow{2}{*}{Larsen 2005 \cite{Larsen:2005}}&cognitive: document &\multirow{2}{*}{heuristics}\\
&query associations&\\
 \arrayrulecolor{gray}\hline
\multirow{2}{*}{Larsen \& Ingwersen 2005 \cite{LarsenI:2005}}&cognitive: document &\multirow{2}{*}{heuristics}\\
&query associations&\\
 \arrayrulecolor{gray}\hline
Medeiros Soares et al. 2005 \cite{SoaresC:2005}&lexical: term associations&graphs/networks\\
 \arrayrulecolor{gray}\hline
\multirow{2}{*}{Metzler \& Croft 2005 \cite{MetzlerC:2005}}&\multirow{2}{*}{lexical: term associations}&graphs/networks\\
&&probabilistic\\
 \arrayrulecolor{gray}\hline
Mishne et al. 2005 \cite{MishneR2005} &lexical: term associations&heuristics\\
 \arrayrulecolor{gray}\hline
Mothe \& Tanguy 2005 \cite{MotheT:2005}&lexical: query difficulty &heuristics\\
 \arrayrulecolor{gray}\hline
\multirow{2}{*}{Plachouras \& Ounis 2005 \cite{PlachourasO05}} &\multirow{2}{*}{discourse: document relevance }&logic\\
&&heuristics\\
 \arrayrulecolor{gray}\hline
Popescu \& Etzioni 2005 \cite{PopescuE:2005}	&lexical: phrase sentiment &heuristics \\
 \arrayrulecolor{gray}\hline
\multirow{2}{*}{Reyna \& Brainerd 2005 \cite{ReynaB:2005}}&\multirow{2}{*}{lexical: term associations}&graphs/networks\\
&&logic\\
 \arrayrulecolor{gray}\hline
\multirow{2}{*}{Steyvers \& Tenenbaum 2005 \cite{SteyversT:2005}	}&discourse, cognitive: term \& &\multirow{2}{*}{graphs/networks }\\
&concept associations&\\
 \arrayrulecolor{gray}\hline
 Venkatapathy \& Joshi 2005 \cite{ZhouC05} &lexical: term compositionality &vector space\\
 \arrayrulecolor{gray}\hline
\multirow{2}{*}{Vitevitch \& Rodriguez 2005 \cite{VitevitchR:2005}}&lexical \& cognitive: &\multirow{2}{*}{graphs/networks}\\
&term associations&\\
 \arrayrulecolor{gray}\hline
Wiebe et al. 2005 \cite{wiebe05} &discourse: document sentiment &heuristics\\
 \arrayrulecolor{gray}\hline
Wilson et al. 2005 \cite{WilsonHSKWCCRP05} &discourse: document subjectivity &heuristics\\
 \arrayrulecolor{gray}\hline
Yom-Tov et al. 2005 \cite{Yom-TovF:2005}&lexical: query difficulty &heuristics\\
 \arrayrulecolor{gray}\hline
Zhou \& Croft 2005 \cite{ZhouC05} &discourse: document quality &heuristics\\
 \arrayrulecolor{gray}\hline
Fung \& Ngai 2006 \cite{FungN06}&discourse: topic cohesion&probabilistic\\
 \arrayrulecolor{gray}\hline
Gamon 2006 \cite{Gamon:2006}&lexical: term associations&graphs/networks\\
 \arrayrulecolor{gray}\hline
Goldberg \& Zhu 2006 \cite{GoldbergZ:2006}	&discourse: document sentiment &graphs/networks \\
 \arrayrulecolor{gray}\hline
Hassan \& Banea 2006 \cite{HassanB:2006}&lexical: term associations&graphs/networks\\
 \arrayrulecolor{gray}\hline
Karamanis 2006 \cite{karamanis:2006}&discourse: document coherence &probabilistic \\
 \arrayrulecolor{gray}\hline
Katz \& Giesbrecht 2006 \cite{Katz:automatic} &lexical: term compositionality &vector space\\
 \arrayrulecolor{gray}\hline
\multirow{2}{*}{Larsen et al. 2006 \cite{LarsenI:2006}}&cognitive: document \& &\multirow{2}{*}{heuristics}\\
 &query associations&\\
 \arrayrulecolor{gray}\hline
Leicht et al. 2006 \cite{LeichtH:2006}	&lexical: term associations &graphs/networks \\
 \arrayrulecolor{gray}\hline
Masucci \& Rodgers 2006 \cite{MasucciR:2006}&lexical: term associations&graphs/networks\\
 \arrayrulecolor{gray}\hline
Muller et al. 2006 \cite{MullerH:2006}&lexical: term associations&graphs/networks\\
 \arrayrulecolor{gray}\hline
Nastase et al. 2006 \cite{NastaseS:2006}&lexical: noun associations&graphs/networks\\
 \arrayrulecolor{gray}\hline
Shanahan et al. 2006 \cite{ShanahanQW06} &discourse: document aspect&heuristics\\
 \arrayrulecolor{gray}\hline
\multirow{2}{*}{Soricut \& Marcu 2006 \cite{SoricutM06a}}	&\multirow{2}{*}{discourse: document coherence }&graphs/networks \\
&&probabilistic\\
 \arrayrulecolor{gray}\hline
Wang et al. 2006 \cite{WangLWK06}	&discourse: rhetorical types &graphs/networks \\
 \arrayrulecolor{gray}\hline
Antiqueira et al. 2007 \cite{AntiqueiraP:2007}	&discourse: author attribution &graphs/networks \\
 \arrayrulecolor{gray}\hline
Blanco \& Lioma 2007 \cite{BlancoL:2007}&lexical: term associations&graphs/networks\\
 \arrayrulecolor{gray}\hline
Chen et al. 2007 \cite{ChenSB07} &discourse: document structure &heuristics\\
 \arrayrulecolor{gray}\hline
Choudhury et al. 2007 \cite{ChoudhuryM:2007}&lexical: term associations&graphs/networks\\
 \arrayrulecolor{gray}\hline
Cook et al. 2007 \cite{cook-fazly-stevenson:2007:ACL07-MWE}&lexical: term associations&linguistics\\
 \arrayrulecolor{gray}\hline
Esuli \& Sebastiani 2007 \cite{EsuliS:2007}&lexical: term associations&graphs/networks\\
 \arrayrulecolor{gray}\hline
Ferrer i Cancho et al. 2007 \cite{CanchoC:2007}&lexical: term associations&graphs/networks\\
 \arrayrulecolor{gray}\hline
Filippova \& Strube 2007 \cite{FilippovaS07} &discourse: document coherence &heuristics\\
 \arrayrulecolor{gray}\hline
Hughes \& Ramage 2007 \cite{HughesR:2007}&lexical: term associations&graphs/networks\\
 \arrayrulecolor{gray}\hline
Lioma \& Ounis 2007 \cite{LiomaO07} &lexical: term quality&heuristics\\
 \arrayrulecolor{gray}\hline
Pado \& Lapata 2007 \cite{PadoL:2007}&lexical: term associations&graphs/networks\\
 \arrayrulecolor{gray}\hline
Takamura et al. 2007 \cite{TakamuraI:2007}	&lexical: phrase sentiment &graphs/networks \\
 \arrayrulecolor{gray}\hline
Tao \& Zhai 2007 \cite{TaoZ07} &lexical: term associations &heuristics\\
 \arrayrulecolor{gray}\hline
\multirow{2}{*}{Tapiero 2007 \cite{Tapiero07}}&cognitive, discourse: document \&  &\multirow{2}{*}{heuristics}\\
&query associations&\\
 \arrayrulecolor{gray}\hline
Barzilay \& Lapata 2008 \cite{Barzilay:2008}&discourse: document coherence &probabilistic \\
 \arrayrulecolor{gray}\hline
Blumenstock 2008 \cite{Blumenstock08} &discourse: document quality&heuristics\\
 \arrayrulecolor{gray}\hline
Elsner \& Charniak 2008 \cite{ElsnerC08a}&discourse: document coherence&probabilistic\\
 \arrayrulecolor{gray}\hline
Etzioni et al. 2008 \cite{EtzioniBSW08} &lexical: term associations &heuristics\\
 \arrayrulecolor{gray}\hline
\multirow{2}{*}{Gamon et al. 2008 \cite{GamonBBFHK08} }&discourse: document orientation \& &\multirow{2}{*}{heuristics}\\
&emotion&\\
 \arrayrulecolor{gray}\hline
Gaume 2008 \cite{Gaume:2008}&lexical: term associations&graphs/networks\\
 \arrayrulecolor{gray}\hline
Joyce \& Miyake 2008 \cite{JoyceM:2008}&lexical: term associations&graphs/networks\\
 \arrayrulecolor{gray}\hline
\multirow{2}{*}{Kittur et al. 2008 \cite{KitturSC08}} &discourse: document  &\multirow{2}{*}{heuristics}\\
&trustworthiness&\\
 \arrayrulecolor{gray}\hline
Kozareva et al. 2008 \cite{KozarevaR:2008}	&lexical: term associations &graphs/networks \\
 \arrayrulecolor{gray}\hline
\multirow{2}{*}{Lau et al. 2008 \cite{LauB:2008} }& cognitive: document \& & \multirow{2}{*}{logic}\\
&query associations&\\
 \arrayrulecolor{gray}\hline
\multirow{2}{*}{Linckels \& Meinel 2008 \cite{LinckelsM:2008} }& cognitive: document \& & \multirow{2}{*}{logic}\\
&query associations&\\
 \arrayrulecolor{gray}\hline
Lioma \& van Rijsbergen 2008 \cite{LiomaK:2008} &lexical: term associations &heuristics\\
 \arrayrulecolor{gray}\hline
Minkov \& Cohen 2008 \cite{MinkovC:2008}&lexical: term associations&graphs/networks\\
 \arrayrulecolor{gray}\hline
\multirow{2}{*}{Oussalah et al. 2008 \cite{OussalahK:2008} }& cognitive: document &\multirow{2}{*}{ logic}\\
&query associations&\\
 \arrayrulecolor{gray}\hline
Plaza et al. 2008 \cite{PlazaD:2008}&lexical: term associations&graphs/networks\\
 \arrayrulecolor{gray}\hline
\multirow{2}{*}{Radhouani \& Falquet 2008 \cite{RadhouaniF:2008} }& cognitive: document \& &\multirow{2}{*}{ logic}\\
&query associations&\\
 \arrayrulecolor{gray}\hline
\multirow{2}{*}{Shi et al. 2008 \cite{ShiN:2008}} &discourse: document \& &\multirow{2}{*}{logic}\\
&query associations&\\
 \arrayrulecolor{gray}\hline
\multirow{2}{*}{Simou et al. 2008 \cite{SimouA:2008} }& cognitive: document \& &\multirow{2}{*}{ logic}\\
&query associations&\\
 \arrayrulecolor{gray}\hline
\multirow{2}{*}{Skov et al. 2008 \cite{SkovL:2008}}&cognitive: document \& &\multirow{2}{*}{heuristics}\\
&query associations&\\
 \arrayrulecolor{gray}\hline
\multirow{2}{*}{Zuccon et al. 2008 \cite{ZucconA:2008} }&cognitive: document \& & logic \\
&query associations&quantum\\
 \arrayrulecolor{gray}\hline
Agirre \& Sorroa 2009 \cite{AgirreS:2009}&lexical: term associations&graphs/networks\\
 \arrayrulecolor{gray}\hline
Antiqueira et al. 2009 \cite{Antiqueira:2009}&lexical: term associations&graphs/networks\\
 \arrayrulecolor{gray}\hline
Banik 2009 \cite{Banik09} &discourse: document coherence &heuristics\\
 \arrayrulecolor{gray}\hline
Bicknell \& Levy 2009 \cite{bicknell:2009}&discourse: document coherence &probabilistic \\
 \arrayrulecolor{gray}\hline
Chen et al. 2009 \cite{ChenBBK09}&discourse: document coherence &probabilistic \\
 \arrayrulecolor{gray}\hline
\multirow{2}{*}{Crestani 2009 \cite{Crestani:2009} }& cognitive: document \& &\multirow{2}{*}{ logic}\\
&query associations&\\
 \arrayrulecolor{gray}\hline
\multirow{2}{*}{Diriye et al. 2009 \cite{DiriyeB:2009}}&cognitive: document \& &\multirow{2}{*}{heuristics}\\
&query associations&\\
 \arrayrulecolor{gray}\hline
Juffinger et al. 2009 \cite{JuffingerGL09} &discourse: document credibility &heuristics\\
 \arrayrulecolor{gray}\hline
\multirow{2}{*}{Karamanis et al. 2009 \cite{KaramanisMPO09}}&\multirow{2}{*}{discourse: document coherence }&probabilistic\\
&&heuristics\\
 \arrayrulecolor{gray}\hline
Korkontzelos \& Manandhar 2009 \cite{Korkontzelos:2009:DCM:1667583.1667605}&lexical: term associations&graphs/networks\\
 \arrayrulecolor{gray}\hline
\multirow{2}{*}{Larsen et al. 2009 \cite{LarsenI:2009}}&cognitive: document \& &\multirow{2}{*}{heuristics}\\
&query associations&\\
 \arrayrulecolor{gray}\hline
\multirow{2}{*}{Lecce \& Amato 2009 \cite{LecceA:2009} }& cognitive: document \& & \multirow{2}{*}{logic}\\
&query associations&\\
 \arrayrulecolor{gray}\hline
\multirow{2}{*}{Lioma et al. 2009 \cite{LiomaB:2009} }& cognitive: document \& &\multirow{2}{*}{ logic}\\ 
&query associations&\\
\arrayrulecolor{gray}\hline
\multirow{2}{*}{Lv \& Zhai 2009 \cite{LvZ09}}&\multirow{2}{*}{lexical: term association}&probabilistic\\
&&heuristics\\
 \arrayrulecolor{gray}\hline
Park et al. 2009 \cite{Park09} &discourse: document bias &heuristics\\
 \arrayrulecolor{gray}\hline
Ramage et al. 2009 \cite{RamageR:2009}&lexical: term associations&graphs/networks\\
 \arrayrulecolor{gray}\hline
Sinha et al. 2009 \cite{SinhaP:2009}&lexical: term associations&graphs/networks\\
 \arrayrulecolor{gray}\hline
\multirow{2}{*}{Somasundaran et al. 2009 \cite{SomaN:2009}}	&discourse: discourse relations \& &classification \\
&opinion polarity&heuristics\\
 \arrayrulecolor{gray}\hline
\multirow{2}{*}{Somasundaran et al. 2009 \cite{Som09}} &discourse: discourse relations \& &classification\\
&opinion polarity&heuristics\\
 \arrayrulecolor{gray}\hline
du Verle \& Prendinger 2009 \cite{duVerle:2009}&discourse: rhetorical relations&vector space\\
 \arrayrulecolor{gray}\hline
Wittek et al. 2009 \cite{Wittek:2009:OTB:1693756.1693780} &lexical: term associations &heuristics\\
 \arrayrulecolor{gray}\hline
Yu et al. 2009 \cite{YuW:2009} &discourse: rhetorical relations &heuristics\\
 \arrayrulecolor{gray}\hline
\multirow{2}{*}{Bendersky et al. 2010 \cite{BenderskyMC10}}&\multirow{2}{*}{lexical: term associations}&graphs/networks\\  
&&probabilistic\\ \arrayrulecolor{gray}\hline
Burstein et al. 2010 \cite{burstein:2010}&discourse: document coherence &probabilistic \\  \arrayrulecolor{gray}\hline
Cheung \& Penn 2010 \cite{CheungP10} &discourse: document coherence &heuristics\\  \arrayrulecolor{gray}\hline
Clarke \& Lapata 2010 \cite{ClarkeL:2010} &discourse: rhetorical relations &heuristics\\  \arrayrulecolor{gray}\hline
\multirow{2}{*}{Efron \& Winget 2010 \cite{EfronW:2010}}&cognitive: document \& &\multirow{2}{*}{heuristics}\\ 
&query associations&\\
 \arrayrulecolor{gray}\hline
Ennals et al. 2010 \cite{EnnalsBAR10} &discourse: document controversy &heuristics\\  \arrayrulecolor{gray}\hline
\multirow{2}{*}{Frommholz et al. 2010 \cite{Frommholz:2010} }& cognitive: document \& & \multirow{2}{*}{quantum}\\ 
&query associations&\\ \arrayrulecolor{gray}\hline
Lex et al. 2010 \cite{LexJG10} &discourse: document objectivity &heuristics\\  \arrayrulecolor{gray}\hline
Lipka \& Stein 2010 \cite{LipkaS10} &discourse: document quality &heuristics\\  \arrayrulecolor{gray}\hline
Mitchell \& Lapata 2010 \cite{COGS:COGS1106}&lexical: term compositionality&vector space\\  \arrayrulecolor{gray}\hline
Suwandaratna \& Perera 2010 \cite{SuP:2010}	&discourse: topic identification &graphs/networks \\  \arrayrulecolor{gray}\hline
\multirow{2}{*}{Yessenalina et al. 2010 \cite{Yes10} }&\multirow{2}{*}{discourse: document subjectivity }&heuristics\\
&&classification\\  \arrayrulecolor{gray}\hline
Bendersky et al. 2011 \cite{BenderskyCD11} &discourse: document quality &heuristics\\  \arrayrulecolor{gray}\hline
Celikyilmaz \& Hakkani-Tur 2011 \cite{CelikyilmazH11}&discourse: document coherence&probabilistic\\  \arrayrulecolor{gray}\hline
\multirow{2}{*}{Elsner \& Charniak 2011 \cite{ElsnerC11}}&\multirow{2}{*}{discourse: document coherence}&probabilistic\\
&&heuristics\\  \arrayrulecolor{gray}\hline
\multirow{2}{*}{Elsner \& Charniak 2011 \cite{Elsner:Charniak:2011}}&\multirow{2}{*}{discourse: document coherence}&probabilistic\\ 
&&heuristics\\ \arrayrulecolor{gray}\hline
\multirow{2}{*}{Ghosh et al. 2011 \cite{ghosh2011b}} &\multirow{2}{*}{discourse: argument segmentation }& linguistic analysis\\
&&classification\\  \arrayrulecolor{gray}\hline
\multirow{2}{*}{Heerschop et al. 2011 \cite{HeerschopG:2011} }&discourse: discourse relations & \multirow{2}{*}{heuristics}\\  
&opinion polarity&\\
\arrayrulecolor{gray}\hline
Herzig et al. 2011 \cite{Herzig11} &discourse: document bias &heuristics\\  \arrayrulecolor{gray}\hline
\multirow{2}{*}{Lin et al. 2011 \cite{LinNK11}}&\multirow{2}{*}{discourse: document coherence}&classification\\
&&heuristics\\  \arrayrulecolor{gray}\hline
\multirow{2}{*}{Lioma et al. 2011 \cite{LiomaKS11} }&\multirow{2}{*}{lexical: term associations}&probabilistic\\  
&&heuristics\\ \arrayrulecolor{gray}\hline
\multirow{2}{*}{Michelbacher et al. 2011 \cite{MichelbacherKFLS11}}&\multirow{2}{*}{lexical: term compositionality}&classification\\  
&&heuristics\\ \arrayrulecolor{gray}\hline
Reddy et al. 2011 \cite{DBLP:conf/ijcnlp/ReddyKMM11}&lexical: term compositionality&vector space\\  \arrayrulecolor{gray}\hline
Schwarz \& Morris 2011 \cite{SchwarzM11} &discourse: document credibility &heuristics\\  \arrayrulecolor{gray}\hline
Socher et al. 2011 \cite{socher-EtAl:2012:EMNLP-CoNLL}&lexical: term compositionality&deep learning\\  \arrayrulecolor{gray}\hline
Wang et al. 2011 \cite{WangLKNB11} &discourse: discourse relations &heuristics\\  \arrayrulecolor{gray}\hline
\multirow{2}{*}{Wiebe \& Riloff 2011 \cite{WiebeR11}} &\multirow{2}{*}{discourse: document subjectivity }&classification\\
&&heuristics\\  \arrayrulecolor{gray}\hline
Zhang 2011 \cite{zhang:2011}	&discourse: document coherence &graphs/networks \\  \arrayrulecolor{gray}\hline
\multirow{2}{*}{Zhou et al. 2011 \cite{ZhouLGWW11} }&discourse: rhetorical relations \& &\multirow{2}{*}{heuristics}\\
&document sentiment&\\
  \arrayrulecolor{gray}\hline
Erk 2012 \cite{LNC3:LNC3362} &lexical: term compositionality &vector space\\  \arrayrulecolor{gray}\hline
Lex et al. 2012 \cite{Lex12} &discourse: document quality &heuristics\\  \arrayrulecolor{gray}\hline
\multirow{2}{*}{Lin et al. 2012 \cite{LinLNK12}}&\multirow{2}{*}{discourse: document coherence}&classification\\
&&heuristics\\  
\arrayrulecolor{gray}\hline
\multirow{2}{*}{Lioma et al. 2012 \cite{Lioma12}}&\multirow{2}{*}{discourse: rhetorical relations}&probabilistic\\
&&heuristics\\  \arrayrulecolor{gray}\hline
Louis \& Nenkova 2012 \cite{LouisN12}&discourse: document coherence&probabilistic\\  \arrayrulecolor{gray}\hline
Morris et al. 2012 \cite{MorrisCRHS12} &discourse: document credibility &heuristics\\  \arrayrulecolor{gray}\hline
Tan et al. 2012 \cite{TanGP12} &discourse: document comprehensibility &heuristics\\  \arrayrulecolor{gray}\hline
Guinaudeau \& Strube 2013 \cite{Guinaudeau:Strube:2013}	&discourse: document coherence &graphs/networks \\  \arrayrulecolor{gray}\hline
Horn et al. 2013 \cite{HornZGKL13} &discourse: document factuality &heuristics\\  \arrayrulecolor{gray}\hline
Kiela \& Clarke 2013 \cite{kiela-clark:2013:EMNLP} &lexical: phrase compositionality &vector space\\  \arrayrulecolor{gray}\hline
Krcmar et al. 2013 \cite{krvcmavr-jevzek-pecina:2013:CVSC} &lexical: term compositionality &vector space\\  \arrayrulecolor{gray}\hline
Mikolov et al. 2013 \cite{MikolovSCCD13}&lexical: term compositionality&deep learning \\  \arrayrulecolor{gray}\hline
Oh et al. 2013 \cite{Oh13} &discourse: document trustworthiness &heuristics\\  \arrayrulecolor{gray}\hline
Rousseau \& Vazirgiannis 2013 \cite{RousseauV13}&lexical: term associations&graph theory\\  \arrayrulecolor{gray}\hline
Salehi \& Cook 2013 \cite{Salehi:2013} &lexical: phrase compositionality &heuristics\\  \arrayrulecolor{gray}\hline
Schulte et al. 2013 \cite{Schulte:2013} &lexical: term compositionality&vector space\\  \arrayrulecolor{gray}\hline
Xiong et al. 2013 \cite{xiong:2013}&discourse: document cohesion&probabilistic\\  \arrayrulecolor{gray}\hline
Banea et al. 2014 \cite{BaneaMW14} &discourse: document subjectivity &heuristics\\  \arrayrulecolor{gray}\hline
Chenlo et al. 2014 \cite{ChenloHL14} &discourse: rhetorical relations&heuristics\\ \arrayrulecolor{gray}\hline
Choi et al. 2014 \cite{ChoiCYKL14}&discourse: rhetorical relations&probabilistic\\ \arrayrulecolor{gray}\hline
\multirow{2}{*}{Frommholz \& Abbasi 2014 \cite{FrommholzA14}}&cognitive: document \& query &heuristics\\  
&associations&statistics\\ \arrayrulecolor{gray}\hline
Hill \& Korhonen 2014 \cite{Hill14} &lexical: phrase subjectivity &heuristics\\  \arrayrulecolor{gray}\hline
Kartsaklis 2014 \cite{Kartsaklis14}&lexical: term compositionality&logic\\  \arrayrulecolor{gray}\hline
Li \& Hovy 2014 \cite{LiH14a}&discourse: document coherence &deep learning\\  \arrayrulecolor{gray}\hline
\multirow{2}{*}{Abbasi \& Frommholz 2015 \cite{AbbasiF15}}&cognitive: document \& query &heuristics\\  
&associations&statistics\\ \arrayrulecolor{gray}\hline
Dima 2015 \cite{Dima15}&lexical: term compositionality & vector space\\  \arrayrulecolor{gray}\hline
Eickhoff et al. 2015 \cite{EickhoffVH15} &lexical: term depencence & logic \\  \arrayrulecolor{gray}\hline
\multirow{2}{*}{Hunter et al. 2015 \cite{HunterAL15} }&\multirow{2}{*}{discourse: rhetorical relations }&graphs/networks\\
&&heuristics\\  \arrayrulecolor{gray}\hline
Kuyten et al. 2015 \cite{KuytenBHPA15} &discourse: rhetorical relations&probabilistic \\ \arrayrulecolor{gray}\hline
Le \& Zuidema 2015 \cite{LeZ15} &lexical: term compositionality &deep learning \\ \arrayrulecolor{gray}\hline
\multirow{2}{*}{Lioma et al. 2015 \cite{LiomaSLH15} }&\multirow{2}{*}{lexical: term compositionality }& vector space\\
&&probabilistic \\ \arrayrulecolor{gray}\hline
Mineshima et al. 2015 \cite{MineshimaMMB15}&lexical: term compositionality &logic \\  \arrayrulecolor{gray}\hline
Neelakantan et al. 2015 \cite{NeelakantanRM15}&lexical: term compositionality&vector space\\ \arrayrulecolor{gray}\hline
\multirow{2}{*}{Petersen et al. 2015 \cite{petersen}}&\multirow{2}{*}{discourse: document coherence}&graphs/networks\\  
&&probabilistic\\ \arrayrulecolor{gray}\hline
\multirow{2}{*}{Qiu et al. 2015 \cite{QiuZL15} }&\multirow{2}{*}{lexical: term associations}&vector space, \\
&&deep learning\\  \arrayrulecolor{gray}\hline
Salehi et al. 2015 \cite{salehi-cook-baldwin:2015:NAACL-HLT} &lexical: term compositionality&deep learning\\ \arrayrulecolor{gray}\hline
Voskarides et al. 2015 \cite{VoskaridesMTRW15}	&discourse: entity relations&graphs/networks \\  \arrayrulecolor{gray}\hline
Xiong et al. 2015 \cite{Xiong:2015}&discourse: document coherence&probabilistic\\  \arrayrulecolor{gray}\hline	
Yazdani et al. 2015 \cite{yazdani-farahmand-henderson:2015:EMNLP}&lexical: term compositionality&deep learning \\ \arrayrulecolor{gray}\hline
Zellh{\"{o}}fer 2015 \cite{Zellhofer15} &cognitive: query associations&heuristics\\ \arrayrulecolor{gray}\hline
\multirow{2}{*}{Zhang et al. 2015 \cite{ZhangFQHLH15}}&\multirow{2}{*}{discourse: document coherence}&graphs/networks\\  
&&probabilistic\\ \arrayrulecolor{gray}\hline
Asher et al. 2016 \cite{AsherCBA16} &lexical: term compositionality & semantic theory\\  \arrayrulecolor{gray}\hline
Cordeiro et al. 2016 \cite{CordeiroRIV16}&lexical: term compositionality & deep learning \\  \arrayrulecolor{gray}\hline
\multirow{2}{*}{Ermakova \& Mothe 2016 \cite{ErmakovaM16}}&\multirow{2}{*}{discourse: document structure}&probabilistic,\\  
&&heuristics\\ \arrayrulecolor{gray}\hline
Gutierrez et al. 2016 \cite{GutierrezSMB16}&lexical: term compositionality &vector space\\  \arrayrulecolor{gray}\hline
Hashimoto \& Tsuruoka 2016 \cite{HashimotoT16}&lexical: term compositionality & deep learning \\  \arrayrulecolor{gray}\hline
Lioma et al. 2016 \cite{LiomaTSPL16}&discourse: document coherence&graphs/networks\\  \arrayrulecolor{gray}\hline
Liu \& Huang 2016 \cite{LiuH16}&discourse: sentence dependence&deep learning\\  \arrayrulecolor{gray}\hline
Monroe et al. 2016 \cite{MonroeGP16}&lexical: term compositionality&deep learning\\  \arrayrulecolor{gray}\hline
Nikolaev et al. 2016 \cite{NikolaevKZ16}&lexical: term associations &heuristics \\ \arrayrulecolor{gray}\hline
Paperno \& Baroni 2016 \cite{PapernoB16}&lexical: term compositionality&vector space \\  \arrayrulecolor{gray}\hline
Pavlick \& Callison-Burch 2016 \cite{PavlickC16}&lexical: term compositionality&logic \\  \arrayrulecolor{gray}\hline
Tian et al. 2016 \cite{TianOI16}&lexical: term compositionality &vector space\\ \arrayrulecolor{gray}\hline
Toutanova et al. 2016 \cite{ToutanovaLYPQ16}&lexical: term compositionality & deep learning \\ \arrayrulecolor{gray}\hline
\multirow{2}{*}{Zhang et al. 2016 \cite{ZhangLR16}}&lexical \& discourse: &{deep learning}\\ 
&term \& sentence associations &\\ \arrayrulecolor{gray}\hline
Zhu et al. 2016 \cite{ZhuSG16}&lexical: term compositionality & deep learning \\ \arrayrulecolor{gray}\hline
Lioma and Hansen 2017 \cite{LiomaH17} &lexical: term compositionality & distance metrics\\
 \end{xtabular}
 \end{center}
\end{scriptsize}

\section{Conclusions}
Building machines that can simulate understanding text like humans is an AI-complete problem\footnote{\textbf{AI-complete} refers to artificial intelligence problems whose solution is non-trivial and cannot be approximated by any simple specific algorithm.}. 
A great deal of
research has already gone into this, with astounding results, allowing everyday people to \textit{discuss} with their
telephones, or orally \textit{instruct} their laptops to select and analyse their reading materials. A
prerequisite for processing text semantics, common to all applications of information retrieval, is having some computational
representation of text as an abstract object. Operations on this representation practically correspond to
making semantic inferences, and by extension simulating understanding text. The complexity and
granularity of semantic processing that can be realised is constrained by the mathematical and computational
robustness, expressiveness, and rigour of the tools used. 

This dissertation presents a series of such tools, diverse in their mathematical formulation, but common in their application to model the semantic dependence of words, sentences and concepts in textual information retrieval. These tools are principally expressed in nine distinct models that capture aspects of semantic dependence in highly interpretable and non-complex ways. 
This dissertation further expands beyond these \textit{separate} nine contributions, and presents \textit{embracive} reflections on the following two levels:

\begin{description}
\item{I. FUTURE ANALYTICAL METHODS.} A great amount of research focus is directed towards refining and improving methods of semantic processing by training them on increasingly larger and more challenging data. This requires additional and powerful parameterised adaptation (this was traditionally manually controlled and interpretable, but is increasingly becoming self-adaptive and harder to interpret). 
While this practice is generally useful and attractive, it \textit{should not be the only one}. Justified as its predominance may be, if unchallenged, it risks leading the area into a standstill (there is only so much parameterisation that can be sustained). 

The amalgamation of the body of work presented in this dissertation shows an alternative way of thinking about our methods of analysis: there is a lot to be gained from taking a step back and considering alternative characterisations, representations and solutions to the general problem of text understanding in information retrieval. Above and beyond the specific methods presented in this dissertation, the overarching message is to start thinking about new methods. This is predicated on our \textit{capacity} and our \textit{need} to do so, or in more plain English, \textit{because we can} and \textit{because we must}.
\begin{itemize}
\item We \textit{can} start thinking of new methods because, as fellow disciplines within and outside computer science and mathematics advance, their output is a stream of inspiration for reasoning about text semantics. Our capacity to think of new methods is anything but constrained by such advances.
\item We \textit{must} start thinking of new methods because, if we do not, the research in the area will eventually close in on itself and lose its innovative drive that has been so far pushing it forward.
\end{itemize}

\item{II. FUTURE RESEARCH QUESTIONS.} The amount of core research questions posed in information retrieval today seems notably smaller than the amount of scenarios (or tasks) within which these questions are examined, without this implying that we have arrived at general solutions of these research questions. There is a lot to be gained from asking new questions about text and its semantic interpretation by computers. Not asking new research questions \textit{cannot} advance semantic processing. New inferences are required to address new and unexpected challenges, such as the novel problems pertaining to discourse dependence that this dissertation argues should be solved in textual information retrieval. After all, reliability and validity are properties of \textit{inferences}, not of \textit{methods}.

For an area that has been vigorously investigated for more than 50 years now, it is alarming to see how little reasoning is done outside word counts. Several of the advances described in this dissertation could, in principle, have been made 20 years ago, in the sense that the theoretical tools needed were already in place; it is just that no one thought of asking these questions before. The time is now ripe for new research questions, instead of continuing to hammer the proverbial same old hammer upon the same old nail.
\end{description}

Collectively, the above reflections pave the way for refined semantic processing in information retrieval systems, where shallow semantic processing is traditionally preferred. These insights have the potential to inspire new models of computational processing and analysis that can improve notably the performance of semantic inferences made by search engines and similar technologies. This type of refined semantic processing is expected to be particularly valuable in the emerging field of content-based retrieval that requires systems enabling people not only to \textit{access} but also to \textit{assess} the information they interact with, for instance in terms of its credibility, transparency, or bias \cite{LiomaLLH16}.

\bibliographystyle{abbrv}
\bibliography{Bibliography.bib}  

\begin{thebibliography}{100}

\bibitem{DBLP:conf/naacl/2005}
{\em {HLT/EMNLP} 2005, Human Language Technology Conference and Conference on
  Empirical Methods in Natural Language Processing, Proceedings of the
  Conference, 6-8 October 2005, Vancouver, British Columbia, Canada}. The
  Association for Computational Linguistics, 2005.

\bibitem{DBLP:conf/emnlp/2011}
{\em Proceedings of the 2011 Conference on Empirical Methods in Natural
  Language Processing, {EMNLP} 2011, 27-31 July 2011, John McIntyre Conference
  Centre, Edinburgh, UK, {A} meeting of SIGDAT, a Special Interest Group of the
  {ACL}}. {ACL}, 2011.

\bibitem{DBLP:conf/emnlp/2013}
{\em Proceedings of the 2013 Conference on Empirical Methods in Natural
  Language Processing, {EMNLP} 2013, 18-21 October 2013, Grand Hyatt Seattle,
  Seattle, Washington, USA, {A} meeting of SIGDAT, a Special Interest Group of
  the {ACL}}. {ACL}, 2013.

\bibitem{DBLP:conf/acl/2015-1}
{\em Proceedings of the 53rd Annual Meeting of the Association for
  Computational Linguistics and the 7th International Joint Conference on
  Natural Language Processing of the Asian Federation of Natural Language
  Processing, {ACL} 2015, July 26-31, 2015, Beijing, China, Volume 1: Long
  Papers}. The Association for Computer Linguistics, 2015.

\bibitem{DBLP:conf/acl/2016-1}
{\em Proceedings of the 54th Annual Meeting of the Association for
  Computational Linguistics, {ACL} 2016, August 7-12, 2016, Berlin, Germany,
  Volume 1: Long Papers}. The Association for Computer Linguistics, 2016.

\bibitem{AbbasiF15}
M.~K. Abbasi and I.~Frommholz.
\newblock Cluster-based polyrepresentation as science modelling approach for
  information retrieval.
\newblock {\em Scientometrics}, 102(3):2301--2322, 2015.

\bibitem{AgirreS:2009}
E.~Agirre and A.~Soroa.
\newblock Personalizing pagerank for word sense disambiguation.
\newblock In A.~Lascarides, C.~Gardent, and J.~Nivre, editors, {\em {EACL}
  2009, 12th Conference of the European Chapter of the Association for
  Computational Linguistics, Proceedings of the Conference, Athens, Greece,
  March 30 - April 3, 2009}, pages 33--41. The Association for Computer
  Linguistics, 2009.

\bibitem{DBLP:conf/ictir/2015}
J.~Allan, W.~B. Croft, A.~P. de~Vries, and C.~Zhai, editors.
\newblock {\em Proceedings of the 2015 International Conference on The Theory
  of Information Retrieval, {ICTIR} 2015, Northampton, Massachusetts, USA,
  September 27-30, 2015}. {ACM}, 2015.

\bibitem{DBLP:conf/ictir/2011}
G.~Amati and F.~Crestani, editors.
\newblock {\em Advances in Information Retrieval Theory - Third International
  Conference, {ICTIR} 2011, Bertinoro, Italy, September 12-14, 2011.
  Proceedings}, volume 6931 of {\em Lecture Notes in Computer Science}.
  Springer, 2011.

\bibitem{Antiqueira:2009}
L.~Antiqueira, O.~N. Oliveira, Jr., L.~d.~F. Costa, and M.~d. G.~V. Nunes.
\newblock A complex network approach to text summarization.
\newblock {\em Inf. Sci.}, 179(5):584--599, 2009.

\bibitem{AsherCBA16}
N.~Asher, T.~V. de~Cruys, A.~Bride, and M.~Abrus{\'{a}}n.
\newblock Integrating type theory and distributional semantics: {A} case study
  on adjective-noun compositions.
\newblock {\em Computational Linguistics}, 42(4):703--725, 2016.

\bibitem{DBLP:conf/ictir/2009}
L.~Azzopardi, G.~Kazai, S.~E. Robertson, S.~M. R{\"{u}}ger, M.~Shokouhi,
  D.~Song, and E.~Yilmaz, editors.
\newblock {\em Advances in Information Retrieval Theory, Second International
  Conference on the Theory of Information Retrieval, {ICTIR} 2009, Cambridge,
  UK, September 10-12, 2009, Proceedings}, volume 5766 of {\em Lecture Notes in
  Computer Science}. Springer, 2009.

\bibitem{DBLP:conf/sigir/2015}
R.~A. Baeza{-}Yates, M.~Lalmas, A.~Moffat, and B.~A. Ribeiro{-}Neto, editors.
\newblock {\em Proceedings of the 38th International {ACM} {SIGIR} Conference
  on Research and Development in Information Retrieval, Santiago, Chile, August
  9-13, 2015}. {ACM}, 2015.

\bibitem{DBLP:conf/sigir/2005}
R.~A. Baeza{-}Yates, N.~Ziviani, G.~Marchionini, A.~Moffat, and J.~Tait,
  editors.
\newblock {\em {SIGIR} 2005: Proceedings of the 28th Annual International {ACM}
  {SIGIR} Conference on Research and Development in Information Retrieval,
  Salvador, Brazil, August 15-19, 2005}. {ACM}, 2005.

\bibitem{Baldwin:2003}
T.~Baldwin, C.~Bannard, T.~Tanaka, and D.~Widdows.
\newblock An empirical model of multiword expression decomposability.
\newblock In {\em Proceedings of the ACL 2003 Workshop on Multiword
  Expressions: Analysis, Acquisition and Treatment}, pages 89--96. Association
  for Computational Linguistics, Sapporo, Japan, July 2003.

\bibitem{BaneaMW14}
C.~Banea, R.~Mihalcea, and J.~Wiebe.
\newblock Sense-level subjectivity in a multilingual setting.
\newblock {\em Computer Speech {\&} Language}, 28(1):7--19, 2014.

\bibitem{Banik09}
E.~Banik.
\newblock Extending a surface realizer to generate coherent discourse.
\newblock In {\em Proceedings of the ACL-IJCNLP 2009 Conference Short Papers},
  pages 305--308, Suntec, Singapore, August 2009. Association for Computational
  Linguistics.

\bibitem{BarzilayEM02}
R.~Barzilay, N.~Elhadad, and K.~McKeown.
\newblock Inferring strategies for sentence ordering in multidocument news
  summarization.
\newblock {\em Journal of Artificial Intelligence Research {(JAIR)}},
  17:35--55, 2002.

\bibitem{Barzilay:2008}
R.~Barzilay and M.~Lapata.
\newblock Modeling local coherence: An entity-based approach.
\newblock {\em Computational Linguistics}, 34(1):1--34, 2008.

\bibitem{BarzilayL04}
R.~Barzilay and L.~Lee.
\newblock Catching the drift: Probabilistic content models, with applications
  to generation and summarization.
\newblock In J.~Hirschberg, S.~T. Dumais, D.~Marcu, and S.~Roukos, editors,
  {\em Human Language Technology Conference of the North American Chapter of
  the Association for Computational Linguistics, {HLT-NAACL} 2004, Boston,
  Massachusetts, USA, May 2-7, 2004}, pages 113--120. The Association for
  Computational Linguistics, 2004.

\bibitem{Baxendale1958}
P.~B. Baxendale.
\newblock Machine-made index for technical literature.
\newblock {\em IBM Journal for R\&D}, 2:354--361, 1958.

\bibitem{Belew:1989}
R.~K. Belew.
\newblock Adaptive information retrieval: Using a connectionist representation
  to retrieve and learn about documents.
\newblock In Belkin and van Rijsbergen \cite{DBLP:conf/sigir/89}, pages 11--20.

\bibitem{BelkinCCC93}
N.~J. Belkin, C.~Cool, W.~B. Croft, and J.~P. Callan.
\newblock Effect of multiple query representations on information retrieval
  system performance.
\newblock In Korfhage et~al. \cite{DBLP:conf/sigir/93}, pages 339--346.

\bibitem{DBLP:conf/sigir/92}
N.~J. Belkin, P.~Ingwersen, and A.~M. Pejtersen, editors.
\newblock {\em Proceedings of the 15th Annual International {ACM} {SIGIR}
  Conference on Research and Development in Information Retrieval. Copenhagen,
  Denmark, June 21-24, 1992}. {ACM}, 1992.

\bibitem{DBLP:conf/iiix/2010}
N.~J. Belkin and D.~Kelly, editors.
\newblock {\em Information Interaction in Context Symposium, IIiX 2010, New
  Brunswick, NJ, USA, August 18-21, 2010}. {ACM}, 2010.

\bibitem{DBLP:conf/sigir/89}
N.~J. Belkin and C.~J. van Rijsbergen, editors.
\newblock {\em SIGIR'89, 12th International Conference on Research and
  Development in Information Retrieval, Cambridge, Massachusetts, USA, June
  25-28, 1989, Proceedings}. {ACM}, 1989.

\bibitem{BenderskyCD11}
M.~Bendersky, W.~B. Croft, and Y.~Diao.
\newblock Quality-biased ranking of web documents.
\newblock In I.~King, W.~Nejdl, and H.~Li, editors, {\em Proceedings of the
  Forth International Conference on Web Search and Web Data Mining, {WSDM}
  2011, Hong Kong, China, February 9-12, 2011}, pages 95--104. {ACM}, 2011.

\bibitem{BenderskyMC10}
M.~Bendersky, D.~Metzler, and W.~B. Croft.
\newblock Learning concept importance using a weighted dependence model.
\newblock In B.~D. Davison, T.~Suel, N.~Craswell, and B.~Liu, editors, {\em
  Proceedings of the Third International Conference on Web Search and Web Data
  Mining, {WSDM} 2010, New York, NY, USA, February 4-6, 2010}, pages 31--40.
  {ACM}, 2010.

\bibitem{bicknell:2009}
K.~Bicknell and R.~Levy.
\newblock A model of local coherence effects in human sentence processing as
  consequences of updates from bottom-up prior to posterior beliefs.
\newblock In {\em Proceedings of Human Language Technologies: The 2009 Annual
  Conference of the North American Chapter of the Association for Computational
  Linguistics}, pages 665--673, Boulder, Colorado, June 2009. Association for
  Computational Linguistics.

\bibitem{DBLP:books/crc/p/BlancoBL14}
R.~Blanco, M.~E.~A. Brea, and C.~Lioma.
\newblock User generated content search.
\newblock In M.~Moens, J.~Li, and T.~Chua, editors, {\em Mining User Generated
  Content.}, pages 167--187. Chapman and Hall/CRC, 2014.

\bibitem{BlancoL:2007}
R.~Blanco and C.~Lioma.
\newblock Random walk term weighting for information retrieval.
\newblock In Kraaij et~al. \cite{DBLP:conf/sigir/2007}, pages 829--830.

\bibitem{DBLP:journals/ir/BlancoL09}
R.~Blanco and C.~Lioma.
\newblock Mixed monolingual homepage finding in 34 languages: the role of
  language script and search domain.
\newblock {\em Inf. Retr.}, 12(3):324--351, 2009.

\bibitem{lioma}
R.~Blanco and C.~Lioma.
\newblock Graph-based term weighting for information retrieval.
\newblock {\em Inf. Retr.}, 15(1):54--92, Feb. 2012.

\bibitem{BlondelG:2004}
V.~D. Blondel, A.~Gajardo, M.~Heymans, P.~Senellart, and P.~V. Dooren.
\newblock A measure of similarity between graph vertices: Applications to
  synonym extraction and web searching.
\newblock {\em SIAM Rev.}, 46(4):647--666, 2004.

\bibitem{Blumenstock08}
J.~E. Blumenstock.
\newblock Size matters: word count as a measure of quality on wikipedia.
\newblock In J.~Huai, R.~Chen, H.~Hon, Y.~Liu, W.~Ma, A.~Tomkins, and X.~Zhang,
  editors, {\em Proceedings of the 17th International Conference on World Wide
  Web, {WWW} 2008, Beijing, China, April 21-25, 2008}, pages 1095--1096. {ACM},
  2008.

\bibitem{DBLP:conf/sigir/91}
A.~Bookstein, Y.~Chiaramella, G.~Salton, and V.~V. Raghavan, editors.
\newblock {\em Proceedings of the 14th Annual International {ACM} {SIGIR}
  Conference on Research and Development in Information Retrieval. Chicago,
  Illinois, USA, October 13-16, 1991 (Special Issue of the {SIGIR} Forum)}.
  {ACM}, 1991.

\bibitem{BordagH:2003}
S.~Bordag, G.~Heyer, and U.~Quasthoff.
\newblock Small worlds of concepts and other principles of semantic search.
\newblock In T.~B{\"{o}}hme, G.~Heyer, and H.~Unger, editors, {\em Innovative
  Internet Community Systems, Third International Workshop, {IICS} 2003,
  Leipzig, Germany, June 19-21, 2003, Revised Papers}, volume 2877 of {\em
  Lecture Notes in Computer Science}, pages 10--19. Springer, 2003.

\bibitem{DBLP:conf/sigir/BrostCSL16}
B.~Brost, I.~J. Cox, Y.~Seldin, and C.~Lioma.
\newblock An improved multileaving algorithm for online ranker evaluation.
\newblock In Perego et~al. \cite{DBLP:conf/sigir/2016}, pages 745--748.

\bibitem{DBLP:conf/cikm/BrostSCL16}
B.~Brost, Y.~Seldin, I.~J. Cox, and C.~Lioma.
\newblock Multi-dueling bandits and their application to online ranker
  evaluation.
\newblock In Mukhopadhyay et~al. \cite{DBLP:conf/cikm/2016}, pages 2161--2166.

\bibitem{BruzaW:1992}
P.~Bruza and T.~P. van~der Weide.
\newblock Stratified hypermedia structures for information disclosure.
\newblock {\em Comput. J.}, 35(3):208--220, 1992.

\bibitem{burstein:2010}
J.~Burstein, J.~Tetreault, and S.~Andreyev.
\newblock Using entity-based features to model coherence in student essays.
\newblock In {\em Human Language Technologies: The 2010 Annual Conference of
  the North American Chapter of the Association for Computational Linguistics},
  pages 681--684, Los Angeles, California, June 2010. Association for
  Computational Linguistics.

\bibitem{CaldeiraP:2006}
S.~M.~G. Caldeira, T.~C.~P. Lob{\~a}o, R.~F.~S. Andrade, A.~Neme, and J.~G.~V.
  Miranda.
\newblock The network of concepts in written texts.
\newblock {\em The European Physical Journal B - Condensed Matter and Complex
  Systems}, 49(4):523--529, 2006.

\bibitem{CelikyilmazH11}
A.~{\c{C}}elikyilmaz and D.~Hakkani{-}T{\"{u}}r.
\newblock Discovery of topically coherent sentences for extractive
  summarization.
\newblock In Lin et~al. \cite{DBLP:conf/acl/2011}, pages 491--499.

\bibitem{ChenSB07}
E.~Chen, B.~Snyder, and R.~Barzilay.
\newblock Incremental text structuring with online hierarchical ranking.
\newblock In {\em Proceedings of the 2007 Joint Conference on Empirical Methods
  in Natural Language Processing and Computational Natural Language Learning
  (EMNLP-CoNLL)}, pages 83--91, Prague, Czech Republic, June 2007. Association
  for Computational Linguistics.

\bibitem{ChenBBK09}
H.~Chen, S.~Branavan, R.~Barzilay, and D.~R. Karger.
\newblock Global models of document structure using latent permutations.
\newblock In {\em Proceedings of Human Language Technologies: The 2009 Annual
  Conference of the North American Chapter of the Association for Computational
  Linguistics}, pages 371--379, Boulder, Colorado, June 2009. Association for
  Computational Linguistics.

\bibitem{ChenloHL14}
J.~M. Chenlo, A.~Hogenboom, and D.~E. Losada.
\newblock Rhetorical structure theory for polarity estimation: An experimental
  study.
\newblock {\em Data Knowl. Eng.}, 94:135--147, 2014.

\bibitem{CheungP10}
J.~C.~K. Cheung and G.~Penn.
\newblock Entity-based local coherence modelling using topological fields.
\newblock In {\em Proceedings of the 48th Annual Meeting of the Association for
  Computational Linguistics}, pages 186--195, Uppsala, Sweden, July 2010.
  Association for Computational Linguistics.

\bibitem{Chevallet:1992}
J.~P. Chevallet.
\newblock {\em {Un Mod\`{e}le Logique de Recherche d'Information Appliqu\'{e}
  au Formalisme des Graphes Conceptuels. Le Prototype ELEN et son
  Experimentation sur un Corpus de Composants Logiciels}}.
\newblock {PhD} in {C}omputer {S}cience, University of Joseph Fourier, Grenoble
  I, 1992.

\bibitem{ChiaramellaC:1992}
Y.~Chiaramella and J.~P. Chevallet.
\newblock {About Retrieval Models and Logic}.
\newblock {\em Comput. J.}, 35(3):233--242, 1992.

\bibitem{ChoiCYKL14}
S.~Choi, J.~Choi, S.~Yoo, H.~Kim, and Y.~Lee.
\newblock Semantic concept-enriched dependence model for medical information
  retrieval.
\newblock {\em Journal of Biomedical Informatics}, 47:18--27, 2014.

\bibitem{ChoudhuryM:2007}
M.~Choudhury, M.~Thomas, A.~Mukherjee, A.~Basu, and N.~Ganguly.
\newblock How difficult is it to develop a perfect spell-checker? a
  cross-linguistic analysis through complex network approach.
\newblock In {\em Proceedings of the Second Workshop on TextGraphs: Graph-Based
  Algorithms for Natural Language Processing}, pages 81--88, Rochester, NY,
  USA, 2007. Association for Computational Linguistics.

\bibitem{ClarkeL:2010}
J.~Clarke and M.~Lapata.
\newblock Discourse constraints for document compression.
\newblock {\em Computational Linguistics}, 36(3):411--441, 2010.

\bibitem{cook-fazly-stevenson:2007:ACL07-MWE}
P.~Cook, A.~Fazly, and S.~Stevenson.
\newblock Pulling their weight: Exploiting syntactic forms for the automatic
  identification of idiomatic expressions in context.
\newblock In {\em Proceedings of the Workshop on A Broader Perspective on
  Multiword Expressions}, pages 41--48, Prague, Czech Republic, June 2007.
  Association for Computational Linguistics.

\bibitem{CordeiroRIV16}
S.~Cordeiro, C.~Ramisch, M.~Idiart, and A.~Villavicencio.
\newblock Predicting the compositionality of nominal compounds: Giving word
  embeddings a hard time.
\newblock In {\em Proceedings of the 54th Annual Meeting of the Association for
  Computational Linguistics, {ACL} 2016, August 7-12, 2016, Berlin, Germany,
  Volume 1: Long Papers\/} \cite{DBLP:conf/acl/2016-1}.

\bibitem{Cox:2006}
D.~R. Cox.
\newblock {\em Principles of Statistical Inference}.
\newblock Cambridge University Press, 2006.

\bibitem{Crestani:2009}
F.~Crestani.
\newblock {Logical Models of Information Retrieval}.
\newblock In L.~Liu and M.~T. {\"O}zsu, editors, {\em Encyclopedia of Database
  Systems}, pages 1652--1658. Springer, 2009.

\bibitem{CrestaniK:1995}
F.~Crestani and C.~J. van Rijsbergen.
\newblock Information retrieval by logical imaging.
\newblock {\em Journal of Documentation}, 51(1):3--17, 1995.

\bibitem{CrestaniR:1998}
F.~Crestani and C.~J. van Rijsbergen.
\newblock A study of probability kinematics in information retrieval.
\newblock {\em {ACM} Trans. Inf. Syst.}, 16(3):225--255, 1998.

\bibitem{CroftT:1991}
W.~B. Croft, H.~R. Turtle, and D.~D. Lewis.
\newblock The use of phrases and structured queries in information retrieval.
\newblock In Bookstein et~al. \cite{DBLP:conf/sigir/91}, pages 32--45.

\bibitem{DBLP:conf/sigir/94}
W.~B. Croft and C.~J. van Rijsbergen, editors.
\newblock {\em Proceedings of the 17th Annual International {ACM-SIGIR}
  Conference on Research and Development in Information Retrieval. Dublin,
  Ireland, 3-6 July 1994 (Special Issue of the {SIGIR} Forum)}. ACM/Springer,
  1994.

\bibitem{DaviesGMW12}
J.~Davies, J.~Gibbons, D.~Milward, and J.~Welch.
\newblock Compositionality and refinement in model-driven engineering.
\newblock In R.~Gheyi and D.~A. Naumann, editors, {\em Formal Methods:
  Foundations and Applications - 15th Brazilian Symposium, {SBMF} 2012, Natal,
  Brazil, September 23-28, 2012. Proceedings}, volume 7498 of {\em Lecture
  Notes in Computer Science}, pages 99--114. Springer, 2012.

\bibitem{dell2013cognition}
P.~Dell'Aversana.
\newblock {\em Cognition in Geosciences: The feeding loop between
  geo-disciplines, cognitive sciences and epistemology}.
\newblock Elsevier Science, 2013.

\bibitem{Devlin:1991}
K.~Devlin.
\newblock {\em Logic and information}.
\newblock Cambridge University Press, 1991.

\bibitem{DBLP:conf/starsem/2013}
M.~T. Diab, T.~Baldwin, and M.~Baroni, editors.
\newblock {\em Proceedings of the Second Joint Conference on Lexical and
  Computational Semantics, *SEM 2013, June 13-14, 2013, Atlanta, Georgia,
  {USA}}. Association for Computational Linguistics, 2013.

\bibitem{DillonG:1983}
M.~Dillon and A.~S. Gray.
\newblock {FASIT:} {A} fully automatic syntactically based indexing system.
\newblock {\em {JASIS}}, 34(2):99--108, 1983.

\bibitem{Dima15}
C.~Dima.
\newblock Reverse-engineering language: {A} study on the semantic
  compositionality of german compounds.
\newblock In M{\`{a}}rquez et~al. \cite{DBLP:conf/emnlp/2015}, pages
  1637--1642.

\bibitem{DiriyeB:2009}
A.~Diriye, A.~Blandford, and A.~Tombros.
\newblock A polyrepresentational approach to interactive query expansion.
\newblock In F.~Heath, M.~L. Rice{-}Lively, and R.~Furuta, editors, {\em
  Proceedings of the 2009 Joint International Conference on Digital Libraries,
  {JCDL} 2009, Austin, TX, USA, June 15-19, 2009}, pages 217--220. {ACM}, 2009.

\bibitem{DorogovtsevM:2001}
S.~N. Dorogovtsev and J.~F.~F. Mendes.
\newblock Language as an evolving word web.
\newblock {\em Proceedings of The Royal Society of London. Series B, Biological
  Sciences}, 268(1485):2603--2606, December 2001.

\bibitem{DoszkocsR:1990}
T.~E. Doszkocs, J.~Reggia, and X.~Lin.
\newblock Connectionist models and information retrieval.
\newblock {\em Annual Review of Information Science and Technology (ARIST)},
  25:209--260, 1990.

\bibitem{Doyle1962}
L.~B. Doyle.
\newblock Indexing and abstracting by association. part i.
\newblock {\em Am. Doc.}, 13:378--390, 1962.

\bibitem{DBLP:journals/ijmi/DragusinPLLJCHIW13}
R.~Dragusin, P.~Petcu, C.~Lioma, B.~Larsen, H.~J{\o}rgensen, I.~J. Cox, L.~K.
  Hansen, P.~Ingwersen, and O.~Winther.
\newblock Findzebra: {A} search engine for rare diseases.
\newblock {\em I. J. Medical Informatics}, 82(6):528--538, 2013.

\bibitem{DragusinPLLJCHIW13}
R.~Dragusin, P.~Petcu, C.~Lioma, B.~Larsen, H.~J{\o}rgensen, I.~J. Cox, L.~K.
  Hansen, P.~Ingwersen, and O.~Winther.
\newblock Specialised tools are needed when searching the web for rare disease
  diagnoses.
\newblock {\em Rare Diseases}, 1(2):e25001--1 -- e25001--4, 2013.

\bibitem{DBLP:conf/ictir/DragusinPLLJW11}
R.~Dragusin, P.~Petcu, C.~Lioma, B.~Larsen, H.~J{\o}rgensen, and O.~Winther.
\newblock Rare disease diagnosis as an information retrieval task.
\newblock In Amati and Crestani \cite{DBLP:conf/ictir/2011}, pages 356--359.

\bibitem{duVerle:2009}
D.~A. duVerle and H.~Prendinger.
\newblock A novel discourse parser based on support vector machine
  classification.
\newblock In {\em Proceedings of the Joint Conference of the 47th Annual
  Meeting of the ACL and the 4th International Joint Conference on Natural
  Language Processing of the AFNLP: Volume 2 - Volume 2}, ACL '09, pages
  665--673, Stroudsburg, PA, USA, 2009. Association for Computational
  Linguistics.

\bibitem{Earl1972}
L.~L. Earl.
\newblock The resolution of syntactic ambiguity in automatic language
  processing.
\newblock {\em Information Storage and Retrieval}, 8:277--308, 1972.

\bibitem{EfronW:2010}
M.~Efron and M.~A. Winget.
\newblock Query polyrepresentation for ranking retrieval systems without
  relevance judgments.
\newblock {\em {JASIST}}, 61(6):1081--1091, 2010.

\bibitem{EickhoffVH15}
C.~Eickhoff, A.~P. de~Vries, and T.~Hofmann.
\newblock Modelling term dependence with copulas.
\newblock In Baeza{-}Yates et~al. \cite{DBLP:conf/sigir/2015}, pages 783--786.

\bibitem{ElsnerC08a}
M.~Elsner and E.~Charniak.
\newblock Coreference-inspired coherence modeling.
\newblock In {\em Proceedings of ACL-08: HLT, Short Papers}, pages 41--44,
  Columbus, Ohio, June 2008. Association for Computational Linguistics.

\bibitem{ElsnerC11}
M.~Elsner and E.~Charniak.
\newblock Disentangling chat with local coherence models.
\newblock In {\em Proceedings of the 49th Annual Meeting of the Association for
  Computational Linguistics: Human Language Technologies}, pages 1179--1189,
  Portland, Oregon, USA, June 2011. Association for Computational Linguistics.

\bibitem{Elsner:Charniak:2011}
M.~Elsner and E.~Charniak.
\newblock Extending the entity grid with entity-specific features.
\newblock In {\em Proceedings of the 49th Annual Meeting of the Association for
  Computational Linguistics: Human Language Technologies}, pages 125--129,
  Portland, Oregon, USA, June 2011. Association for Computational Linguistics.

\bibitem{EnnalsBAR10}
R.~Ennals, D.~Byler, J.~M. Agosta, and B.~Rosario.
\newblock What is disputed on the web?
\newblock In K.~Tanaka, X.~Zhou, M.~Zhang, and A.~Jatowt, editors, {\em
  Proceedings of the 4th {ACM} Workshop on Information Credibility on the Web,
  {WICOW} 2010, Raleigh, North Carolina, USA, April 27, 2010}, pages 67--74.
  {ACM}, 2010.

\bibitem{LNC3:LNC3362}
K.~Erk.
\newblock Vector space models of word meaning and phrase meaning: A survey.
\newblock {\em Language and Linguistics Compass}, 6(10):635--653, 2012.

\bibitem{ErkanR:2004}
G.~Erkan and D.~R. Radev.
\newblock Lexrank: Graph-based lexical centrality as salience in text
  summarization.
\newblock {\em J. Artif. Intell. Res. {(JAIR)}}, 22:457--479, 2004.

\bibitem{ErmakovaM16}
L.~Ermakova and J.~Mothe.
\newblock Document re-ranking based on topic-comment structure.
\newblock In {\em Tenth {IEEE} International Conference on Research Challenges
  in Information Science, {RCIS} 2016, Grenoble, France, June 1-3, 2016}, pages
  1--10. {IEEE}, 2016.

\bibitem{EsuliS:2007}
A.~Esuli and F.~Sebastiani.
\newblock Pageranking wordnet synsets: An application to opinion mining.
\newblock In {\em Proceedings of the 45th Annual Meeting of the Association of
  Computational Linguistics}, pages 424--431, Prague, Czech Republic, June
  2007. Association for Computational Linguistics.

\bibitem{EtzioniBSW08}
O.~Etzioni, M.~Banko, S.~Soderland, and D.~S. Weld.
\newblock Open information extraction from the web.
\newblock {\em Commun. {ACM}}, 51(12):68--74, 2008.

\bibitem{EvansZ:1996}
D.~A. Evans and C.~Zhai.
\newblock Noun-phrase analysis in unrestricted text for information retrieval.
\newblock In A.~K. Joshi and M.~Palmer, editors, {\em 34th Annual Meeting of
  the Association for Computational Linguistics, 24-27 June 1996, University of
  California, Santa Cruz, California, USA, Proceedings.}, pages 17--24. Morgan
  Kaufmann Publishers / {ACL}, 1996.

\bibitem{Fagan1989}
J.~L. Fagan.
\newblock The effectiveness of a nonsyntactic approach to automatic phrase
  indexing for document retrieval.
\newblock {\em {JASIS}}, 40(2):115--132, 1989.

\bibitem{farahmand-smith-nivre:2015:MWE}
M.~Farahmand, A.~Smith, and J.~Nivre.
\newblock A multiword expression data set: Annotating non-compositionality and
  conventionalization for english noun compounds.
\newblock In V.~Kordoni, K.~Cholakov, M.~Egg, S.~Markantonatou, and S.~Wintner,
  editors, {\em Proceedings of the 11th Workshop on Multiword Expressions,
  MWE@NAACL-HLT 2015, June 4, 2015, Denver, Colorado, {USA}}, pages 29--33. The
  Association for Computer Linguistics, 2015.

\bibitem{Ferrer:2005}
R.~{Ferrer-i-Cancho}.
\newblock The structure of syntactic dependency networks: insights from recent
  advances in network theory.
\newblock {\em The Problems of Quantitative Linguistics}, pages 60--75, 2005.

\bibitem{CanchoC:2007}
R.~Ferrer{-}i{-}Cancho, A.~Capocci, and G.~Caldarelli.
\newblock Spectral methods cluster words of the same class in a syntactic
  dependency network.
\newblock {\em International Journal of Bifurcation and Chaos},
  17(7):2453--2463, 2007.

\bibitem{FerrerS:2001}
R.~Ferrer{-}i{-}Cancho and R.~V. Sol{\'{e}}.
\newblock Two regimes in the frequency of words and the origins of complex
  lexicons: Zipf's law revisited.
\newblock {\em Journal of Quantitative Linguistics}, 8(3):165--173, 2001.

\bibitem{FerrerS:2004}
R.~{Ferrer i Cancho}, R.~V. Sol\'e, and R.~K\"ohler.
\newblock Patterns in syntactic dependency networks.
\newblock {\em Phys. Rev. E}, 69(5):051915, 2004.

\bibitem{FilippovaS07}
K.~Filippova and M.~Strube.
\newblock The german vorfeld and local coherence.
\newblock {\em Journal of Logic, Language and Information}, 16(4):465--485,
  2007.

\bibitem{Firth:1968}
J.~R. Firth.
\newblock {A} synopsis of {L}inguistic {T}heory.
\newblock In F.~R. Palmer, editor, {\em Selected papers of J.R. Firth
  1952-1959}, pages 168--205. London:Longmans, 1968b.

\bibitem{FoltzKL98}
P.~Foltz, W.~Kintsch, and T.~Landauer.
\newblock The measurement of textual coherence with latent semantic analysis.
\newblock {\em Discourse Processes}, 25(2\&3):285--307, 1998.

\bibitem{FrommholzA14}
I.~Frommholz and M.~K. Abbasi.
\newblock On clustering and polyrepresentation.
\newblock In M.~de~Rijke, T.~Kenter, A.~P. de~Vries, C.~Zhai, F.~de~Jong,
  K.~Radinsky, and K.~Hofmann, editors, {\em Advances in Information Retrieval
  - 36th European Conference on {IR} Research, {ECIR} 2014, Amsterdam, The
  Netherlands, April 13-16, 2014. Proceedings}, volume 8416 of {\em Lecture
  Notes in Computer Science}, pages 618--623. Springer, 2014.

\bibitem{Frommholz:2010}
I.~Frommholz, B.~Larsen, B.~Piwowarski, M.~Lalmas, P.~Ingwersen, and K.~van
  Rijsbergen.
\newblock Supporting polyrepresentation in a quantum-inspired geometrical
  retrieval framework.
\newblock In Belkin and Kelly \cite{DBLP:conf/iiix/2010}, pages 115--124.

\bibitem{Fujita:2001}
S.~Fujita.
\newblock More reflections on "aboutness" {TREC-2001} evaluation experiments at
  justsystem.
\newblock In E.~M. Voorhees and D.~K. Harman, editors, {\em Proceedings of The
  Tenth Text REtrieval Conference, {TREC} 2001, Gaithersburg, Maryland, USA,
  November 13-16, 2001}, volume Special Publication 500-250. National Institute
  of Standards and Technology {(NIST)}, 2001.

\bibitem{FungN06}
P.~Fung and G.~Ngai.
\newblock One story, one flow: Hidden markov story models for multilingual
  multidocument summarization.
\newblock {\em {TSLP}}, 3(2):1--16, 2006.

\bibitem{Gamon:2006}
M.~Gamon.
\newblock Graph-based text representation for novelty detection.
\newblock In {\em Proceedings of TextGraphs: the First Workshop on Graph Based
  Methods for Natural Language Processing}, pages 17--24, New York City, June
  2006. Association for Computational Linguistics.

\bibitem{GamonBBFHK08}
M.~Gamon, S.~Basu, D.~Belenko, D.~Fisher, M.~Hurst, and A.~C. K{\"{o}}nig.
\newblock {BLEWS:} using blogs to provide context for news articles.
\newblock In E.~Adar, M.~Hurst, T.~Finin, N.~S. Glance, N.~Nicolov, and B.~L.
  Tseng, editors, {\em Proceedings of the Second International Conference on
  Weblogs and Social Media, {ICWSM} 2008, Seattle, Washington, USA, March 30 -
  April 2, 2008}. The {AAAI} Press, 2008.

\bibitem{Gaume:2008}
B.~Gaume.
\newblock Mapping the forms of meaning in small worlds.
\newblock {\em Int. J. Intell. Syst.}, 23(7):848--862, 2008.

\bibitem{ghosh2011b}
S.~Ghosh, R.~Johansson, G.~Riccardi, and S.~Tonelli.
\newblock Shallow discourse parsing with conditional random fields.
\newblock In {\em Proceedings of 5th International Joint Conference on Natural
  Language Processing}, pages 1071--1079, Chiang Mai, Thailand, November 2011.
  Asian Federation of Natural Language Processing.

\bibitem{Giuliano1963}
V.~E. Giuliano and P.~E. Jones.
\newblock Linear associative ir.
\newblock {\em Vistas in Information Handling: The Augmentation of Man's
  Intellect by Machine}, 1:30--54, 1963.

\bibitem{GoldbergZ:2006}
A.~Goldberg and X.~Zhu.
\newblock Seeing stars when there aren't many stars: Graph-based
  semi-supervised learning for sentiment categorization.
\newblock In {\em Proceedings of TextGraphs: the First Workshop on Graph Based
  Methods for Natural Language Processing}, pages 45--52, New York City, June
  2006. Association for Computational Linguistics.

\bibitem{grosz:1995}
B.~J. Grosz, A.~K. Joshi, and S.~Weinstein.
\newblock Centering: {A} framework for modeling the local coherence of
  discourse.
\newblock {\em Computational Linguistics}, 21(2):203--225, 1995.

\bibitem{Guinaudeau:Strube:2013}
C.~Guinaudeau and M.~Strube.
\newblock Graph-based local coherence modeling.
\newblock In {\em Proceedings of the 51st Annual Meeting of the Association for
  Computational Linguistics, {ACL} 2013, 4-9 August 2013, Sofia, Bulgaria,
  Volume 1: Long Papers}, pages 93--103. The Association for Computer
  Linguistics, 2013.

\bibitem{DBLP:journals/ipm/GunsLL12}
R.~Guns, C.~Lioma, and B.~Larsen.
\newblock The tipping point: F-score as a function of the number of retrieved
  items.
\newblock {\em Inf. Process. Manage.}, 48(6):1171--1180, 2012.

\bibitem{GutierrezSMB16}
E.~D. Guti{\'{e}}rrez, E.~Shutova, T.~Marghetis, and B.~Bergen.
\newblock Literal and metaphorical senses in compositional distributional
  semantic models.
\newblock In {\em Proceedings of the 54th Annual Meeting of the Association for
  Computational Linguistics, {ACL} 2016, August 7-12, 2016, Berlin, Germany,
  Volume 1: Long Papers\/} \cite{DBLP:conf/acl/2016-1}.

\bibitem{HallidayH76}
M.~A.~K. Halliday and R.~Hasan.
\newblock {\em Cohesion in English}.
\newblock Longman, London, 1976.

\bibitem{DBLP:conf/irfc/HansenLLA14}
N.~D. Hansen, C.~Lioma, B.~Larsen, and S.~Alstrup.
\newblock Temporal context for authorship attribution - {A} study of danish
  secondary schools.
\newblock In D.~Lamas and P.~Buitelaar, editors, {\em Multidisciplinary
  Information Retrieval - 7th Information Retrieval Facility Conference, {IRFC}
  2014, Copenhagen, Denmark, November 10-12, 2014, Proceedings}, volume 8849 of
  {\em Lecture Notes in Computer Science}, pages 22--40. Springer, 2014.

\bibitem{DBLP:conf/cikm/HansenLM16}
N.~D. Hansen, C.~Lioma, and K.~M{\o}lbak.
\newblock Ensemble learned vaccination uptake prediction using web search
  queries.
\newblock In Mukhopadhyay et~al. \cite{DBLP:conf/cikm/2016}, pages 1953--1956.

\bibitem{hansen2017time}
N.~D. Hansen, K.~M{\o}lbak, I.~J. Cox, and C.~Lioma.
\newblock Time-series adaptive estimation of vaccination uptake using web
  search queries.
\newblock {\em International ACM Conference of the World Wide Web}, page in
  press, 2017.

\bibitem{HarperK1978}
D.~J. Harper and C.~J.~K. van Rijsbergen.
\newblock An evaluation of feedback in document retrieval using concurrence
  data.
\newblock {\em Journal of Documentation}, 34:189--216, 1978.

\bibitem{harris:1981}
Z.~S. Harris.
\newblock {\em Distributional structure}.
\newblock Springer, 1981.

\bibitem{HashimotoT16}
K.~Hashimoto and Y.~Tsuruoka.
\newblock Adaptive joint learning of compositional and non-compositional phrase
  embeddings.
\newblock In {\em Proceedings of the 54th Annual Meeting of the Association for
  Computational Linguistics, {ACL} 2016, August 7-12, 2016, Berlin, Germany,
  Volume 1: Long Papers\/} \cite{DBLP:conf/acl/2016-1}.

\bibitem{HassanB:2006}
S.~Hassan and C.~Banea.
\newblock Random-walk term weighting for improved text classification.
\newblock In {\em Proceedings of TextGraphs: the First Workshop on Graph Based
  Methods for Natural Language Processing}, pages 53--60, New York City, June
  2006. Association for Computational Linguistics.

\bibitem{Heo:2004}
B.~He and I.~Ounis.
\newblock Inferring query performance using pre-retrieval predictors.
\newblock In A.~Apostolico and M.~Melucci, editors, {\em String Processing and
  Information Retrieval, 11th International Conference, {SPIRE} 2004, Padova,
  Italy, October 5-8, 2004, Proceedings}, volume 3246 of {\em Lecture Notes in
  Computer Science}, pages 43--54. Springer, 2004.

\bibitem{HeerschopG:2011}
B.~Heerschop, F.~Goossen, A.~Hogenboom, F.~Frasincar, U.~Kaymak, and
  F.~de~Jong.
\newblock Polarity analysis of texts using discourse structure.
\newblock In {\em Proceedings of the 20th ACM international conference on
  Information and knowledge management}, CIKM '11, pages 1061--1070, New York,
  NY, USA, 2011. ACM.

\bibitem{DBLP:conf/sigir/2012}
W.~R. Hersh, J.~Callan, Y.~Maarek, and M.~Sanderson, editors.
\newblock {\em The 35th International {ACM} {SIGIR} conference on research and
  development in Information Retrieval, {SIGIR} '12, Portland, OR, USA, August
  12-16, 2012}. {ACM}, 2012.

\bibitem{Herzig11}
L.~Herzig, A.~Nunes, and B.~Snir.
\newblock An annotation scheme for automated bias detection in wikipedia.
\newblock In {\em Proceedings of the Fifth Linguistic Annotation Workshop,
  {LAW} 2011, June 23-24, 2011, Portland, Oregon, {USA}}, pages 47--55. The
  Association for Computer Linguistics, 2011.

\bibitem{Hill14}
F.~Hill and A.~Korhonen.
\newblock Concreteness and subjectivity as dimensions of lexical meaning.
\newblock In {\em Proceedings of the 52nd Annual Meeting of the Association for
  Computational Linguistics (Volume 2: Short Papers)}, pages 725--731,
  Baltimore, Maryland, June 2014. Association for Computational Linguistics.

\bibitem{HoC:2004}
N.~Ho and C.~Fairon.
\newblock Lexical similarity based on quantity of information exchanged -
  synonym extraction.
\newblock In {\em Actes de la Deuxieme Conference Internationale Associant
  Chercheurs Vietnamiens et Francophones en Informatique, Hano{\"{\i}} Vietnam,
  2-5 F{\'{e}}vrier 2004}, pages 193--198, 2004.

\bibitem{Hoey:1991}
M.~Hoey.
\newblock {\em Patterns of Lexis in Text}.
\newblock Oxford University Press, Oxford, UK, 1991.

\bibitem{Hopfield:1982}
J.~J. Hopfield.
\newblock Neural networks and physical systems with emergent collective
  computational abilities.
\newblock In {\em Proceedings of the National Academy of Sciences}, pages
  2554--2558. National Academy of Sciences, 1982.

\bibitem{HopfieldT:1986}
J.~J. Hopfield and D.~W. Tank.
\newblock Computing with neural circuits: a model.
\newblock {\em Science}, 233:625--633, 1986.

\bibitem{HornZGKL13}
C.~Horn, A.~Zhila, A.~F. Gelbukh, R.~Kern, and E.~Lex.
\newblock Using factual density to measure informativeness of web documents.
\newblock In S.~Oepen, K.~Hagen, and J.~B. Johannessen, editors, {\em
  Proceedings of the 19th Nordic Conference of Computational Linguistics,
  {NODALIDA} 2013, May 22-24, 2013, Oslo University, Norway}, volume~85 of {\em
  Link{\"{o}}ping Electronic Conference Proceedings}, pages 227--238.
  Link{\"{o}}ping University Electronic Press, 2013.

\bibitem{HuangL:1987}
W.~Y. Huang and R.~Lippmann.
\newblock Neural net and traditional classifiers.
\newblock In D.~Z. Anderson, editor, {\em Neural Information Processing
  Systems, Denver, Colorado, USA, 1987}, pages 387--396. American Institue of
  Physics, 1987.

\bibitem{HughesR:2007}
T.~Hughes and D.~Ramage.
\newblock Lexical semantic relatedness with random graph walks.
\newblock In {\em Proceedings of the 2007 Joint Conference on Empirical Methods
  in Natural Language Processing and Computational Natural Language Learning
  (EMNLP-CoNLL)}, pages 581--589, Prague, Czech Republic, June 2007.
  Association for Computational Linguistics.

\bibitem{HuibersL:1996}
T.~W.~C. Huibers, M.~Lalmas, and C.~J. van Rijsbergen.
\newblock Information retrieval and situation theory.
\newblock {\em {SIGIR} Forum}, 30(1):11--25, 1996.

\bibitem{HunterAL15}
J.~Hunter, N.~Asher, and A.~Lascarides.
\newblock Integrating non-linguistic events into discourse structure.
\newblock In M.~Purver, M.~Sadrzadeh, and M.~Stone, editors, {\em Proceedings
  of the 11th International Conference on Computational Semantics, {IWCS} 2015,
  15-17 April, 2015, Queen Mary University of London, London, {UK}}, pages
  184--194. The Association for Computer Linguistics, 2015.

\bibitem{Schulte:2013}
S.~S. im~Walde, S.~M{\"{u}}ller, and S.~Roller.
\newblock Exploring vector space models to predict the compositionality of
  german noun-noun compounds.
\newblock In Diab et~al. \cite{DBLP:conf/starsem/2013}, pages 255--265.

\bibitem{DBLP:books/tg/Ingwersen92}
P.~Ingwersen.
\newblock {\em Information Retrieval Interaction}.
\newblock Taylor Graham, 1992.

\bibitem{DBLP:conf/sigir/Ingwersen94}
P.~Ingwersen.
\newblock Polyrepresentation of information needs and semantic entities:
  Elements of a cognitive theory for information retrieval interaction.
\newblock In Croft and van Rijsbergen \cite{DBLP:conf/sigir/94}, pages
  101--110.

\bibitem{Ingwersen:1996}
P.~Ingwersen.
\newblock Cognitive perspectives of information retrieval interaction: Elements
  of a cognitive {IR} theory.
\newblock {\em Journal of Documentation}, 52(1):3--50, 1996.

\bibitem{ingwersen99}
P.~Ingwersen.
\newblock {Cognitive Information Retrieval}.
\newblock {\em Annual Review of Information Science and Technology {(ARIST)}},
  34:3--52, 1999.

\bibitem{IngwersenJ:2005}
P.~Ingwersen and K.~J{\"{a}}rvelin.
\newblock {\em The Turn - Integration of Information Seeking and Retrieval in
  Context}, volume~18 of {\em The Kluwer International Series on Information
  Retrieval}.
\newblock Kluwer, 2005.

\bibitem{DBLP:conf/iiix/IngwersenLLW12}
P.~Ingwersen, C.~Lioma, B.~Larsen, and P.~Wang.
\newblock An exploratory study into perceived task complexity, topic
  specificity and usefulness for integrated search.
\newblock In Kamps et~al. \cite{DBLP:conf/iiix/2012}, pages 302--305.

\bibitem{DBLP:conf/irfc/JochimLS11}
C.~Jochim, C.~Lioma, and H.~Sch{\"{u}}tze.
\newblock Expanding queries with term and phrase translations in patent
  retrieval.
\newblock In A.~Hanbury, A.~Rauber, and A.~P. de~Vries, editors, {\em
  Multidisciplinary Information Retrieval - Second Information Retrieval
  Facility Conference, {IRFC} 2011, Vienna, Austria, June 6, 2011.
  Proceedings}, volume 6653 of {\em Lecture Notes in Computer Science}, pages
  16--29. Springer, 2011.

\bibitem{Jochim:2010:PSQ:1871888.1871899}
C.~Jochim, C.~Lioma, H.~Sch\"{u}tze, S.~Koch, and T.~Ertl.
\newblock Preliminary study into query translation for patent retrieval.
\newblock In {\em Proceedings of the 3rd International Workshop on Patent
  Information Retrieval}, PaIR '10, pages 57--66, New York, NY, USA, 2010. ACM.

\bibitem{JonesRHHM99}
S.~L.~P. Jones, A.~Reid, F.~Henderson, C.~A.~R. Hoare, and S.~Marlow.
\newblock A semantics for imprecise exceptions.
\newblock In B.~G. Ryder and B.~G. Zorn, editors, {\em Proceedings of the 1999
  {ACM} {SIGPLAN} Conference on Programming Language Design and Implementation
  (PLDI), Atlanta, Georgia, USA, May 1-4, 1999}, pages 25--36. {ACM}, 1999.

\bibitem{Josang:2001}
A.~J{\o}sang.
\newblock A logic for uncertain probabilities.
\newblock {\em International Journal of Uncertainty, Fuzziness and
  Knowledge-Based Systems}, 9(3):279--212, 2001.

\bibitem{JoyceM:2008}
T.~Joyce and M.~Miyake.
\newblock Capturing the structures in association knowledge: Application of
  network analyses to large-scale databases of japanese word associations.
\newblock In T.~Tokunaga and A.~Ortega, editors, {\em Large-Scale Knowledge
  Resources. Construction and Application, Third International Conference on
  Large-Scale Knowledge Resources, {LKR} 2008, Tokyo, Japan, March 3-5, 2008,
  Proceedings}, volume 4938 of {\em Lecture Notes in Computer Science}, pages
  116--131. Springer, 2008.

\bibitem{JuffingerGL09}
A.~Juffinger, M.~Granitzer, and E.~Lex.
\newblock Blog credibility ranking by exploiting verified content.
\newblock In K.~Tanaka, X.~Zhou, and A.~Jatowt, editors, {\em Proceedings of
  the 3rd {ACM} Workshop on Information Credibility on the Web, {WICOW} 2008,
  Madrid, Spain, April 20, 2009}, pages 51--58. {ACM}, 2009.

\bibitem{DBLP:conf/iiix/2012}
J.~Kamps, W.~Kraaij, and N.~Fuhr, editors.
\newblock {\em Information Interaction in Context: 2012, IIix'12, Nijmegen, The
  Netherlands, August 21-24, 2012}. {ACM}, 2012.

\bibitem{karamanis:2006}
N.~Karamanis.
\newblock Evaluating centering for sentence ordering in two new domains.
\newblock In {\em Proceedings of the Human Language Technology Conference of
  the NAACL, Companion Volume: Short Papers}, pages 65--68, New York City, USA,
  June 2006. Association for Computational Linguistics.

\bibitem{KaramanisMPO09}
N.~Karamanis, C.~Mellish, M.~Poesio, and J.~Oberlander.
\newblock Evaluating centering for information ordering using corpora.
\newblock {\em Computational Linguistics}, 35(1):29--46, 2009.

\bibitem{Kartsaklis14}
D.~Kartsaklis.
\newblock {\em Compositional distributional semantics with compact closed
  categories and Frobenius algebras}.
\newblock PhD thesis, University of Oxford, {UK}, 2014.

\bibitem{Katz:automatic}
G.~Katz and E.~Giesbrecht.
\newblock Automatic identification of non-compositional multi-word expressions
  using latent semantic analysis.
\newblock In {\em Proceedings of the Workshop on Multiword Expressions:
  Identifying and Exploiting Underlying Properties}, pages 12--19. Association
  for Computational Linguistics, Sydney, Australia, July 2006.

\bibitem{Kehler02}
A.~Kehler.
\newblock {\em Coherence, Reference and the Theory of Grammar}.
\newblock CSLI Publications, California, 2002.

\bibitem{Kibble01}
R.~Kibble.
\newblock A reformulation of rule 2 of centering theory.
\newblock {\em Computational Linguistics}, 27(4):579--587, 2001.

\bibitem{KibbleP04}
R.~Kibble and R.~Power.
\newblock Optimizing referential coherence in text generation.
\newblock {\em Computational Linguistics}, 30(4):401--416, 2004.

\bibitem{kiela-clark:2013:EMNLP}
D.~Kiela and S.~Clark.
\newblock Detecting compositionality of multi-word expressions using nearest
  neighbours in vector space models.
\newblock In {\em Proceedings of the 2013 Conference on Empirical Methods in
  Natural Language Processing, {EMNLP} 2013, 18-21 October 2013, Grand Hyatt
  Seattle, Seattle, Washington, USA, {A} meeting of SIGDAT, a Special Interest
  Group of the {ACL}\/} \cite{DBLP:conf/emnlp/2013}, pages 1427--1432.

\bibitem{KitturSC08}
A.~Kittur, B.~Suh, and E.~H. Chi.
\newblock Can you ever trust a wiki?: impacting perceived trustworthiness in
  wikipedia.
\newblock In B.~Begole and D.~W. McDonald, editors, {\em Proceedings of the
  2008 {ACM} Conference on Computer Supported Cooperative Work, {CSCW} 2008,
  San Diego, CA, USA, November 8-12, 2008}, pages 477--480. {ACM}, 2008.

\bibitem{DBLP:conf/naacl/2016}
K.~Knight, A.~Nenkova, and O.~Rambow, editors.
\newblock {\em {NAACL} {HLT} 2016, The 2016 Conference of the North American
  Chapter of the Association for Computational Linguistics: Human Language
  Technologies, San Diego California, USA, June 12-17, 2016}. The Association
  for Computational Linguistics, 2016.

\bibitem{DBLP:conf/sigir/93}
R.~Korfhage, E.~M. Rasmussen, and P.~Willett, editors.
\newblock {\em Proceedings of the 16th Annual International {ACM-SIGIR}
  Conference on Research and Development in Information Retrieval. Pittsburgh,
  PA, USA, June 27 - July 1, 1993}. {ACM}, 1993.

\bibitem{Korkontzelos:2009:DCM:1667583.1667605}
I.~Korkontzelos and S.~Manandhar.
\newblock Detecting compositionality in multi-word expressions.
\newblock In {\em Proceedings of the ACL-IJCNLP 2009 Conference Short Papers},
  pages 65--68, Suntec, Singapore, August 2009. Association for Computational
  Linguistics.

\bibitem{KozarevaR:2008}
Z.~Kozareva, E.~Riloff, and E.~Hovy.
\newblock Semantic class learning from the web with hyponym pattern linkage
  graphs.
\newblock In {\em Proceedings of ACL-08: HLT}, pages 1048--1056, Columbus,
  Ohio, June 2008. Association for Computational Linguistics.

\bibitem{Kozima:1993}
H.~Kozima.
\newblock Similarity between words computed by spreading activation on an
  english dictionary.
\newblock In S.~Krauwer, M.~Moortgat, and L.~des Tombe, editors, {\em Sixth
  Conference of the European Chapter of the Association for Computational
  Linguistics, Proceedings of the Conference, 21-23 April 1993, Utrecht, The
  Netherlands}, pages 232--239. The Association for Computer Linguistics, 1993.

\bibitem{DBLP:conf/sigir/2007}
W.~Kraaij, A.~P. de~Vries, C.~L.~A. Clarke, N.~Fuhr, and N.~Kando, editors.
\newblock {\em {SIGIR} 2007: Proceedings of the 30th Annual International {ACM}
  {SIGIR} Conference on Research and Development in Information Retrieval,
  Amsterdam, The Netherlands, July 23-27, 2007}. {ACM}, 2007.

\bibitem{krvcmavr-jevzek-pecina:2013:CVSC}
L.~Kr\v{c}m\'{a}\v{r}, K.~Je\v{z}ek, and P.~Pecina.
\newblock Determining compositionality of expresssions using various word space
  models and methods.
\newblock In {\em Proceedings of the Workshop on Continuous Vector Space Models
  and their Compositionality}, pages 64--73, Sofia, Bulgaria, August 2013.
  Association for Computational Linguistics.

\bibitem{KuytenBHPA15}
P.~Kuyten, D.~Bollegala, B.~Hollerit, H.~Prendinger, and K.~Aizawa.
\newblock A discourse search engine based on rhetorical structure theory.
\newblock In A.~Hanbury, G.~Kazai, A.~Rauber, and N.~Fuhr, editors, {\em
  Advances in Information Retrieval - 37th European Conference on {IR}
  Research, {ECIR} 2015, Vienna, Austria, March 29 - April 2, 2015.
  Proceedings}, volume 9022 of {\em Lecture Notes in Computer Science}, pages
  80--91, 2015.

\bibitem{Kwok:1989}
K.~L. Kwok.
\newblock A neural network for probabilistic information retrieval.
\newblock In Belkin and van Rijsbergen \cite{DBLP:conf/sigir/89}, pages 21--30.

\bibitem{AntiqueiraP:2007}
L.~L.~Antiqueira, T.~A.~S. Pardo, M.~Nunes, and J.~O.~N. Oliveira.
\newblock Some issues on complex networks for author characterization.
\newblock {\em Inteligencia Artificial, Revista Iberoamericana de IA},
  11(36):51--58, 2007.

\bibitem{Lalmas:1998}
M.~Lalmas.
\newblock Logical models in information retrieval: Introduction and overview.
\newblock {\em Inf. Process. Manage.}, 34(1):19--33, 1998.

\bibitem{Lapata03}
M.~Lapata.
\newblock Probabilistic text structuring: Experiments with sentence ordering.
\newblock In {\em Proceedings of the 41st Annual Meeting of the Association for
  Computational Linguistics}, pages 545--552, Sapporo, Japan, July 2003.
  Association for Computational Linguistics.

\bibitem{Larsen:2005}
B.~Larsen.
\newblock Practical implications of handling multiple contexts in the principle
  of polyrepresentation.
\newblock In F.~Crestani and I.~Ruthven, editors, {\em Context: Nature, Impact,
  and Role - 5th International Conference on Conceptions of Library and
  Information Sciences, CoLIS 2005, Glasgow, UK, June 4-8, 2005, Proceedings},
  volume 3507 of {\em Lecture Notes in Computer Science}, pages 20--31.
  Springer, 2005.

\bibitem{LarsenI:2005}
B.~Larsen and P.~Ingwersen.
\newblock {Cognitive Overlaps along the Polyrepresentation Continuum}.
\newblock {\em New Directions in Cognitive Information Retrieval}, pages
  43--60, 2005.

\bibitem{LarsenI:2006}
B.~Larsen, P.~Ingwersen, and J.~Kek{\"{a}}l{\"{a}}inen.
\newblock The polyrepresentation continuum in {IR}.
\newblock In I.~Ruthven, editor, {\em Proceedings of the 1st International
  Conference on Information Interaction in Context, IIiX 2006, Copenhagen,
  Denmark, October 18-20, 2006}, pages 88--96. {ACM}, 2006.

\bibitem{LarsenI:2009}
B.~Larsen, P.~Ingwersen, and B.~Lund.
\newblock Data fusion according to the principle of polyrepresentation.
\newblock {\em {JASIST}}, 60(4):646--654, 2009.

\bibitem{DBLP:conf/ecir/LarsenL16}
B.~Larsen and C.~Lioma.
\newblock On the need for and provision for an 'ideal' scholarly information
  retrieval test collection.
\newblock In P.~Mayr, I.~Frommholz, and G.~Cabanac, editors, {\em Proceedings
  of the Third Workshop on Bibliometric-enhanced Information Retrieval
  co-located with the 38th European Conference on Information Retrieval {(ECIR}
  2016), Padova, Italy, March 20, 2016.}, volume 1567 of {\em {CEUR} Workshop
  Proceedings}, pages 73--81. CEUR-WS.org, 2016.

\bibitem{LarsenL12}
B.~Larsen, C.~Lioma, I.~Frommholz, and H.~Sch{\"{u}}tze.
\newblock Preliminary study of technical terminology for the retrieval of
  scientific book metadata records.
\newblock In Hersh et~al. \cite{DBLP:conf/sigir/2012}, pages 1131--1132.

\bibitem{LauB:2008}
R.~Y.~K. Lau, P.~D. Bruza, and D.~Song.
\newblock Towards a belief-revision-based adaptive and context-sensitive
  information retrieval system.
\newblock {\em {ACM} Trans. Inf. Syst.}, 26(2), 2008.

\bibitem{LeZ15}
P.~Le and W.~Zuidema.
\newblock The forest convolutional network: Compositional distributional
  semantics with a neural chart and without binarization.
\newblock In M{\`{a}}rquez et~al. \cite{DBLP:conf/emnlp/2015}, pages
  1155--1164.

\bibitem{LecceA:2009}
V.~D. Lecce and A.~Amato.
\newblock A fuzzy logic based approach to feedback reinforcement in image
  retrieval.
\newblock In D.~Huang, K.~Jo, H.~Lee, H.~Kang, and V.~Bevilacqua, editors, {\em
  Emerging Intelligent Computing Technology and Applications, 5th International
  Conference on Intelligent Computing, {ICIC} 2009, Ulsan, South Korea,
  September 16-19, 2009. Proceedings}, volume 5754 of {\em Lecture Notes in
  Computer Science}, pages 939--947. Springer, 2009.

\bibitem{LeichtH:2006}
E.~A. Leicht, P.~Holme, and M.~E.~J. Newman.
\newblock Vertex similarity in networks.
\newblock {\em Physical Review E}, (73):026120, 2006.

\bibitem{Lesk1969}
M.~E. Lesk.
\newblock Word-word associations in document retrieval systems.
\newblock {\em Am. Doc.}, 20:27--38, 1969.

\bibitem{Lewis1992}
D.~D. Lewis.
\newblock An evaluation of phrasal and clustered representations on a text
  categorization task.
\newblock In Belkin et~al. \cite{DBLP:conf/sigir/92}, pages 37--50.

\bibitem{LewisC:1990}
D.~D. Lewis and W.~B. Croft.
\newblock Term clustering of syntactic phrases.
\newblock In J.~Vidick, editor, {\em SIGIR'90, 13th International Conference on
  Research and Development in Information Retrieval, Brussels, Belgium, 5-7
  September 1990, Proceedings}, pages 385--404. {ACM}, 1990.

\bibitem{LexJG10}
E.~Lex, A.~Juffinger, and M.~Granitzer.
\newblock Objectivity classification in online media.
\newblock In M.~H. Chignell and E.~G. Toms, editors, {\em HT'10, Proceedings of
  the 21st {ACM} Conference on Hypertext and Hypermedia, Toronto, Ontario,
  Canada, June 13-16, 2010}, pages 293--294. {ACM}, 2010.

\bibitem{Lex12}
E.~Lex, M.~Voelske, M.~Errecalde, E.~Ferretti, L.~Cagnina, C.~Horn, B.~Stein,
  and M.~Granitzer.
\newblock Measuring the quality of web content using factual information.
\newblock In {\em WebQuality}, pages 7--10, 2012.

\bibitem{LiH14a}
J.~Li and E.~H. Hovy.
\newblock A model of coherence based on distributed sentence representation.
\newblock In A.~Moschitti, B.~Pang, and W.~Daelemans, editors, {\em {EMNLP}},
  pages 2039--2048. {ACL}, 2014.

\bibitem{DBLP:conf/acl/2011}
D.~Lin, Y.~Matsumoto, and R.~Mihalcea, editors.
\newblock {\em The 49th Annual Meeting of the Association for Computational
  Linguistics: Human Language Technologies, Proceedings of the Conference,
  19-24 June, 2011, Portland, Oregon, {USA}}. The Association for Computer
  Linguistics, 2011.

\bibitem{Lin2001}
J.~Lin.
\newblock {Indexing \& Retrieving Natural Language Using Ternary Expressions}.
\newblock Master's thesis, U. of Maryland, USA, 2001.

\bibitem{LinLNK12}
Z.~Lin, C.~Liu, H.~T. Ng, and M.~Kan.
\newblock Combining coherence models and machine translation evaluation metrics
  for summarization evaluation.
\newblock In {\em The 50th Annual Meeting of the Association for Computational
  Linguistics, Proceedings of the Conference, July 8-14, 2012, Jeju Island,
  Korea - Volume 1: Long Papers}, pages 1006--1014. The Association for
  Computer Linguistics, 2012.

\bibitem{LinNK11}
Z.~Lin, H.~T. Ng, and M.~Kan.
\newblock Automatically evaluating text coherence using discourse relations.
\newblock In Lin et~al. \cite{DBLP:conf/acl/2011}, pages 997--1006.

\bibitem{LinckelsM:2008}
S.~Linckels and C.~Meinel.
\newblock Applications of description logics to improve multimedia information
  retrieval for efficient educational tools.
\newblock In M.~S. Lew, A.~D. Bimbo, and E.~M. Bakker, editors, {\em
  Proceedings of the 1st {ACM} {SIGMM} International Conference on Multimedia
  Information Retrieval, {MIR} 2008, Vancouver, British Columbia, Canada,
  October 30-31, 2008}, pages 321--328. {ACM}, 2008.

\bibitem{lind:2005}
P.~G. Lind, M.~C. Gonzalez, and H.~J. Herrmann.
\newblock Cycles and clustering in bipartite networks.
\newblock {\em Phys. Rev. E}, page 72:056127, 2005.

\bibitem{LiomaB:2009ecir}
C.~Lioma and R.~Blanco.
\newblock Part of speech based term weighting for information retrieval.
\newblock In M.~Boughanem, C.~Berrut, J.~Mothe, and C.~Soul{\'{e}}{-}Dupuy,
  editors, {\em Advances in Information Retrieval, 31th European Conference on
  {IR} Research, {ECIR} 2009, Toulouse, France, April 6-9, 2009. Proceedings},
  volume 5478 of {\em Lecture Notes in Computer Science}, pages 412--423.
  Springer, 2009.

\bibitem{DBLP:conf/ictir/LiomaBM09}
C.~Lioma, R.~Blanco, and M.~Moens.
\newblock A logical inference approach to query expansion with social tags.
\newblock In Azzopardi et~al. \cite{DBLP:conf/ictir/2009}, pages 358--361.

\bibitem{LiomaB:2009}
C.~Lioma, R.~Blanco, R.~M. Palau, and M.~Moens.
\newblock A belief model of query difficulty that uses subjective logic.
\newblock In Azzopardi et~al. \cite{DBLP:conf/ictir/2009}, pages 92--103.

\bibitem{LiomaH17}
C.~Lioma and N.~D. Hansen.
\newblock A study of metrics of distance and correlation between ranked lists
  for compositionality detection.
\newblock {\em Cognitive Systems Research}, page in press, 2017.

\bibitem{LiomaKS11}
C.~Lioma, A.~Kothari, and H.~Sch{\"{u}}tze.
\newblock Sense discrimination for physics retrieval.
\newblock In W.~Ma, J.~Nie, R.~A. Baeza{-}Yates, T.~Chua, and W.~B. Croft,
  editors, {\em Proceeding of the 34th International {ACM} {SIGIR} Conference
  on Research and Development in Information Retrieval, {SIGIR} 2011, Beijing,
  China, July 25-29, 2011}, pages 1101--1102. {ACM}, 2011.

\bibitem{LiomaLI12}
C.~Lioma, B.~Larsen, and P.~Ingwersen.
\newblock Preliminary experiments using subjective logic for the
  polyrepresentation of information needs.
\newblock In Kamps et~al. \cite{DBLP:conf/iiix/2012}, pages 174--183.

\bibitem{Lioma12}
C.~Lioma, B.~Larsen, and W.~Lu.
\newblock Rhetorical relations for information retrieval.
\newblock In Hersh et~al. \cite{DBLP:conf/sigir/2012}, pages 931--940.

\bibitem{LiomaLLH16}
C.~Lioma, B.~Larsen, W.~Lu, and Y.~Huang.
\newblock A study of factuality, objectivity and relevance: three desiderata in
  large-scale information retrieval?
\newblock In A.~Anjum and X.~Zhao, editors, {\em Proceedings of the 3rd
  {IEEE/ACM} International Conference on Big Data Computing, Applications and
  Technologies, {BDCAT} 2016, Shanghai, China, December 6-9, 2016}, pages
  107--117. {ACM}, 2016.

\bibitem{DBLP:journals/corr/LiomaLPS16}
C.~Lioma, B.~Larsen, C.~Petersen, and J.~G. Simonsen.
\newblock Deep learning relevance: Creating relevant information (as opposed to
  retrieving it).
\newblock {\em CoRR}, abs/1606.07660, 2016.

\bibitem{DBLP:conf/ictir/LiomaLS11}
C.~Lioma, B.~Larsen, and H.~Sch{\"{u}}tze.
\newblock User perspectives on query difficulty.
\newblock In Amati and Crestani \cite{DBLP:conf/ictir/2011}, pages 3--14.

\bibitem{Lioma:2010}
C.~Lioma, B.~Larsen, H.~Sch{\"{u}}tze, and P.~Ingwersen.
\newblock A subjective logic formalisation of the principle of
  polyrepresentation for information needs.
\newblock In Belkin and Kelly \cite{DBLP:conf/iiix/2010}, pages 125--134.

\bibitem{LiomaO07}
C.~Lioma and I.~Ounis.
\newblock Extending weighting models with a term quality measure.
\newblock In N.~Ziviani and R.~A. Baeza{-}Yates, editors, {\em String
  Processing and Information Retrieval, 14th International Symposium, {SPIRE}
  2007, Santiago, Chile, October 29-31, 2007, Proceedings}, volume 4726 of {\em
  Lecture Notes in Computer Science}, pages 205--216. Springer, 2007.

\bibitem{LiomaSLH15}
C.~Lioma, J.~G. Simonsen, B.~Larsen, and N.~D. Hansen.
\newblock Non-compositional term dependence for information retrieval.
\newblock In Baeza{-}Yates et~al. \cite{DBLP:conf/sigir/2015}, pages 595--604.

\bibitem{LiomaTSPL16}
C.~Lioma, F.~Tarissan, J.~G. Simonsen, C.~Petersen, and B.~Larsen.
\newblock Exploiting the bipartite structure of entity grids for document
  coherence and retrieval.
\newblock In B.~Carterette, H.~Fang, M.~Lalmas, and J.~Nie, editors, {\em
  Proceedings of the 2016 {ACM} on International Conference on the Theory of
  Information Retrieval, {ICTIR} 2016, Newark, DE, USA, September 12- 6, 2016},
  pages 11--20. {ACM}, 2016.

\bibitem{LiomaK:2008}
C.~Lioma and C.~J.~K. van Rijsbergen.
\newblock Part of speech n-grams and information retrieval.
\newblock {\em Revue fran\c{c}aise de linguistique appliqu\'{e}e},
  XIII(1):9--11, 2008.

\bibitem{LipkaS10}
N.~Lipka and B.~Stein.
\newblock Identifying featured articles in wikipedia: writing style matters.
\newblock In M.~Rappa, P.~Jones, J.~Freire, and S.~Chakrabarti, editors, {\em
  Proceedings of the 19th International Conference on World Wide Web, {WWW}
  2010, Raleigh, North Carolina, USA, April 26-30, 2010}, pages 1147--1148.
  {ACM}, 2010.

\bibitem{LiuH16}
B.~Liu and M.~Huang.
\newblock A sentence interaction network for modeling dependence between
  sentences.
\newblock In {\em Proceedings of the 54th Annual Meeting of the Association for
  Computational Linguistics, {ACL} 2016, August 7-12, 2016, Berlin, Germany,
  Volume 1: Long Papers\/} \cite{DBLP:conf/acl/2016-1}.

\bibitem{LoganR:1994}
B.~Logan, S.~Reece, and K.~S. Jones.
\newblock Modelling information retrieval agents with belief revision.
\newblock In Croft and van Rijsbergen \cite{DBLP:conf/sigir/94}, pages 91--100.

\bibitem{LosadaB:2001}
D.~E. Losada and A.~Barreiro.
\newblock A logical model for information retrieval based on propositional
  logic and belief revision.
\newblock {\em Comput. J.}, 44(5):410--424, 2001.

\bibitem{Losee:1994}
R.~M. Losee.
\newblock Term dependence: Truncating the bahadur lazarsfeld expansion.
\newblock {\em Information Processing \& Management}, 30(2):293--303, 1994.

\bibitem{LouisN12}
A.~Louis and A.~Nenkova.
\newblock A coherence model based on syntactic patterns.
\newblock In Tsujii et~al. \cite{DBLP:conf/emnlp/2012}, pages 1157--1168.

\bibitem{DBLP:conf/sigir/LuCL12}
W.~Lu, Q.~Cheng, and C.~Lioma.
\newblock Fixed versus dynamic co-occurrence windows in textrank term weights
  for information retrieval.
\newblock In Hersh et~al. \cite{DBLP:conf/sigir/2012}, pages 1079--1080.

\bibitem{LvZ09}
Y.~Lv and C.~Zhai.
\newblock Positional language models for information retrieval.
\newblock In J.~Allan, J.~A. Aslam, M.~Sanderson, C.~Zhai, and J.~Zobel,
  editors, {\em Proceedings of the 32nd Annual International {ACM} {SIGIR}
  Conference on Research and Development in Information Retrieval, {SIGIR}
  2009, Boston, MA, USA, July 19-23, 2009}, pages 299--306. {ACM}, 2009.

\bibitem{MacLeodR:1991}
K.~J. Macleod and W.~Robertson.
\newblock A neural algorithm for document clustering.
\newblock {\em Inf. Process. Manage.}, 27(4):337--346, 1991.

\bibitem{MannT88}
W.~C. Mann and S.~A. Thompson.
\newblock Rhetorical structure theory: Toward a functional theory of text
  organization.
\newblock Number~3, pages 243--281, 1998.

\bibitem{DBLP:conf/emnlp/2015}
L.~M{\`{a}}rquez, C.~Callison{-}Burch, J.~Su, D.~Pighin, and Y.~Marton,
  editors.
\newblock {\em Proceedings of the 2015 Conference on Empirical Methods in
  Natural Language Processing, {EMNLP} 2015, Lisbon, Portugal, September 17-21,
  2015}. The Association for Computational Linguistics, 2015.

\bibitem{MasucciR:2006}
A.~P. Masucci and G.~J. Rodgers.
\newblock Network properties of written human language.
\newblock {\em Phys. Rev. E}, 74(2):026102, Aug 2006.

\bibitem{McCarthy:2003}
D.~McCarthy, B.~Keller, and J.~Carroll.
\newblock Detecting a continuum of compositionality in phrasal verbs.
\newblock In {\em Proceedings of the ACL 2003 Workshop on Multiword
  Expressions: Analysis, Acquisition and Treatment}, pages 73--80. Association
  for Computational Linguistics, Sapporo, Japan, July 2003.

\bibitem{MeghiniS:1993}
C.~Meghini, F.~Sebastiani, U.~Straccia, and C.~Thanos.
\newblock A model of information retrieval based on a terminological logic.
\newblock In Korfhage et~al. \cite{DBLP:conf/sigir/93}, pages 298--307.

\bibitem{MetzlerC:2005}
D.~Metzler and W.~B. Croft.
\newblock A markov random field model for term dependencies.
\newblock In Baeza{-}Yates et~al. \cite{DBLP:conf/sigir/2005}, pages 472--479.

\bibitem{MetzlerN:1984}
D.~P. Metzler, T.~Noreault, L.~Richey, and P.~B. Heidorn.
\newblock Dependency parsing for information retrieval.
\newblock In {\em {SIGIR}}, pages 313--324, 1984.

\bibitem{MichelbacherKFLS11}
L.~Michelbacher, A.~Kothari, M.~Forst, C.~Lioma, and H.~Sch{\"{u}}tze.
\newblock A cascaded classification approach to semantic head recognition.
\newblock In {\em Proceedings of the 2011 Conference on Empirical Methods in
  Natural Language Processing, {EMNLP} 2011, 27-31 July 2011, John McIntyre
  Conference Centre, Edinburgh, UK, {A} meeting of SIGDAT, a Special Interest
  Group of the {ACL}\/} \cite{DBLP:conf/emnlp/2011}, pages 793--803.

\bibitem{DBLP:conf/naacl/2015}
R.~Mihalcea, J.~Y. Chai, and A.~Sarkar, editors.
\newblock {\em {NAACL} {HLT} 2015, The 2015 Conference of the North American
  Chapter of the Association for Computational Linguistics: Human Language
  Technologies, Denver, Colorado, USA, May 31 - June 5, 2015}. The Association
  for Computational Linguistics, 2015.

\bibitem{MihalceaT:2004}
R.~Mihalcea and P.~Tarau.
\newblock Textrank: Bringing order into text.
\newblock In {\em Proceedings of the 2004 Conference on Empirical Methods in
  Natural Language Processing , {EMNLP} 2004, {A} meeting of SIGDAT, a Special
  Interest Group of the ACL, held in conjunction with {ACL} 2004, 25-26 July
  2004, Barcelona, Spain}, pages 404--411. {ACL}, 2004.

\bibitem{Mikk:2001}
J.~Mikk.
\newblock Prior knowledge of text content and values of text characteristics.
\newblock {\em Journal of Quantitative Linguistics}, 8(1):67--80, 2001.

\bibitem{MikolovSCCD13}
T.~Mikolov, I.~Sutskever, K.~Chen, G.~S. Corrado, and J.~Dean.
\newblock Distributed representations of words and phrases and their
  compositionality.
\newblock In C.~J.~C. Burges, L.~Bottou, Z.~Ghahramani, and K.~Q. Weinberger,
  editors, {\em Advances in Neural Information Processing Systems 26: 27th
  Annual Conference on Neural Information Processing Systems 2013. Proceedings
  of a meeting held December 5-8, 2013, Lake Tahoe, Nevada, United States.},
  pages 3111--3119, 2013.

\bibitem{MiloI:2004}
R.~Milo, S.~Itzkovitz, N.~Kashtan, R.~Levitt, S.~Shen-Orr, I.~Ayzenshtat,
  M.~Sheffer, and U.~Alon.
\newblock Superfamilies of evolved and designed networks.
\newblock {\em Science}, 303(5663):1538--1542, 2004.

\bibitem{MiltsakakiK04}
E.~Miltsakaki and K.~Kukich.
\newblock Evaluation of text coherence for electronic essay scoring systems.
\newblock {\em Natural Language Engineering}, 10(1):25--55, 2004.

\bibitem{MineshimaMMB15}
K.~Mineshima, P.~Mart{\'{\i}}nez{-}G{\'{o}}mez, Y.~Miyao, and D.~Bekki.
\newblock Higher-order logical inference with compositional semantics.
\newblock In M{\`{a}}rquez et~al. \cite{DBLP:conf/emnlp/2015}, pages
  2055--2061.

\bibitem{MinkovC:2008}
E.~Minkov and W.~W. Cohen.
\newblock Learning graph walk based similarity measures for parsed text.
\newblock In {\em 2008 Conference on Empirical Methods in Natural Language
  Processing, {EMNLP} 2008, Proceedings of the Conference, 25-27 October 2008,
  Honolulu, Hawaii, USA, {A} meeting of SIGDAT, a Special Interest Group of the
  {ACL}}, pages 907--916. {ACL}, 2008.

\bibitem{MishneR2005}
G.~Mishne and M.~de~Rijke.
\newblock Boosting web retrieval through query operations.
\newblock In {\em ECIR}, pages 502--516, 2005.

\bibitem{COGS:COGS1106}
J.~Mitchell and M.~Lapata.
\newblock Composition in distributional models of semantics.
\newblock {\em Cognitive Science}, 34(8):1388--1429, 2010.

\bibitem{MonroeGP16}
W.~Monroe, N.~D. Goodman, and C.~Potts.
\newblock Learning to generate compositional color descriptions.
\newblock In J.~Su, X.~Carreras, and K.~Duh, editors, {\em Proceedings of the
  2016 Conference on Empirical Methods in Natural Language Processing, {EMNLP}
  2016, Austin, Texas, USA, November 1-4, 2016}, pages 2243--2248. The
  Association for Computational Linguistics, 2016.

\bibitem{MoratoL:2003}
J.~Morato, J.~Llorens, G.~Genova, and J.~A. Moreiro.
\newblock Experiments in discourse analysis impact on information
  classification and retrieval algorithms.
\newblock {\em Inf. Process. Manage.}, 39:825--851, November 2003.

\bibitem{MorrisCRHS12}
M.~R. Morris, S.~Counts, A.~Roseway, A.~Hoff, and J.~Schwarz.
\newblock Tweeting is believing?: understanding microblog credibility
  perceptions.
\newblock In S.~E. Poltrock, C.~Simone, J.~Grudin, G.~Mark, and J.~Riedl,
  editors, {\em {CSCW} '12 Computer Supported Cooperative Work, Seattle, WA,
  USA, February 11-15, 2012}, pages 441--450. {ACM}, 2012.

\bibitem{MotheT:2005}
J.~Mothe and L.~Tanguy.
\newblock Linguistic features to predict query difficulty - a case study on
  previous {TREC} campaigns.
\newblock In {\em ACM Conference on research and Development in Information
  Retrieval, SIGIR, Predicting query difficulty - methods and applications
  workshop}, pages 7--10, 2005.

\bibitem{MotterM:2002}
A.~E. Motter, A.~P. S.~D. Moura, Y.~C. Lai, and P.~Dasgupta.
\newblock Topology of the conceptual network of language.
\newblock {\em Phys. Rev. E}, 65(6).

\bibitem{DBLP:conf/cikm/2016}
S.~Mukhopadhyay, C.~Zhai, E.~Bertino, F.~Crestani, J.~Mostafa, J.~Tang, L.~Si,
  X.~Zhou, Y.~Chang, Y.~Li, and P.~Sondhi, editors.
\newblock {\em Proceedings of the 25th {ACM} International on Conference on
  Information and Knowledge Management, {CIKM} 2016, Indianapolis, IN, USA,
  October 24-28, 2016}. {ACM}, 2016.

\bibitem{MullerK:1995}
A.~M{\"u}ller and S.~Kutschekmanesch.
\newblock {Using Abductive Inference and Dynamic Indexing to Retrieve
  Multimedia SGML Documents}.
\newblock In I.~Ruthven, editor, {\em MIRO}, Workshops in Computing, page~11.
  BCS, 1995.

\bibitem{MullerH:2006}
P.~Muller, N.~Hathout, and B.~Gaume.
\newblock Synonym extraction using a semantic distance on a dictionary.
\newblock In {\em Proceedings of TextGraphs: the First Workshop on Graph Based
  Methods for Natural Language Processing}, pages 65--72, New York City, June
  2006. Association for Computational Linguistics.

\bibitem{NallapatiA:2002}
R.~Nallapati and J.~Allan.
\newblock Capturing term dependencies using a language model based on sentence
  trees.
\newblock In {\em Proceedings of the 2002 {ACM} {CIKM} International Conference
  on Information and Knowledge Management, McLean, VA, USA, November 4-9,
  2002}, pages 383--390. {ACM}, 2002.

\bibitem{NaritaO2000}
M.~Narita and Y.~Ogawa.
\newblock The use of phrases from query texts in information retrieval.
\newblock In {\em {SIGIR}}, pages 318--320, 2000.

\bibitem{NastaseS:2006}
V.~Nastase, J.~Sayyad{-}Shirabad, M.~Sokolova, and S.~Szpakowicz.
\newblock Learning noun-modifier semantic relations with corpus-based and
  wordnet-based features.
\newblock In {\em Proceedings, The Twenty-First National Conference on
  Artificial Intelligence and the Eighteenth Innovative Applications of
  Artificial Intelligence Conference, July 16-20, 2006, Boston, Massachusetts,
  {USA}}, pages 781--787. {AAAI} Press, 2006.

\bibitem{NeelakantanRM15}
A.~Neelakantan, B.~Roth, and A.~McCallum.
\newblock Compositional vector space models for knowledge base completion.
\newblock In {\em Proceedings of the 53rd Annual Meeting of the Association for
  Computational Linguistics and the 7th International Joint Conference on
  Natural Language Processing of the Asian Federation of Natural Language
  Processing, {ACL} 2015, July 26-31, 2015, Beijing, China, Volume 1: Long
  Papers\/} \cite{DBLP:conf/acl/2015-1}, pages 156--166.

\bibitem{Nie:1992}
J.~Nie.
\newblock Towards a probabilistic modal logic for semantic-based information
  retrieval.
\newblock In Belkin et~al. \cite{DBLP:conf/sigir/92}, pages 140--151.

\bibitem{NieL:1996}
J.~Nie, M.~Brisebois, and F.~Lepage.
\newblock Information retrieval as counterfactual.
\newblock {\em Comput. J.}, 38(8):643--657, 1995.

\bibitem{NikolaevKZ16}
F.~Nikolaev, A.~Kotov, and N.~Zhiltsov.
\newblock Parameterized fielded term dependence models for ad-hoc entity
  retrieval from knowledge graph.
\newblock In Perego et~al. \cite{DBLP:conf/sigir/2016}, pages 435--444.

\bibitem{Oh13}
H.~Oh, Y.~Yoon, and H.~Kim.
\newblock Finding more trustworthy answers: Various trustworthiness factors in
  question answering.
\newblock {\em J. Information Science}, 39(4):509--522, 2013.

\bibitem{OussalahK:2008}
M.~Oussalah, S.~Khan, and S.~Nefti.
\newblock Personalized information retrieval system in the framework of fuzzy
  logic.
\newblock {\em Expert Syst. Appl.}, 35(1-2):423--433, 2008.

\bibitem{PadoL:2007}
S.~Pad{\'{o}} and M.~Lapata.
\newblock Dependency-based construction of semantic space models.
\newblock {\em Computational Linguistics}, 33(2):161--199, 2007.

\bibitem{Pang04}
B.~Pang and L.~Lee.
\newblock A sentimental education: Sentiment analysis using subjectivity
  summarization based on minimum cuts.
\newblock In D.~Scott, W.~Daelemans, and M.~A. Walker, editors, {\em
  Proceedings of the 42nd Annual Meeting of the Association for Computational
  Linguistics, 21-26 July, 2004, Barcelona, Spain.}, pages 271--278. {ACL},
  2004.

\bibitem{PapernoB16}
D.~Paperno and M.~Baroni.
\newblock When the whole is less than the sum of its parts: How composition
  affects {PMI} values in distributional semantic vectors.
\newblock {\em Computational Linguistics}, 42(2):345--350, 2016.

\bibitem{Park09}
S.~Park, S.~Kang, S.~Chung, and J.~Song.
\newblock Newscube: delivering multiple aspects of news to mitigate media bias.
\newblock In D.~R.~O. Jr., R.~B. Arthur, K.~Hinckley, M.~R. Morris, S.~E.
  Hudson, and S.~Greenberg, editors, {\em Proceedings of the 27th International
  Conference on Human Factors in Computing Systems, {CHI} 2009, Boston, MA,
  USA, April 4-9, 2009}, pages 443--452. {ACM}, 2009.

\bibitem{PavlickC16}
E.~Pavlick and C.~Callison{-}Burch.
\newblock Most "babies" are "little" and most "problems" are "huge":
  Compositional entailment in adjective-nouns.
\newblock In {\em Proceedings of the 54th Annual Meeting of the Association for
  Computational Linguistics, {ACL} 2016, August 7-12, 2016, Berlin, Germany,
  Volume 1: Long Papers\/} \cite{DBLP:conf/acl/2016-1}.

\bibitem{Pearl:1988}
J.~Pearl.
\newblock {\em Probabilistic reasoning in intelligent systems: networks of
  plausible inference}.
\newblock Morgan Kaufmann Publishers Inc., San Francisco, CA, USA, 1988.

\bibitem{PedersonS:1997}
J.~O. Pedersen, C.~Silverstein, and C.~C. Vogt.
\newblock Verity at {TREC-6:} out-of-the-box and beyond.
\newblock In Voorhees and Harman \cite{DBLP:conf/trec/1997}, pages 259--273.

\bibitem{PedersenP:2004}
T.~Pedersen, S.~Patwardhan, and J.~Michelizzi.
\newblock Wordnet: : Similarity - measuring the relatedness of concepts.
\newblock In D.~L. McGuinness and G.~Ferguson, editors, {\em Proceedings of the
  Nineteenth National Conference on Artificial Intelligence, Sixteenth
  Conference on Innovative Applications of Artificial Intelligence, July 25-29,
  2004, San Jose, California, {USA}}, pages 1024--1025. {AAAI} Press / The
  {MIT} Press, 2004.

\bibitem{DBLP:conf/sigir/2016}
R.~Perego, F.~Sebastiani, J.~A. Aslam, I.~Ruthven, and J.~Zobel, editors.
\newblock {\em Proceedings of the 39th International {ACM} {SIGIR} conference
  on Research and Development in Information Retrieval, {SIGIR} 2016, Pisa,
  Italy, July 17-21, 2016}. {ACM}, 2016.

\bibitem{DBLP:conf/eurohcir/PetersenLS13}
C.~Petersen, C.~Lioma, and J.~G. Simonsen.
\newblock Comparative study of search engine result visualisation: Ranked lists
  versus graphs.
\newblock In M.~L. Wilson, T.~Russell{-}Rose, B.~Larsen, P.~Hansen, and
  K.~Norling, editors, {\em Proceedings of the 3rd European Workshop on
  Human-Computer Interaction and Information Retrieval, Dublin, Ireland, August
  1, 2013}, volume 1033 of {\em {CEUR} Workshop Proceedings}, pages 27--30.
  CEUR-WS.org, 2013.

\bibitem{petersen}
C.~Petersen, C.~Lioma, J.~G. Simonsen, and B.~Larsen.
\newblock Entropy and graph based modelling of document coherence using
  discourse entities: An application to {IR}.
\newblock In Allan et~al. \cite{DBLP:conf/ictir/2015}, pages 191--200.

\bibitem{DBLP:conf/cikm/PetersenSJL16}
C.~Petersen, J.~G. Simonsen, K.~J{\"{a}}rvelin, and C.~Lioma.
\newblock Adaptive distributional extensions to {DFR} ranking.
\newblock In Mukhopadhyay et~al. \cite{DBLP:conf/cikm/2016}, pages 2005--2008.

\bibitem{DBLP:conf/airs/PetersenSL15}
C.~Petersen, J.~G. Simonsen, and C.~Lioma.
\newblock The impact of using combinatorial optimisation for static caching of
  posting lists.
\newblock In G.~Zuccon, S.~Geva, H.~Joho, F.~Scholer, A.~Sun, and P.~Zhang,
  editors, {\em Information Retrieval Technology - 11th Asia Information
  Retrieval Societies Conference, {AIRS} 2015, Brisbane, QLD, Australia,
  December 2-4, 2015. Proceedings}, volume 9460 of {\em Lecture Notes in
  Computer Science}, pages 420--425. Springer, 2015.

\bibitem{DBLP:journals/tois/PetersenSL16}
C.~Petersen, J.~G. Simonsen, and C.~Lioma.
\newblock Power law distributions in information retrieval.
\newblock {\em {ACM} Trans. Inf. Syst.}, 34(2):8:1--8:37, 2016.

\bibitem{PlachourasO05}
V.~Plachouras and I.~Ounis.
\newblock Dempster-shafer theory for a query-biased combination of evidence on
  the web.
\newblock {\em Information Retrieval}, 8(2):197--218, 2005.

\bibitem{PlazaD:2008}
L.~Plaza, A.~Diaz, and P.~Gervas.
\newblock Concept-graph based biomedical automatic summarization using
  ontologies.
\newblock In {\em Coling 2008: Proceedings of the 3rd Textgraphs workshop on
  Graph-based Algorithms for Natural Language Processing}, pages 53--56,
  Manchester, UK, August 2008. Coling 2008 Organizing Committee.

\bibitem{PoesioSEH04}
M.~Poesio, R.~Stevenson, B.~D. Eugenio, and J.~Hitzeman.
\newblock Centering: {A} parametric theory and its instantiations.
\newblock {\em Computational Linguistics}, 30(3):309--363, 2004.

\bibitem{PopescuE:2005}
A.~Popescu and O.~Etzioni.
\newblock Extracting product features and opinions from reviews.
\newblock In {\em {HLT/EMNLP} 2005, Human Language Technology Conference and
  Conference on Empirical Methods in Natural Language Processing, Proceedings
  of the Conference, 6-8 October 2005, Vancouver, British Columbia, Canada\/}
  \cite{DBLP:conf/naacl/2005}.

\bibitem{QiuZL15}
L.~Qiu, Y.~Zhang, and Y.~Lu.
\newblock Syntactic dependencies and distributed word representations for
  analogy detection and mining.
\newblock In M{\`{a}}rquez et~al. \cite{DBLP:conf/emnlp/2015}, pages
  2441--2450.

\bibitem{RadhouaniF:2008}
S.~Radhouani and G.~Falquet.
\newblock Description logics-based modelling for precise information retrieval.
\newblock In F.~Baader, C.~Lutz, and B.~Motik, editors, {\em Proceedings of the
  21st International Workshop on Description Logics (DL2008), Dresden, Germany,
  May 13-16, 2008}, volume 353 of {\em {CEUR} Workshop Proceedings}.
  CEUR-WS.org, 2008.

\bibitem{RamageR:2009}
D.~Ramage, A.~N. Rafferty, and C.~D. Manning.
\newblock Random walks for text semantic similarity.
\newblock In {\em Proceedings of the 2009 Workshop on Graph-based Methods for
  Natural Language Processing (TextGraphs-4)}, pages 23--31, Suntec, Singapore,
  August 2009. Association for Computational Linguistics.

\bibitem{DBLP:conf/ijcnlp/ReddyKMM11}
S.~Reddy, I.~P. Klapaftis, D.~McCarthy, and S.~Manandhar.
\newblock Dynamic and static prototype vectors for semantic composition.
\newblock In {\em Fifth International Joint Conference on Natural Language
  Processing, {IJCNLP} 2011, Chiang Mai, Thailand, November 8-13, 2011}, pages
  705--713. The Association for Computer Linguistics, 2011.

\bibitem{McCarthy:2011}
S.~Reddy, D.~McCarthy, S.~Manandhar, and S.~Gella.
\newblock Exemplar-based word-space model for compositionality detection:
  Shared task system description.
\newblock In {\em Proceedings of the Workshop on Distributional Semantics and
  Compositionality}, pages 54--60. Association for Computational Linguistics,
  Portland, Oregon, USA, June 2011.

\bibitem{Reppy93}
J.~H. Reppy.
\newblock Concurrent {ML:} design, application and semantics.
\newblock In P.~E. Lauer, editor, {\em Functional Programming, Concurrency,
  Simulation and Automated Reasoning: International Lecture Series 1991-1992,
  McMaster University, Hamilton, Ontario, Canada}, volume 693 of {\em Lecture
  Notes in Computer Science}, pages 165--198. Springer, 1993.

\bibitem{ReynaB:2005}
V.~F. Reyna and C.~J. Brainerd.
\newblock Fuzzy processing in transitivity development.
\newblock {\em Annals of Operations Research}, 23(1):37--63, 2005.

\bibitem{Roscoe:1997:TPC:550448}
A.~W. Roscoe, C.~A.~R. Hoare, and R.~Bird.
\newblock {\em The Theory and Practice of Concurrency}.
\newblock Prentice Hall PTR, Upper Saddle River, NJ, USA, 1997.

\bibitem{RousseauV13}
F.~Rousseau and M.~Vazirgiannis.
\newblock Graph-of-word and {TW-IDF:} new approach to ad hoc {IR}.
\newblock In Q.~He, A.~Iyengar, W.~Nejdl, J.~Pei, and R.~Rastogi, editors, {\em
  22nd {ACM} International Conference on Information and Knowledge Management,
  CIKM'13, San Francisco, CA, USA, October 27 - November 1, 2013}, pages
  59--68. {ACM}, 2013.

\bibitem{Salehi:2013}
B.~Salehi and P.~Cook.
\newblock Predicting the compositionality of multiword expressions using
  translations in multiple languages.
\newblock In Diab et~al. \cite{DBLP:conf/starsem/2013}, pages 266--275.

\bibitem{salehi-cook-baldwin:2015:NAACL-HLT}
B.~Salehi, P.~Cook, and T.~Baldwin.
\newblock A word embedding approach to predicting the compositionality of
  multiword expressions.
\newblock In Mihalcea et~al. \cite{DBLP:conf/naacl/2015}, pages 977--983.

\bibitem{Salton:1966}
G.~Salton.
\newblock Automatic phrase matching.
\newblock {\em Readings in Automatic Language Processing}, pages 169--188,
  1966.

\bibitem{SaltonBY82}
G.~Salton, C.~Buckley, and C.~T. Yu.
\newblock An evaluation of term dependence models in information retrieval.
\newblock In {\em {SIGIR}}, pages 151--173, 1982.

\bibitem{ASI:ASI4}
T.~Saracevic and P.~Kantor.
\newblock A study of information seeking and retrieving. iii. searchers,
  searches, and overlap.
\newblock {\em Journal of the American Society for Information Science},
  39(3):197--216, 1988.

\bibitem{SchwarzM11}
J.~Schwarz and M.~R. Morris.
\newblock Augmenting web pages and search results to support credibility
  assessment.
\newblock In D.~S. Tan, S.~Amershi, B.~Begole, W.~A. Kellogg, and M.~Tungare,
  editors, {\em Proceedings of the International Conference on Human Factors in
  Computing Systems, {CHI} 2011, Vancouver, BC, Canada, May 7-12, 2011}, pages
  1245--1254. {ACM}, 2011.

\bibitem{SegataB08}
N.~Segata and E.~Blanzieri.
\newblock Stochastic \emph{pi}-calculus modelling of multisite phosphorylation
  based signaling: The {PHO} pathway in saccharomyces cerevisiae.
\newblock {\em Trans. Computational Systems Biology}, 10:163--196, 2008.

\bibitem{ShanahanQW06}
J.~G. Shanahan, Y.~Qu, and J.~Wiebe.
\newblock {\em Computing Attitude and Affect in Text}.
\newblock Springer, 2006.

\bibitem{ShiN:2008}
L.~Shi, J.~Nie, and G.~Cao.
\newblock Relating dependent indexes using dempster-shafer theory.
\newblock In J.~G. Shanahan, S.~Amer{-}Yahia, I.~Manolescu, Y.~Zhang, D.~A.
  Evans, A.~Kolcz, K.~Choi, and A.~Chowdhury, editors, {\em Proceedings of the
  17th {ACM} Conference on Information and Knowledge Management, {CIKM} 2008,
  Napa Valley, California, USA, October 26-30, 2008}, pages 429--438. {ACM},
  2008.

\bibitem{shin:2000}
H.~Shin and J.~F. Stach.
\newblock Using long runs as predictors of semantic coherence in a partial
  document retrieval system.
\newblock In {\em {NAACL-ANLP} 2000 Workshop: Syntactic and Semantic Complexity
  in Natural Language Processing Systems}, pages 6--13, 2000.

\bibitem{SigmanC:2002}
M.~Sigman and G.~A. Cecchi.
\newblock Global organization of the {WordNet} lexicon.
\newblock {\em Proceedings of the National Academy of Sciences},
  3(99):1742--1747, 2002.

\bibitem{SimouA:2008}
N.~Simou, T.~Athanasiadis, G.~Stoilos, and S.~D. Kollias.
\newblock Image indexing and retrieval using expressive fuzzy description
  logics.
\newblock {\em Signal, Image and Video Processing}, 2(4):321--335, 2008.

\bibitem{Sinclair:1991}
J.~Sinclair.
\newblock {\em {C}orpus, {C}oncordance, {C}ollocation}.
\newblock Oxford University Press, Oxford, 1991.

\bibitem{SinhaP:2009}
S.~Sinha, R.~K. Pan, N.~Yadav, M.~Vahia, and I.~Mahadevan.
\newblock Network analysis reveals structure indicative of syntax in the corpus
  of undeciphered indus civilization inscriptions.
\newblock In {\em Proceedings of the 2009 Workshop on Graph-based Methods for
  Natural Language Processing (TextGraphs-4)}, pages 5--13, Suntec, Singapore,
  August 2009. Association for Computational Linguistics.

\bibitem{SkovL:2008}
M.~Skov, B.~Larsen, and P.~Ingwersen.
\newblock Inter and intra-document contexts applied in polyrepresentation for
  best match {IR}.
\newblock {\em Inf. Process. Manage.}, 44(5):1673--1683, 2008.

\bibitem{Smeaton:1986}
A.~F. Smeaton.
\newblock Incorporating syntactic information into a document retrieval
  strategy: An investigation.
\newblock In L.~R. Bernardi and F.~Rabitti, editors, {\em SIGIR'86, Proceedings
  of the 9th Annual International {ACM} {SIGIR} Conference on Research and
  Development in Information Retrieval, Pisa, Italy, September 8-10, 1986},
  pages 103--113. {ACM}, 1986.

\bibitem{SmeatonK:1988}
A.~F. Smeaton and C.~J. van Rijsbergen.
\newblock Experiment on incorporation syntactic processing of user queries into
  a document retrieval strategy.
\newblock In Y.~Chiaramella, editor, {\em SIGIR'88, Proceedings of the 11th
  Annual International {ACM} {SIGIR} Conference on Research and Development in
  Information Retrieval, Grenoble, France, June 13-15, 1988}, pages 31--51.
  {ACM}, 1988.

\bibitem{SmithD:1985}
F.~J. Smith and K.~Devine.
\newblock Storing and retrieving word phrases.
\newblock {\em Information Processing and Management}, 21(3):215--224, 1985.

\bibitem{SoaresC:2005}
M.~M. Soares, C.~Corso, and L.~S. Lucena.
\newblock Network of syllables in portuguese.
\newblock {\em Physica A: Statistical Mechanics and its Applications},
  355(2-4):678--684, September 2005.

\bibitem{socher-EtAl:2012:EMNLP-CoNLL}
R.~Socher, B.~Huval, C.~D. Manning, and A.~Y. Ng.
\newblock Semantic compositionality through recursive matrix-vector spaces.
\newblock In Tsujii et~al. \cite{DBLP:conf/emnlp/2012}, pages 1201--1211.

\bibitem{SomaN:2009}
S.~Somasundaran, G.~Namata, L.~Getoor, and J.~Wiebe.
\newblock Opinion graphs for polarity and discourse classification.
\newblock In {\em Proceedings of the 2009 Workshop on Graph-based Methods for
  Natural Language Processing (TextGraphs-4)}, pages 66--74, Suntec, Singapore,
  August 2009. Association for Computational Linguistics.

\bibitem{Som09}
S.~Somasundaran, G.~Namata, J.~Wiebe, and L.~Getoor.
\newblock Supervised and unsupervised methods in employing discourse relations
  for improving opinion polarity classification.
\newblock In {\em Proceedings of the 2009 Conference on Empirical Methods in
  Natural Language Processing, {EMNLP} 2009, 6-7 August 2009, Singapore, {A}
  meeting of SIGDAT, a Special Interest Group of the {ACL}}, pages 170--179.
  {ACL}, 2009.

\bibitem{SongC1999}
F.~Song and W.~B. Croft.
\newblock A general language model for information retrieval.
\newblock In {\em Proceedings of the 1999 {ACM} {CIKM} International Conference
  on Information and Knowledge Management, Kansas City, Missouri, USA, November
  2-6, 1999}, pages 316--321. {ACM}, 1999.

\bibitem{DBLP:conf/cikm/SordoniBVLSN15}
A.~Sordoni, Y.~Bengio, H.~Vahabi, C.~Lioma, J.~G. Simonsen, and J.~Nie.
\newblock A hierarchical recurrent encoder-decoder for generative context-aware
  query suggestion.
\newblock In J.~Bailey, A.~Moffat, C.~C. Aggarwal, M.~de~Rijke, R.~Kumar,
  V.~Murdock, T.~K. Sellis, and J.~X. Yu, editors, {\em Proceedings of the 24th
  {ACM} International Conference on Information and Knowledge Management,
  {CIKM} 2015, Melbourne, VIC, Australia, October 19 - 23, 2015}, pages
  553--562. {ACM}, 2015.

\bibitem{SoricutM06a}
R.~Soricut and D.~Marcu.
\newblock Discourse generation using utility-trained coherence models.
\newblock In N.~Calzolari, C.~Cardie, and P.~Isabelle, editors, {\em {ACL}
  2006, 21st International Conference on Computational Linguistics and 44th
  Annual Meeting of the Association for Computational Linguistics, Proceedings
  of the Conference, Sydney, Australia, 17-21 July 2006}. The Association for
  Computer Linguistics, 2006.

\bibitem{Sowa:1984}
J.~F. Sowa.
\newblock {\em {Conceptual Structures: Information Processing in Mind and
  Machine}}.
\newblock Addison-Wesley, 1984.

\bibitem{SrikanthS:2003}
M.~Srikanth and R.~K. Srihari.
\newblock Incorporating query term dependencies in language models for document
  retrieval.
\newblock In {\em {SIGIR} 2003: Proceedings of the 26th Annual International
  {ACM} {SIGIR} Conference on Research and Development in Information
  Retrieval, July 28 - August 1, 2003, Toronto, Canada}, pages 405--406. {ACM},
  2003.

\bibitem{SteyversT:2005}
M.~Steyvers and J.~B. Tenenbaum.
\newblock The large-scale structure of semantic networks: Statistical analyses
  and a model of semantic growth.
\newblock {\em Cognitive Science}, 29(1):41--78, 2005.

\bibitem{Stiles1961}
H.~E. Stiles.
\newblock The association factor in information retrieval.
\newblock {\em Journal of the ACM}, 8(2):271--279, Apr. 1961.

\bibitem{Str1997}
T.~Strzalkowski, F.~Lin, and J.~P. Carballo.
\newblock Natural language information retrieval {TREC-6} report.
\newblock In Voorhees and Harman \cite{DBLP:conf/trec/1997}, pages 347--366.

\bibitem{SuP:2010}
N.~Suwandarathna and U.~Perera.
\newblock Discourse marker based topic identification and search results
  refining.
\newblock In {\em 5thIEEE International Conference on Information and
  Automation for Sustainability {(ICIAfS?10)}, Colombo, Sri Lanka}, pages
  119--125, 2010.

\bibitem{TakamuraI:2007}
H.~Takamura, T.~Inui, and M.~Okumura.
\newblock Extracting semantic orientations of phrases from dictionary.
\newblock In C.~L. Sidner, T.~Schultz, M.~Stone, and C.~Zhai, editors, {\em
  Human Language Technology Conference of the North American Chapter of the
  Association of Computational Linguistics, Proceedings, April 22-27, 2007,
  Rochester, New York, {USA}}, pages 292--299. The Association for
  Computational Linguistics, 2007.

\bibitem{TanGP12}
C.~Tan, E.~Gabrilovich, and B.~Pang.
\newblock To each his own: personalized content selection based on text
  comprehensibility.
\newblock In E.~Adar, J.~Teevan, E.~Agichtein, and Y.~Maarek, editors, {\em
  Proceedings of the Fifth International Conference on Web Search and Web Data
  Mining, {WSDM} 2012, Seattle, WA, USA, February 8-12, 2012}, pages 233--242.
  {ACM}, 2012.

\bibitem{TaoZ07}
T.~Tao and C.~Zhai.
\newblock An exploration of proximity measures in information retrieval.
\newblock In Kraaij et~al. \cite{DBLP:conf/sigir/2007}, pages 295--302.

\bibitem{Tapiero07}
I.~Tapiero.
\newblock {\em Situation Models and Levels of Coherence: Towards a Definition
  of Comprehension}.
\newblock Lawrence Erlbaum Associates, Manwah, New Jersey, 2007.

\bibitem{TeufelM02}
S.~Teufel and M.~Moens.
\newblock Summarizing scientific articles: Experiments with relevance and
  rhetorical status.
\newblock {\em Computational Linguistics}, 28(4):409--445, 2002.

\bibitem{frege}
R.~H. Thomason.
\newblock {\em Formal Philosophy. Selected Papers of Richard Montague}.
\newblock Yale University Press, 1974.

\bibitem{TianOI16}
R.~Tian, N.~Okazaki, and K.~Inui.
\newblock Learning semantically and additively compositional distributional
  representations.
\newblock In {\em Proceedings of the 54th Annual Meeting of the Association for
  Computational Linguistics, {ACL} 2016, August 7-12, 2016, Berlin, Germany,
  Volume 1: Long Papers\/} \cite{DBLP:conf/acl/2016-1}.

\bibitem{Tomlinson:2004}
S.~Tomlinson.
\newblock Robust, web and terabyte retrieval with hummingbird searchserver at
  {TREC} 2004.
\newblock In E.~M. Voorhees and L.~P. Buckland, editors, {\em Proceedings of
  the Thirteenth Text REtrieval Conference, {TREC} 2004, Gaithersburg,
  Maryland, USA, November 16-19, 2004}, volume Special Publication 500-261.
  National Institute of Standards and Technology {(NIST)}, 2004.

\bibitem{Zhai1997}
X.~Tong, C.~Zhai, N.~Milic{-}Frayling, and D.~A. Evans.
\newblock Evaluation of syntactic phrase indexing -- {CLARIT} {NLP} track
  report.
\newblock In E.~M. Voorhees and D.~K. Harman, editors, {\em Proceedings of The
  Fifth Text REtrieval Conference, {TREC} 1996, Gaithersburg, Maryland, USA,
  November 20-22, 1996}, volume Special Publication 500-238. National Institute
  of Standards and Technology {(NIST)}, 1996.

\bibitem{ToutanovaLYPQ16}
K.~Toutanova, V.~Lin, W.~Yih, H.~Poon, and C.~Quirk.
\newblock Compositional learning of embeddings for relation paths in knowledge
  base and text.
\newblock In {\em Proceedings of the 54th Annual Meeting of the Association for
  Computational Linguistics, {ACL} 2016, August 7-12, 2016, Berlin, Germany,
  Volume 1: Long Papers\/} \cite{DBLP:conf/acl/2016-1}.

\bibitem{TsikrikaL:2004}
T.~Tsikrika and M.~Lalmas.
\newblock Combining evidence for web retrieval using the inference network
  model: an experimental study.
\newblock {\em Inf. Process. Manage.}, 40(5):751--772, 2004.

\bibitem{DBLP:conf/emnlp/2012}
J.~Tsujii, J.~Henderson, and M.~Pasca, editors.
\newblock {\em Proceedings of the 2012 Joint Conference on Empirical Methods in
  Natural Language Processing and Computational Natural Language Learning,
  EMNLP-CoNLL 2012, July 12-14, 2012, Jeju Island, Korea}. {ACL}, 2012.

\bibitem{TurtleC91}
H.~R. Turtle and W.~B. Croft.
\newblock Evaluation of an inference network-based retrieval model.
\newblock {\em {ACM} Transactions on Information Systems}, 9(3):187--222, 1991.

\bibitem{keith:1986}
C.~J. van Rijsbergen.
\newblock A non-classical logic for information retrieval.
\newblock {\em Comput. J.}, 29(6):481--485, 1986.

\bibitem{keith:2004}
C.~J. van Rijsbergen.
\newblock {\em The geometry of information retrieval}.
\newblock Cambridge University Press, 2004.

\bibitem{K1977}
C.~J.~K. van Rijsbergen.
\newblock A theoretical basis for the use of co-occurrence data in information
  retrieval.
\newblock {\em J. Doc.}, 33:106--119, 1977.

\bibitem{keithC:1998}
C.~J.~K. van Rijsbergen, F.~Crestani, and M.~Lalmas.
\newblock {\em {Information Retrieval: Uncertainty and Logics}}.
\newblock Springer, 1998.

\bibitem{VeronisI:1990}
J.~V{\'{e}}ronis and N.~Ide.
\newblock Word sense disambiguation with very large neural networks extracted
  from machine readable dictionaries.
\newblock In {\em 13th International Conference on Computational Linguistics,
  {COLING} 1990, University of Helsinki, Finland, August 20-25, 1990}, pages
  389--394, 1990.

\bibitem{VitevitchR:2005}
M.~S. Vitevitch and E.~Rodriguez.
\newblock Neighborhood density effects in spoken word recognition in spanish.
\newblock {\em J. Multilingual Communication Disorders}, 3:64--73, 2005.

\bibitem{DBLP:conf/trec/1997}
E.~M. Voorhees and D.~K. Harman, editors.
\newblock {\em Proceedings of The Sixth Text REtrieval Conference, {TREC} 1997,
  Gaithersburg, Maryland, USA, November 19-21, 1997}, volume Special
  Publication 500-240. National Institute of Standards and Technology {(NIST)},
  1997.

\bibitem{VoskaridesMTRW15}
N.~Voskarides, E.~Meij, M.~Tsagkias, M.~de~Rijke, and W.~Weerkamp.
\newblock Learning to explain entity relationships in knowledge graphs.
\newblock In {\em Proceedings of the 53rd Annual Meeting of the Association for
  Computational Linguistics and the 7th International Joint Conference on
  Natural Language Processing of the Asian Federation of Natural Language
  Processing, {ACL} 2015, July 26-31, 2015, Beijing, China, Volume 1: Long
  Papers\/} \cite{DBLP:conf/acl/2015-1}, pages 564--574.

\bibitem{WangLWK06}
D.~Y. Wang, R.~W.~P. Luk, K.~Wong, and K.-L. Kwok.
\newblock An information retrieval approach based on discourse type.
\newblock In C.~Kop, G.~Fliedl, H.~C. Mayr, and E.~M{\'{e}}tais, editors, {\em
  Natural Language Processing and Information Systems, 11th International
  Conference on Applications of Natural Language to Information Systems, {NLDB}
  2006, Klagenfurt, Austria, May 31 - June 2, 2006, Proceeding}, volume 3999 of
  {\em Lecture Notes in Computer Science}, pages 197--202. Springer, 2006.

\bibitem{WangLKNB11}
L.~Wang, M.~Lui, S.~N. Kim, J.~Nivre, and T.~Baldwin.
\newblock Predicting thread discourse structure over technical web forums.
\newblock In {\em Proceedings of the 2011 Conference on Empirical Methods in
  Natural Language Processing, {EMNLP} 2011, 27-31 July 2011, John McIntyre
  Conference Centre, Edinburgh, UK, {A} meeting of SIGDAT, a Special Interest
  Group of the {ACL}\/} \cite{DBLP:conf/emnlp/2011}, pages 13--25.

\bibitem{LinS:1991}
E.~B. Wendlandt and J.~R. Driscoll.
\newblock Incorporating a semantic analysis into a document retrieval strategy.
\newblock In Bookstein et~al. \cite{DBLP:conf/sigir/91}, pages 270--279.

\bibitem{WiddowsD:2002}
D.~Widdows and B.~Dorow.
\newblock A graph model for unsupervised lexical acquisition.
\newblock In {\em 19th International Conference on Computational Linguistics,
  {COLING} 2002, Howard International House and Academia Sinica, Taipei,
  Taiwan, August 24 - September 1, 2002}, 2002.

\bibitem{WiebeR11}
J.~Wiebe and E.~Riloff.
\newblock Finding mutual benefit between subjectivity analysis and information
  extraction.
\newblock {\em {IEEE} Trans. Affective Computing}, 2(4):175--191, 2011.

\bibitem{wiebe05}
J.~Wiebe, T.~Wilson, and C.~Cardie.
\newblock Annotating expressions of opinions and emotions in language.
\newblock {\em Language Resources and Evaluation}, 39(2-3):165--210, 2005.

\bibitem{WilkinsonH:1991}
R.~Wilkinson and P.~Hingston.
\newblock Using the cosine measure in a neural network for document.
\newblock In Bookstein et~al. \cite{DBLP:conf/sigir/91}, pages 202--210.

\bibitem{WilsonHSKWCCRP05}
T.~Wilson, P.~Hoffmann, S.~Somasundaran, J.~Kessler, J.~Wiebe, Y.~Choi,
  C.~Cardie, E.~Riloff, and S.~Patwardhan.
\newblock Opinionfinder: {A} system for subjectivity analysis.
\newblock In {\em {HLT/EMNLP} 2005, Human Language Technology Conference and
  Conference on Empirical Methods in Natural Language Processing, Proceedings
  of the Conference, 6-8 October 2005, Vancouver, British Columbia, Canada\/}
  \cite{DBLP:conf/naacl/2005}.

\bibitem{Wittek:2009:OTB:1693756.1693780}
P.~Wittek, S.~Dar\'{a}nyi, and C.~L. Tan.
\newblock An ordering of terms based on semantic relatedness.
\newblock In {\em Proceedings of the Eighth International Conference on
  Computational Semantics}, IWCS-8 '09, pages 235--247, Stroudsburg, PA, USA,
  2009. Association for Computational Linguistics.

\bibitem{xiong:2013}
D.~Xiong, Y.~Ding, M.~Zhang, and C.~L. Tan.
\newblock Lexical chain based cohesion models for document-level statistical
  machine translation.
\newblock In {\em Proceedings of the 2013 Conference on Empirical Methods in
  Natural Language Processing, {EMNLP} 2013, 18-21 October 2013, Grand Hyatt
  Seattle, Seattle, Washington, USA, {A} meeting of SIGDAT, a Special Interest
  Group of the {ACL}\/} \cite{DBLP:conf/emnlp/2013}, pages 1563--1573.

\bibitem{Xiong:2015}
D.~Xiong, M.~Zhang, and X.~Wang.
\newblock Topic-based coherence modeling for statistical machine translation.
\newblock {\em Trans. Audio, Speech and Lang. Proc.}, 23(3):483--493, Mar.
  2015.

\bibitem{yazdani-farahmand-henderson:2015:EMNLP}
M.~Yazdani, M.~Farahmand, and J.~Henderson.
\newblock Learning semantic composition to detect non-compositionality of
  multiword expressions.
\newblock In M{\`{a}}rquez et~al. \cite{DBLP:conf/emnlp/2015}, pages
  1733--1742.

\bibitem{Yes10}
A.~Yessenalina, Y.~Yue, and C.~Cardie.
\newblock Multi-level structured models for document-level sentiment
  classification.
\newblock In {\em Proceedings of the 2010 Conference on Empirical Methods in
  Natural Language Processing, {EMNLP} 2010, 9-11 October 2010, {MIT} Stata
  Center, Massachusetts, USA, {A} meeting of SIGDAT, a Special Interest Group
  of the {ACL}}, pages 1046--1056. {ACL}, 2010.

\bibitem{Yom-TovF:2005}
E.~Yom{-}Tov, S.~Fine, D.~Carmel, and A.~Darlow.
\newblock Learning to estimate query difficulty: including applications to
  missing content detection and distributed information retrieval.
\newblock In Baeza{-}Yates et~al. \cite{DBLP:conf/sigir/2005}, pages 512--519.

\bibitem{Yu1983}
C.~T. Yu, C.~Buckley, K.~Lam, and G.~Salton.
\newblock A generalised term dependence model in {IR}.
\newblock {\em Information Technology: R\&D}, 2:129--154, 1983.

\bibitem{YuW:2009}
L.-C. Yu, C.-H. Wu, and F.-L. Jang.
\newblock Psychiatric document retrieval using a discourse-aware model.
\newblock {\em Artif. Intell.}, 173:817--829, May 2009.

\bibitem{Zellhofer15}
D.~Zellh{\"{o}}fer.
\newblock Predicting relevance feedback effectiveness with the help of the
  principle of polyrepresentation in {MIR}.
\newblock In Allan et~al. \cite{DBLP:conf/ictir/2015}, pages 345--348.

\bibitem{ZhangFQHLH15}
M.~Zhang, V.~W. Feng, B.~Qin, G.~Hirst, T.~Liu, and J.~Huang.
\newblock Encoding world knowledge in the evaluation of local coherence.
\newblock In Mihalcea et~al. \cite{DBLP:conf/naacl/2015}, pages 1087--1096.

\bibitem{zhang:2011}
R.~Zhang.
\newblock Sentence ordering driven by local and global coherence for summary
  generation.
\newblock In {\em The 49th Annual Meeting of the Association for Computational
  Linguistics: Human Language Technologies, Proceedings of the Conference,
  19-24 June, 2011, Portland, Oregon, {USA} - Student Session}, pages 6--11.
  The Association for Computer Linguistics, 2011.

\bibitem{ZhangLR16}
R.~Zhang, H.~Lee, and D.~R. Radev.
\newblock Dependency sensitive convolutional neural networks for modeling
  sentences and documents.
\newblock In Knight et~al. \cite{DBLP:conf/naacl/2016}, pages 1512--1521.

\bibitem{ZhouLGWW11}
L.~Zhou, B.~Li, W.~Gao, Z.~Wei, and K.~Wong.
\newblock Unsupervised discovery of discourse relations for eliminating
  intra-sentence polarity ambiguities.
\newblock In {\em Proceedings of the 2011 Conference on Empirical Methods in
  Natural Language Processing, {EMNLP} 2011, 27-31 July 2011, John McIntyre
  Conference Centre, Edinburgh, UK, {A} meeting of SIGDAT, a Special Interest
  Group of the {ACL}\/} \cite{DBLP:conf/emnlp/2011}, pages 162--171.

\bibitem{ZhouC05}
Y.~Zhou and W.~B. Croft.
\newblock Document quality models for web ad hoc retrieval.
\newblock In O.~Herzog, H.~Schek, N.~Fuhr, A.~Chowdhury, and W.~Teiken,
  editors, {\em Proceedings of the 2005 {ACM} {CIKM} International Conference
  on Information and Knowledge Management, Bremen, Germany, October 31 -
  November 5, 2005}, pages 331--332. {ACM}, 2005.

\bibitem{ZhuG00}
X.~Zhu and S.~Gauch.
\newblock Incorporating quality metrics in centralized/distributed information
  retrieval on the world wide web.
\newblock In {\em {SIGIR}}, pages 288--295, 2000.

\bibitem{ZhuSG16}
X.~Zhu, P.~Sobhani, and H.~Guo.
\newblock Dag-structured long short-term memory for semantic compositionality.
\newblock In Knight et~al. \cite{DBLP:conf/naacl/2016}, pages 917--926.

\bibitem{ZucconA:2008}
G.~Zuccon, L.~Azzopardi, and C.~J. van Rijsbergen.
\newblock A formalization of logical imaging for information retrieval using
  quantum theory.
\newblock In {\em 19th International Workshop on Database and Expert Systems
  Applications {(DEXA} 2008), 1-5 September 2008, Turin, Italy}, pages 3--8.
  IEEE Computer Society, 2008.

\end{thebibliography}

\end{document}